\documentclass[11pt]{article}

\usepackage{amsmath, amsfonts, amsthm, amssymb, graphicx}
\usepackage{xspace}
\usepackage{psfrag}
\usepackage[dvips]{color}
\usepackage{hyperref}
\numberwithin{equation}{subsection}

\newcommand{\Complex}{\ensuremath{\mathbb{C}}\xspace}
\newcommand{\Real}{\ensuremath{\mathbb{R}}\xspace}

\newcommand{\op}{\ensuremath{\mathcal{O}}\xspace}
\newcommand{\rpp}{r_+}
\newcommand{\rmm}{r_-}

\newcommand{\fdel}[2][]{\ensuremath{\frac{\delta #1}{\delta #2}}}
\newcommand{\bra}[1]{\ensuremath{\langle #1 |}\xspace}
\newcommand{\ket}[1]{\ensuremath{| #1 \rangle}\xspace}
\newcommand{\inprod}[2]{\ensuremath{\langle #1 | #2 \rangle}\xspace}
\newcommand{\vev}[1]{\ensuremath{\langle #1 \rangle}\xspace}

\def\cN{{\cal N}}

\def\cL{{\cal L}}

\def\0{{\sst{(0)}}}
\def\1{{\sst{(1)}}}
\def\2{{\sst{(2)}}}
\def\3{{\sst{(3)}}}
\def\4{{\sst{(4)}}}
\def\5{{\sst{(5)}}}
\def\6{{\sst{(6)}}}
\def\7{{\sst{(7)}}}
\def\8{{\sst{(8)}}}

 \let\b=\beta

   \let\f=\phi  
  \let\D=\Delta

\def\ds{\documentstyle}

\let\na=\nabla
 
\newcommand{\be}{\begin{equation}}
\newcommand{\ee}{\end{equation}}
\def\ba{\begin{array}}
\def\ea{\end{array}}

\def\sst#1{{\scriptscriptstyle #1}}
 
\def\ie{{\it i.e.\ }}

\def\<{ \langle }
\def\>{ \rangle }

\def\cD{{\cal D}}

\def\cK{{\cal K}}
\def\cO{{\cal O}}

\def\cT{{\cal T}}

\def\hh{H}

\newcommand{\bea}{\begin{eqnarray}}
\newcommand{\eea}{\end{eqnarray}}

\def\Tr{{\rm Tr}}

\usepackage{fancyhdr}

\begin{document}

\title{Real-time gauge/gravity duality: \\
Prescription, Renormalization and Examples}
\author{Kostas Skenderis and Balt C. van Rees\\
\it Institute for Theoretical Physics,
University of Amsterdam, \\
Valckenierstraat 65, 1018 XE Amsterdam, The Netherlands
}
\date{\today}
\maketitle

\thispagestyle{fancy}
\vskip 6em
\begin{abstract}
We present a comprehensive analysis of the
prescription we recently put forward for the computation
of real-time correlation functions
using gauge/gravity duality. The
prescription is valid for any holographic supergravity background and it
naturally maps initial and final data in the bulk to initial and
final states or density matrices in the field theory. We show in
detail how the technique of holographic renormalization can be
applied in this setting and we provide numerous illustrative
examples, including the computation of time-ordered, Wightman and
retarded 2-point functions in Poincar\'e and global coordinates,
thermal correlators and higher-point functions.
\end{abstract}

\newpage
\tableofcontents
\newpage
\section{Introduction}
Since the advent of the AdS/CFT correspondence
\cite{Maldacena:1997re,Gubser:1998bc,Witten:1998qj}
considerable amount of work has been devoted to developing
holographic dualities leading to a very precise understanding of
the holographic dictionary in Euclidean signature,
see \cite{Aharony:1999ti,D'Hoker:2002aw,Skenderis:2002wp} for reviews.
In this paper, continuing our recent work \cite{us},
we aim at developing a real-time prescription
that is applicable at the same level of generality
as the corresponding Euclidean
prescription. More precisely, we would like to have a setup that is
valid for all QFTs that have a holographic dual and is
applicable for the holographic computation of
$n$-point functions of gauge invariant operators in non-trivial states.
Furthermore, this prescription should be fully
holographic, \ie only boundary data and regularity in the interior
should be needed for the computation, and within the supergravity approximation
all information should be encoded in classical bulk dynamics.
Such a prescription is an integral part of the definition
of the holographic correspondence and as such it is important on general
grounds. Furthermore, there is a wide range of applications
for such a general real-time prescription.
To mention a few:
one would like to understand better
holography for time dependent backgrounds, to have a holographic
description of non-equilibrium QFT and to be able to compute
correlators in non-trivial states. Such a development is also becoming
urgent as potential current and future applications of holography to
modelling of the
quark-gluon plasma in RHIC and LHC require real-time techniques,
see \cite{Son:2007vk} for a review.

From a more theoretical perspective, one would like to understand better
the interplay between causality and holography. Since
bulk and boundary lightcones are different, it is not {\it a priori}
clear that a bulk computation will produce the correct causal
structure for boundary correlators, for example the correct $i \epsilon$
insertions. Conversely, one can ask how the bulk causal structure emerges
from boundary correlators. A related question is to understand how
black hole horizons are encoded in boundary correlators. More
generally one would like to study holographically the process
of gravitational collapse. All of these applications
require a formalism that can handle the general case,
rather than being tied up to particularities of specific examples.

In many applications, the Euclidean holographic techniques for obtaining correlators are sufficient, since one can often analytically continue from Lorentzian to Euclidean signature.
While such a Wick rotation is often the most direct way
of arriving at the result, there are also many cases
where the analytic continuation is technically difficult,
even though possible in principle.
For example, in the case of a thermal correlator one would need
to continue from a discrete set of Matsubara frequencies
which is technically not very easy. More importantly, the Wick rotation
obscures the bulk and boundary dynamics and none of the questions raised
above can be answered in this setting.

There have been several earlier works discussing holography in Lorentzian
signature, including \cite{Balasubramanian:1998sn,Balasubramanian:1998de,
Maldacena:2001kr,Son:2002sd, Herzog:2002pc,Satoh:2002bc,
Kraus:2002iv,Marolf:2004fy,Festuccia:2005pi,Lawrence:2006ze,Iqbal:2008by}. One set of these papers is based
on semi-classical  bulk quantization around a classical bulk solution.
For example, a case often  discussed is that of the quantization
of a free scalar field in AdS and the computation of the associated
boundary 2-point function.
Such results are clearly difficult to
extend to cases where bulk interactions are essential because of the
difficulty in quantizing the bulk gravitational theory. For example,
higher point functions, correlators of the stress energy
tensor and holographic RG flows are outside the remit of
these works. Moreover, for the computation of the correlators
in the large $N$ and strong 't Hooft limit, one should
not have to consider the quantization of the bulk theory at all -- classical
bulk dynamics should suffice.

A Lorentzian prescription that has been used widely in the literature
is that of Son and Starinets \cite{Son:2002sd}. This prescription
leads to the computation of retarded 2-point functions and is based
on imposing specific boundary conditions in the interior of the spacetime.
More specifically, this prescription assumes that one deals with a
spacetime with a horizon and imposes incoming-wave boundary conditions
at the horizon.
It was later shown
in \cite{Herzog:2002pc} that these conditions
are related to boundary conditions discussed earlier in
the black hole literature \cite{Hartle:1976tp,Unruh:1976db}
and that (with such boundary conditions understood) the prescription
follows from taking functional derivatives of the on-shell action,
although the authors did not take into account contributions to the
on-shell action from timelike infinity, which are generically non-zero.

This prescription leads to correct results (provided the infinities
have been subtracted correctly, see below). It is somewhat unsatisfactory,
however, from the holographic point of view, at least in
view of the general applications we have in mind. For holography
one would want to have all information encoded in boundary data,
so that boundary data is sufficient on its own to reconstruct the
bulk dynamics. The prescription in \cite{Son:2002sd,Herzog:2002pc}, 
on the other hand,
presumes the existence of a horizon and uses specific behavior of the
bulk fields there. Furthermore, as mentioned above there are additional
contributions in the on-shell action from initial and final surfaces
within the setup of \cite{Son:2002sd,Herzog:2002pc}.
We will see that these additional terms cancel in our prescription
and we will also derive the behavior of the fields at the horizon
used in \cite{Son:2002sd,Herzog:2002pc} from a fully holographic
prescription. 

Another issue that has not been discussed adequately in the past
is that of renormalization. First, there are the infinities
due to the non-compactness of the bulk spacetime
in the radial direction. These are the same
infinities appearing in the Euclidean setup and can be dealt
with by a straightforward adaptation of the Euclidean discussion,
namely by introducing boundary counterterms etc.,
see \cite{Skenderis:2002wp} for a review. 
Such counterterms
not only remove infinities but also
in general affect the finite part of correlators,
even in such simple cases as the 2-point functions of scalar operators.
For example, the naive computation of 2-point functions of scalar operators
of scaling dimension $\D \neq d$, 
where the infinite terms are simply dropped,
leads to results that are inconsistent with
Ward identities,  see
\cite{deHaro:2000xn,Bianchi:2001de,Bianchi:2001kw,Skenderis:2002wp,Karch:2005ms,Skenderis:2006uy}
for examples and discussions of this point.
A second issue that is specific to the Lorentzian setup is that there may
also be new infinities due to the non-compactness in the time direction
and one would also have to understand how to deal with those.

The main difference between the Euclidean and the Lorentzian cases is that in the latter case one also has to specify initial and final conditions for the bulk fields. It has long been appreciated
\cite{Balasubramanian:1998sn} that these conditions should be related
to a choice of in- and out-state in the Lorentzian boundary QFT, but the precise relation was given only recently in \cite{us}, which we now review.
The starting point in \cite{us} was the fact that in QFT the
initial and final conditions can be implemented using
a time contour in a complex time plane.
For example, one can compute the expectation values in a non-trivial state
in the `in-in' formalism by choosing a closed time contour that
starts from the operator that creates the state,
runs along the real time axis and returns to the operator
\cite{Schwinger:1960qe,Bakshi:1962dv,Bakshi:1963bn,Keldysh:1964ud}.
Thermal correlators can be obtained by having the
contour run also along the imaginary axis.

Since the gauge/gravity duality is believed to be an exact equivalence,
one should be able to holographically compute
correlators in non-trivial states and the choice of
a contour in the complex time plane should be reflected on the
gravitational side too.
The prescription of \cite{us} is therefore to start from the QFT contour
and `fill it in'  with a bulk manifold. Real
segments of the contour are associated with Lorentzian spacetimes,
and purely imaginary segments with Euclidean
solutions\footnote{With Euclidean solutions we mean a solution of
the field equations after Wick rotation to positive definite signature.
If the solution is real, it should be more properly called
`Riemannian'. We will see in examples, however, that the `Euclidean'
solutions  can also be complex.}.
The Euclidean bulk solution which is associated with
the initial state on the QFT side can also be thought
of as providing a Hartle-Hawking wave function \cite{Hartle:1983ai}
for the bulk theory. Thus our prescription is not
only QFT inspired but also in line  with standard considerations
on wave functions in quantum gravity, see also
\cite{Maldacena:2001kr,Marolf:2004fy} for related discussions.
There has been considerable discussion in the literature
over the choice of contour in the Euclidean
path integrals and
the reality conditions of the the semi-classical saddle point evaluation,
see for example
\cite{Halliwell:1989dy}.
In our case, the bulk reality conditions are dictated by
the boundary theory and, in particular, for a generic complex boundary contour
the bulk manifold would have a complex metric (but in all cases the
boundary correlators would satisfy standard reality conditions).

In this paper we extend and further develop the framework
discussed in \cite{us} by presenting a comprehensive discussion of all
issues involved. Although the discussion below is meant to be
self-contained, to fix ideas it may be helpful to read the concrete
example presented in \cite{us} first. The organization of this 
paper is the following. In the next section we present the
general prescription in detail.  Holographic renormalization is
discussed in section \ref{ren}. This section contains important but
rather technical results and can be skipped on a first reading. In
section \ref{examples}, we discuss a range of different examples that
complement the example of \cite{us}. In particular, we discuss the
holographic computation of a Wightman function,
compute a real-time
two-point function in thermal AdS, in eternal BTZ and in AdS in
Poincar\'{e} coordinates.
We also illustrate how to
compute higher-point functions
and discuss the prescription for rotating black holes.
Finally, we summarize
some relevant background QFT material in the appendix.

\section{Real-time prescription} \label{prescription}
In this section we discuss the real-time gauge/gravity prescription in detail. We begin with a discussion of basic QFT results that will be the springboard for the holographic prescription in subsection \ref{sec:prescription2}.

\subsection{QFT preliminaries}
\begin{figure}
\centering
\psfrag{T}{$T$}
\psfrag{-T}{$-T$}
\psfrag{t}{$t$}
\psfrag{C0}{$C_0$}
\psfrag{C1}{$C_1$}
\psfrag{C2}{$C_2$}
\psfrag{C3}{$C_3$}
\psfrag{a}{(a)}
\psfrag{b}{(b)}
\psfrag{c}{(c)}
\includegraphics[width=12cm]{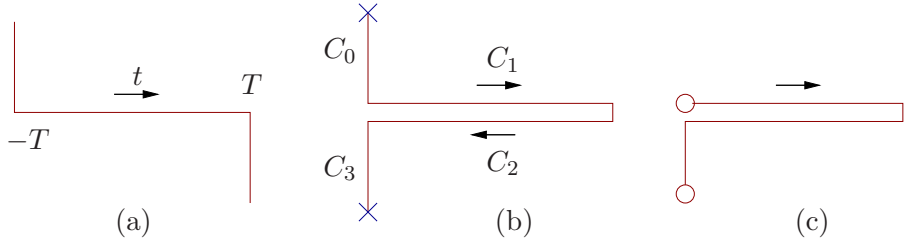}
\caption{\label{fig:contour}(a) Vacuum-to-vacuum contour. (b) In-in contour. (c) Real-time thermal contour.}
\end{figure}

Consider a field configuration with initial condition $\phi_-(\vec{x})$ at
$t=-T$ and final condition $\phi_+(\vec{x})$ at $t=T$. The path integral
with fields constrained to satisfy these conditions produces the
transition amplitude $\langle \phi_+,T|\phi_-, -T\rangle$.
If we are interested in vacuum amplitudes we should multiply this
expression by the vacuum wave functions $\langle 0 |\phi_+, T\rangle$
and $\langle \phi_-,-T| 0\rangle$ and integrate over $\phi_+,\phi_-$.
The insertion of these wave functions is equivalent to extending the
fields in the path integral to live along a contour in the complex time
plane as sketched in Fig.~\ref{fig:contour}a. Indeed, the infinite
vertical segment starting at $-T$ corresponds to a transition amplitude
$\lim_{\beta \to \infty} \langle \phi_-,-T | e^{-\beta H} | \Psi \rangle$
for some state $|\Psi\rangle$, which is however irrelevant since taking the
limit projects it onto the vacuum wave function $\langle \phi_-,-T | 0\rangle$. Similarly, we obtain $\langle 0 | \phi_+, T\rangle$ from the vertical
segment starting at $t=T$. As is reviewed in appendix \ref{sec:app:fieldthy},
these wave function insertions ultimately lead to the
$i \epsilon$ factors in the Feynman propagators.

In this discussion, we used the Euclidean path integral to create the
vacuum state which is then fed into the Lorentzian path integral
as the initial and final state. More generally, one can
use the Euclidean path integral to generate other states that can serve
as initial/final states for the Lorentzian path integral.
In the context of a conformal field theory on $\Real ^d$ the relation between
Euclidean path integrals and states is the basis for
the operator-state correspondence: inserting a local operator $\cO$,
say at the origin of $\Real^{d}$, and then performing the path
integral over the interior of the sphere $S^{d-1}$
that surrounds the origin results in the corresponding quantum
state $|\Psi_{\cO} \>$ on $S^{d-1}$. In particular, the vacuum
state is generated by inserting the identity operator.

Suppose now that we are interested in computing real-time 
correlation functions in a given initial state,
$\< \Psi| \cO_1(x_1) \cdots \cO_n(x_n) |\Psi \>$.
One can do this by using a closed time
contour \cite{Schwinger:1960qe,Bakshi:1962dv,Bakshi:1963bn,Keldysh:1964ud},
as sketched in Fig.~\ref{fig:contour}b.
In the figure, the vertical pieces $C_0, C_3$ represent Euclidean
path integrals, with the crosses representing operator insertions. 
As described in the
previous paragraph, these segments create the chosen initial state $|\Psi \>$.
We then evolve this state forward and backward in time following the
horizontal segments $C_1$ and $C_2$.
To compute real-time expectation values, one
may insert operators either in $C_1$
or $C_2$. For example, one can have 2-point functions with
operators inserted both in $C_1$, or one in $C_1$ and one in $C_2$, or
both in $C_2$. This leads to
a $2 \times 2$ matrix of 2-point functions, the Schwinger-Keldysh
propagators, discussed in more detail in appendix \ref{sec:app:fieldthy}.

For real-time thermal correlators one can use the closed time path
contour in Fig.~\ref{fig:contour}c. The vertical segment now
represents the thermal density matrix, $\hat \rho = \exp(-\beta \hat
H)$, with $\beta = 1/T$ and $\hat H$ the Hamiltonian. The circles
indicate points that should be identified and reflect the fact that
thermal correlators satisfy appropriate periodicity conditions
in imaginary time
(bosonic/fermionic correlators are periodic/antiperiodic).
As in the discussion in the previous paragraph, one
can insert operators at any point in the horizontal segments. Other
density matrices, for example a thermal density matrix with
chemical potentials, may be obtained in a similar manner. 

For all of the contours above, one can write a generating functional
of correlation functions of gauge invariant operators in non-trivial states
with the following path integral representation:
\be \label{zqft}
Z_{QFT}[\f^I_{(0)}; C] = \int_C [\cD \varphi] \exp \left( - i \int_{C}dt \int d^{d-1}x \sqrt{-g_{(0)}}
\left(\cL_{QFT}[\varphi] + \f^I_{(0)} \cO_I[\varphi] \right) \right)
\, .
\ee
Here $\varphi$ denotes collectively all QFT fields, $\f^I_{(0)}$ are sources that couple to
gauge invariant operators $\cO^I$, $g_{(0)ij}$ is the spacetime metric (and also the source for the stress energy tensor $T_{ij}$). The path integral is performed for fields living on the contour $C$ in the complex time plane. Therefore, we think of $t$ as a \emph{complex} coordinate and $\int_C dt$ is then a contour integral. By the usual `slicing' arguments in deriving a path integral, one obtains {\it contour-time-ordered} correlators after functionally differentiating w.r.t. sources.

Let us exemplify (\ref{zqft}) using the contour in Fig. \ref{fig:contour}b.
In this case the contour has four segments and the path integral splits into four corresponding parts.
The segments $C_0, C_3$ are associated with Euclidean path integrals,
as discussed above.
In these segments, we can parametrize the contour
using $t=- i \tau$ with $\tau$ a real coordinate along $C$.
This substitution leads to the usual signs in the Euclidean path integral.
The source terms in these segments are related to the choice of
initial and final states. The segments $C_1$ and $C_2$ form a closed time path.
Let us parametrize this path using a contour time coordinate $t_c$ that
increases monotonically along the contour.
In the segment $C_1$ we can simply set $t=t_c$,
where now $t_c$ ranges from 0 to $T$ (where $T$ may be $\infty$),
and we can integrate along $C_2$ using $t =2 T-t_c$, with $T < t_c < 2T$.
(Notice that $dt = -dt_c$ on $C_2$, giving rise to an important extra sign
for the action on this segment.)
The source terms in these segments are the usual sources, which upon functional differentiation
lead to contour-time-ordered $n$-point functions. For operator insertions on $C_1$ and $C_2$
contour-time-ordering coincides with $t_c$-time-ordering, see appendix \ref{sec:app:fieldthy} 
for further discussion. 

Notice that in quantum field theory, one may consider deforming the
contour in the complex time plane in any direction. In fact, one may
deform it into any other direction in complex coordinate
space. Such deformations are allowed, provided the contour does not
run upward in the complex time plane (so that the path integral converges)
and similar restrictions apply for deformations in complex coordinate space. In
general, the `metric' along such a deformed contour would be complex,
which may lead to a complex bulk metric as well. We will consider
such an example when we analyze the rotating BTZ black hole
in section \ref{sec:rotatingbh}.

\subsection{Prescription}
\label{sec:prescription2}
We are now ready to present the real-time gauge/gravity prescription.
Our starting point is that
the contour dependence we discussed in the previous section
should be reflected in the bulk string theory, and in the low
energy approximation it should be part of the supergravity
description. Within the saddle-point approximation, our prescription
is to associate supergravity solutions with QFT
contours, or, more figuratively, to `fill in' the QFT contour with a
bulk solution. We have sketched several examples of such a construction in Fig.~\ref{fig:contoursintplane} on page \pageref{fig:contoursintplane}.

One can think of the field theory contour $C$ as a
$d$-dimensional subspace of a complexified boundary spacetime. In most
cases, as we saw above,
this would be a line in the complexified time plane times a real
space, $\Real \times \Sigma^{d-1}$. The bulk solution should
have $C$ as its conformal boundary and the bulk fields $\Phi^I$ should
satisfy boundary conditions parametrized by fields $\phi^I_{(0)}(x)$
living on $C$. This means that horizontal segments of $C$ will be
filled in with Lorentzian solutions, while vertical segments
will be filled in with
Euclidean solutions. These segments are then glued together along bulk
hypersurfaces that end on the corners of the contour. The total
manifold consisting of all these segments is denoted by $M_C$ and it
has a metric whose signature jumps at the `corner'
hypersurfaces where a vertical segment meets with a horizontal one.
Below we show how appropriate \emph{matching
  conditions} control the behavior of the fields at these
hypersurfaces.

Note that the bulk manifold is \emph{not} necessarily of the form
$\Real \times X^d$ with $\partial X^d = \Sigma^{d-1}$. Instead, we can
have more general bulk solutions that may `interpolate' between
various parts of the contour. An important example is the eternal BTZ
black hole we consider below.

Given such a solution $M_C$ that fills in the entire field theory contour $C$,
the next step in the prescription is to compute the corresponding on-shell supergravity action. This action is then identified with the generating functional of correlators
in non-trivial states discussed in the previous subsection,
\be
\label{eq:zz}
Z_{QFT}[\phi^I_{(0)};C] =\exp \left( i \int_{M_C}d^{d{+}1}x \sqrt{-G} \cL^{{\rm on-shell}}_{\rm bulk}
  [\phi_{(0)}^I] \right)\, .
\ee
Vertical segments of the contour involve
the Euclidean on-shell action and horizontal segments the Lorentzian
on-shell action, with factors of $i$ and signs becoming standard when
one passes from the complex coordinate $t$ to the corresponding real
contour time variable $\tau$ or $t_c$.  Notice that this discussion
does not require that  $t_c$ and $\tau$ extend globally on $M$, as the asymptotic analysis suffices to fix all signs.
The sources $\phi_{(0)}^I$ that are localized in the
conformal boundary of the Euclidean part of the solution are
associated with the initial and final state,
whereas sources on the conformal boundary of the Lorentzian solution
lead to $n$-point functions upon differentiation. Note that
(\ref{eq:zz}) is a bare relation, as both sides diverge. The
holographic renormalization needed to render the on-shell
supergravity action finite will be described in section \ref{ren}.

\subsubsection{Corners}
\label{sec:corners}
Piecewise straight contours have corners, where either a
  horizontal and a vertical segment meet or two horizontal
  segments join. These corners extend to hypersurfaces $S$ in the
  bulk. The signature of the metric changes at the hypersurface
corresponding to a corner of a horizontal and a vertical segment, but
otherwise it remains unchanged.
Modulo subtleties at the boundary of $S$, which we discuss in the next section, we impose
the following two matching conditions at $S$:
\begin{enumerate}
\item We impose continuity of the fields across $S$. That is, we require the induced metric, the values of the scalars, and induced values of the other fields to be continuous;
\item If the contour passes from a segment $M_-$ to $M_+$, then
we impose appropriate continuity of the conjugate momenta across $S$:
\begin{equation}
\label{eq:2ndmatching}
\pi_- = \eta \pi_+\,,
\end{equation}
\end{enumerate}
where $\pi_\pm$ denote collectively the conjugate momenta of all
fields on the two sides $M_\pm$ of $S$ (defined using $t_c$ or $\tau$ as the
time coordinate), and $\eta = -i$
when we consider a Euclidean to a Lorentzian corner like for
example from $C_0$ to
$C_1$ in Fig. \ref{fig:contour}b, whereas $\eta=-1$ if we have a
(non-trivial) Lorentzian to Lorentzian corner as from $C_1$ to $C_2$
in Fig. \ref{fig:contour}b. In all cases, the matching condition is
equivalent to
\be
\hat{\pi}_+ = \hat{\pi}_-\,,
\ee
where $\hat{\pi}$ is defined using the complex time variable $t$.
In other words, if we use analytic continuation of the fields in
the complex $t$ coordinate to smooth out the corner by bending the
contour, then the matching conditions dictate that the solution would
be at least $C^1$. In section \ref{examples} we illustrate with
examples how these matching conditions determine the bulk
solution for a given contour.

The on-shell supergravity action can be regarded as the saddle point approximation of the
`bulk path integral' and the matching conditions can also be justified
in the same way. Recall that a path integral for fields living on a certain manifold can always
be split in two by cutting the manifold in two halves and imposing boundary conditions for the fields on the cut surface. Afterwards, one can glue the pieces back together by imposing the same boundary condition on either side and then integrate over these boundary conditions.

The saddle-point approximation can similarly be performed in steps. After cutting the manifold, one first finds a saddle-point approximation on either side with arbitrary initial data at the cut surface. This replaces the partial path integrals on either side by an on-shell action which in particular depends on the initial data. Then, one imposes continuity of the initial data, which is the first matching condition, and performs a second saddle-point approximation with respect to the initial data. Since the first variation of an on-shell action with respect to boundary data yields the conjugate momentum, this second saddle-point precisely yields \eqref{eq:2ndmatching}. The matching conditions should then be viewed as an equation determining the initial data. One may verify that the signs come out right, too. This formalism presupposes a two-derivative action, but higher-derivative terms can be dealt with perturbatively.

\section{Holographic renormalization} \label{ren}
The fundamental holographic relation (\ref{eq:zz}) is a bare relation
because both sides are divergent: there are UV divergences on the QFT side
and IR infinite volume divergences on the gravitational side.
So appropriate renormalization is needed to make this relation well-defined. In this section, we will show that the procedure of holographic renormalization for the spaces under consideration is {\it a priori} more complicated, but that none of these complications enter in the final result. Therefore, the formulas presented in for example \cite{Skenderis:2002wp} for the holographically computed correlation functions remain valid in the context of our real-time prescription as well. As the precise derivation of this result is not essential for the rest of the paper, the reader may wish to skip this section on a first reading and proceed directly to the examples of section \ref{examples}.

The holographic renormalization in the Euclidean case
is done by introducing
a set of local covariant boundary counterterms. These counterterms are needed
not only for finiteness of the on-shell action
\cite{Henningson:1998gx,Henningson:1998ey,Balasubramanian:1999re,deHaro:2000xn,Papadimitriou:2004ap}
but also for the variational problem for AdS
gravity to be well-posed \cite{Papadimitriou:2005ii}.
In the Lorentzian setup, in addition to the
infinities due to the non-compactness
of the radial direction, there are also new infinities because of the
non-compactness of the time direction. Correspondingly, in checking the
variational problem one now has to deal both with boundary terms at
spatial infinity and also
at timelike infinity. Thus, in generalizing
the Euclidean analysis to the Lorentzian case there are
two issues to be discussed. First, one has to check that
the Euclidean analysis that leads to the radial counterterms goes through
when we move from Euclidean to Lorentzian signature.
This is indeed the case because all steps involved in the derivation of the
radial counterterms are algebraic and hold irrespectively of the
signature of spacetime.
The second issue one needs to analyze are the infinities due to the
non-compactness of the time directions and the
new boundary terms at timelike infinity.
A complete analysis of this issue requires knowledge of the
asymptotic structure of the solutions near timelike infinity,
which as far as we know is not available. Our prescription
bypasses this problem by gluing in Euclidean AdS manifolds
near timelike infinity. This effectively pushes the
asymptotic region to the (radial) boundary of the Euclidean AdS
manifold, whose asymptotic structure is well known.

What remains to analyze is whether there are any  problems at the
`corners', \ie at the hypersurfaces where the Lorentzian and
Euclidean solutions are joined. In principle, there can be new
corner divergences which would require new counterterms.
In this section we show that such corner divergences are absent in
two examples: a free massive scalar field in a fixed background and
pure gravity. We expect such corner divergences to be absent in general.



\subsection{Scalar field}
\label{sec:scalarfield}
This subsection serves to illustrate the problems at hand, and we will therefore adopt the simplest possible setting. As indicated in Fig.~\ref{fig:onecorner}, we consider a single corner where the contour makes a right angle, passing from a vertical segment to a horizontal segment. The spacelike manifold is taken to be $\Real^{d-1}$. In the absence of sources, we explain below how this contour can be `filled' with empty AdS$_{d+1}$, with a metric that jumps along a spacelike hypersurface from Euclidean to Lorentzian. On this background, we consider a massive scalar field which propagates freely and without backreaction, and we compute the renormalized one-point function of the dual operator.

\subsubsection{Background manifold}
For the bulk manifold under consideration, we take one copy $M_1$ of empty Lorentzian AdS$_{d+1}$ in the Poincar\'e coordinate system $(r,x^i)$ with the metric
\be
ds^2 = dr^2 + e^{2r} \eta_{ij}dx^i dx^j\,,
\ee
and one copy $M_0$ of empty Euclidean AdS in similar coordinates and metric
\be
ds^2 = dr^2 + e^{2r} \delta_{ij} dx^i dx^j\,.
\ee
We will take $x^0$ to be the time coordinate, denoting it by $t$ on $M_1$ and $\tau$ on $M_0$. We use the notation $x^a$ for the other boundary coordinates, so for example $x^i = (t,x^a)$ on $M_1$, and we also introduce $x^A = (r,x^a)$. The conformal boundaries of the spacetimes lie at $r \to \infty$ and are denoted $\partial_r M_1$ and $\partial_r M_0$.

Next, we perform the gluing and the matching. To this end, we cut off the spacetimes across the surface $t = 0$ and $\tau = 0$ such that $t > 0$ and $\tau < 0$, and glue them together along the cut surface which we call $\partial_t M$. This surface is the extension of the corner in the boundary to the bulk. The induced metric on $\partial_t M$ is the same on both sides,
\be
h_{AB}dx^A dx^B = dr^2 + e^{2r}\delta_{ab}dx^a dx^b\,,
\ee
and the extrinsic curvature $\cK_{AB}$ vanishes on both sides. Therefore, both the conjugate momentum $\pi_{AB} = \cK \, h_{AB} - \cK_{AB}$ as well as the induced metric $h_{AB}$ are continuous across $\partial_t M$ and all the matching conditions of section \ref{sec:corners} are satisfied for this background. (We elaborate on the matching conditions for gravity in the next subsection.) The unit normals to $\partial_t M$ on either side are given by
\be
n_{[1]\mu} dx^\mu = - e^{r} dt\,, \qquad \qquad n_{[0]\mu}dx^\mu = e^{r}d\tau\,,
\ee
where we used subscripts in square brackets to indicate whether we are on $M_1$ or on $M_0$. We will use this notation throughout the paper.

\begin{figure}
\centering
\psfrag{M0}{$\partial_r M_0$}
\psfrag{M1}{$\partial_r M_1$}
\psfrag{tau}{$\tau$}
\psfrag{t}{$t$}
\includegraphics[width=3cm]{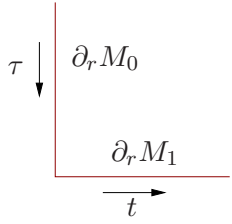}
\caption{\label{fig:onecorner}A single corner in the contour in the
complex time plane. We use this part of a field theory contour
to illustrate the holographic renormalization.}
\end{figure}

Notice that the contour of Fig.~\ref{fig:onecorner} is not complete, since there
is no out state specified at the right end of the contour. This should
be remedied, for example by gluing a Euclidean segment at $t=T$ which
would result in the vacuum-to-vacuum contour of
Fig.~\ref{fig:contour}a. In the bulk, this incompleteness means that
we should also glue another solution to some `final' hypersurface lying in
$M_1$. To obtain the contour of Fig.~\ref{fig:contour}a, for example,
one should glue in half a Euclidean solution $M_2$. With $\Real^{d-1}$ as the spacelike manifold, this would result in a spacetime as sketched in Fig.~\ref{fig:pacman}a.

In this section we will focus on a single corner. We will therefore omit any contributions from such an $M_2$, as well as some terms defined on the final matching surface for $M_1$. Since the matching
between $M_1$ and $M_2$ is a word-for-word
repetition of the matching between $M_0$ and $M_1$, these terms can be
easily reinstated.

\subsubsection{Scalar field setup}
In the background we just described, we consider a scalar field $\Phi$ of mass $m$, dual to a scalar operator $\op$ of dimension $\Delta$ such that $m^2 = \Delta (\Delta - d)$. We will consider the case where $\Delta = d/2 + k$ with $k \in \{0,1,2,3,\ldots\}$, and sometimes we will specialize to $k=2$. The actions for $\Phi$ on $M_1$ and $M_0$ are respectively given by:
\be
\label{eq:2actionsfreescalar}
\begin{split}
S_1 &= \frac{1}{2}\int_{M_1}\sqrt{-G}(-\partial_\mu \Phi \partial^\mu\Phi  - m^2 \Phi^2)\,,\\
S_0 &= \frac{1}{2}\int_{M_0}\sqrt{G}(\partial_\mu \Phi \partial^\mu\Phi + m^2 \Phi^2)\,.
\end{split}
\ee
Suppose $\Phi$ is a solution on $M_0$ and $M_1$ of the equations of motion derived from these actions, with asymptotic value corresponding to the radial boundary data and furthermore satisfies the aforementioned matching conditions (which we discuss in more detail below) on the gluing surface. Our aim is then to compute the corresponding on-shell action,
\begin{equation}
\label{eq:stot}
i S_1 - S_0\,,
\end{equation}
while using the method of holographic renormalization to make it finite.
Note that \eqref{eq:stot} can alternatively be written as:
\begin{equation}
\label{eq:scont}
\frac{i}{2}\int_C dt \int dr \, d^{d-1}x \,\sqrt{-G}(-\partial_\mu \Phi \partial^\mu\Phi  - m^2 \Phi^2)\,,
\end{equation}
with a path $C$ in the complex time plane as in Fig.~\ref{fig:onecorner}, which goes down at first (yielding $-S_0$ after substituting $t = -i\tau$) and then makes a corner and lies along the real $t$ axis. We will not use this notation in this example, but it will be relevant when we consider gravity below.

The holographic renormalization relies on the fact that the solution $\Phi$ can (both on $M_0$ and on $M_1$) be written as a Fefferman-Graham expansion:
\be
\label{eq:fgphi}
\Phi = e^{(k-d/2)r} (\phi_{(0)} + e^{-2r} \phi_{(2)} + \ldots + e^{- 2k r}[ \phi_{(2k)} + \tilde \phi_{(2k)} \log e^{-2r}] + \ldots)\,.
\ee
In this expansion, the radial boundary data is given by specification of $\phi_{(0)}(x^i)$. As one can verify using the equation of motion for $\Phi$, the coefficients $\phi_{(2n)}$ with $2n < 2k$, as well as $\tilde \phi_{(2k)}$, are \emph{locally} determined by $\phi_{(0)}$. For example, for $k \neq 1$, we find
\be
\label{eq:phi2}
\phi_{(2)} = \frac{\square \phi_{(0)}}{4(k-1)}\,,
\ee
with $\square$ the Laplacian of the boundary metric on $\partial_r M$, which in the case at hand is either $\eta_{ij}$ or $\delta_{ij}$. Similarly, all coefficients $\phi_{(2n)} \propto \square^n \phi_{(0)}$ for $n < k$ and $\tilde \phi_{(2k)} \propto \square^k \phi_{(0)}$, all with some $k$-dependent coefficients. The coefficient $\phi_{(2k)}$ is normally nonlocally determined by $\phi_{(0)}$, but in our case it also depends on the initial data that one may specify at $\partial_t M$. In Euclidean backgrounds without corners, this coefficient (times a factor $-2k$) is precisely the renormalized one-point function \cite{deHaro:2000xn}. Below, we show this is still the case in Lorentzian signature and in the presence of corners.

\subsubsection{Matching conditions}
Let us first discuss the matching conditions of section \ref{sec:corners} in more detail. Consider two solutions $\Phi_1$ and $\Phi_0$ on $M_1$ and $M_0$ that satisfy the given radial boundary data, but have arbitrary initial data. The first matching condition is continuity of $\Phi$ across $\partial_t M$, that is:
\begin{equation}
\label{eq:1stmatch}
\Phi_0(\tau = 0,r,x^a) = \Phi_1(t=0,r,x^a)
\end{equation}
for all $r$ and $x^a$.

To derive the second matching condition, we compute the on-shell action for $\Phi_0$ and $\Phi_1$ satisfying the equation of motion and the first condition \eqref{eq:1stmatch}. This action is divergent and we regulate it by cutting off the radial integrals at some large but finite $r_0$. We then consider the variation of the regulated version of the total action \eqref{eq:stot} as we vary the initial data $\Phi(t=0,r,x^a)$ and obtain
\be
\delta (i S_1 - S_0) = \int_{\partial_t M} \sqrt{h} e^{-r}(-i \partial_t \Phi_1 - \partial_\tau \Phi_0) \delta \Phi_1\,,
\ee
where we used that $\delta \Phi_1 = \delta \Phi_0$ by \eqref{eq:1stmatch}. As explained in section \ref{prescription}, we then request that the total action is also at an extremum with respect to the initial data. The second matching condition thus becomes:
\be
\label{eq:2ndmatch}
i \partial_t \Phi_1 + \partial_\tau \Phi_0 = 0 \qquad \text{on }\partial_t M\,.
\ee
As we mentioned before (and as one may check easily using $t = -i\tau$), this second matching condition can be read as $C^1$-continuity in the complex time plane of $\Phi$ across the corner. In the remainder of this section, whenever we write $C^n$-continuity, we always mean continuity in the complex time plane.

Now let us substitute the Fefferman-Graham expansion \eqref{eq:fgphi} of $\Phi_1$ and $\Phi_2$ in the matching conditions \eqref{eq:1stmatch} and \eqref{eq:2ndmatch}. The matching conditions imply the $C^1$-continuity of all coefficients $\phi_{(2l)}$, which, in turn, implies higher-order continuity of the source $\phi_{(0)}$. For example, the first matching condition for $\phi_{(2)}$ becomes, via \eqref{eq:phi2},
\be
\square_{[1]}\phi_{(0)[1]} =  \square_{[0]}\phi_{(0)[0]}\qquad \text{on }\partial_t M\,,
\ee
which shows that $\phi_{(0)}$ has to be at least $C^2$-continuous across the matching surface. Notice that  this is again continuity in the complex time plane, since $\square_{[1]}$ is not equal to $\square_{[0]}$. Next, the second matching condition applied to $\phi_{(2)}$ actually implies $C^3$-continuity for $\phi_{(0)}$:
\be
i \partial_t \square_{[1]} \phi_{(0)[1]} + \partial_\tau \square_{[0]} \phi_{(0)[0]} = 0\,.
\ee
A similar story holds for the subsequent terms. Since the highest number of derivatives is always in $\tilde \phi_{(2k)} \propto \square^k \phi_{(0)}$, applying the second matching condition to this term results eventually in a $C^{2k+1}$-continuity condition for $\phi_{(0)}$ in the complex time plane. Below, we will see the relevance of these high-order continuity conditions.

A comment considering this smoothness condition for $\phi_{(0)}$ is in
order. Namely, this continuity condition essentially follows
from the requirement of the existence of a Fefferman-Graham expansion
at the matching surface. In that light, this higher-order smoothness
condition for $\phi_{(0)}$ is not surprising, since without
it the Fefferman-Graham expansion would fail even in the case without
a corner. Although it would be interesting to study what happens for
discontinuous boundary data, such an investigation can be undertaken
independently of the presence of corners and shall not be pursued
here.

\subsubsection{Holographic renormalization}
The on-shell action \eqref{eq:stot}, evaluated on the solution that satisfies the matching conditions, is of the form:
\begin{equation}
\label{eq:onshellradial}
\begin{split}
i S_1 - S_0 &= - \frac{i}{2} \int_{\partial_r M_1} \sqrt{-\gamma} \, \Phi_1 \partial_r \Phi_1 - \frac{1}{2} \int_{\partial_r M_0} \sqrt{\gamma} \, \Phi_0 \partial_r \Phi_0 \\&\qquad - \frac{1}{2} \int_{\partial_t M} \sqrt{h} [-i \Phi_1 \partial_t \Phi_1 +  \Phi_0 \partial_\tau \Phi_0 ] \,.\\
\end{split}
\end{equation}
The contributions from $\partial_t M$, \ie the second line in \eqref{eq:onshellradial}, vanish by virtue of the matching conditions. Recall that we are omitting the contribution from any `final' surface for $M_1$, which will however by the same mechanism cancel against a matching solution.

The remainder of the action is defined on the cutoff surface $r=r_0$ and
it would diverge if $r_0 \to \infty$. Therefore, a counterterm action has to be added before removing the cutoff. Since the radial terms in \eqref{eq:onshellradial} have a familiar form, one can use the usual procedures of holographic renormalization to find the counterterm action \cite{Skenderis:2002wp}. Let us for example take $k = 2$, for which
\begin{equation}
\label{eq:sctscalar}
S_{\text{ct}} = \frac{1}{2}\int_{\partial_r M} \sqrt{|\gamma|} \, \Big((k - \frac{d}{2}) \Phi^2 + \frac{\Phi \square_\gamma \Phi}{2(1-k)} + \frac{1}{4}\Phi\square_\gamma^2 \Phi \log e^{-r}\Big)
\end{equation}
is the counterterm action. The first two terms are actually valid for any $k \geq 2$ and we used the notation $\square_\gamma$ for the Laplacian of the induced metric $\gamma$ at $r = r_0$. In our case, we simply have $\square_\gamma = e^{-2r}\square$, both on $M_1$ and on $M_0$. Taking care of the signs, we find that
\be
\label{eq:stotalscalar}
iS_1 - S_0 + i S_{\text{ct},1} + S_{\text{ct},0}
\ee
is finite as $r_0 \to \infty$. We see that the usual procedure of holographic renormalization yields a finite on-shell action and possible initial or final terms (which might have caused corner divergences) are absent exactly by the matching conditions.

\subsubsection{One-point functions}
One-point functions are computed by taking variational derivatives of the on-shell action with respect to the boundary data. Let us compute the one-point function $\vev{\op_{[1]}(x)}$, where the subscript indicates that $x$ lies on $\partial_r M_1$. In QFT on a background with a Lorentzian metric $g_{(0)ij}$, the coupling between a source $\phi_{(0)}$ and an operator $\op$ in the partition function is as in \eqref{zqft}. Therefore, the one-point function is
\be
\vev{\op_{[1]}(x)} = \frac{i}{\sqrt{-g_{(0)}}} \fdel{\phi_{(0)}(x)} Z[\phi_{(0)}]\,.
\ee
In our case, the partition function $Z[\phi_{(0)}]$ is given by the renormalized on-shell supergravity action. The easiest way to take care of the divergences is by taking the functional derivative before removing the regulator, resulting in:
\be
\label{eq:vevscalarbare}
\vev{\op_{[1]}(x)} = \lim_{r_0 \to \infty} \frac{ie^{(k + d/2) r_0}}{\sqrt{-\gamma}} \fdel{\Phi_1(x,r_0)}\Big[iS_1 - S_0 + i S_{ct,1} + S_{ct,0}\Big]\,,
\ee
where the extra factor $e^{(k+d/2) r_0}$ converts $\Phi$ to $\phi_{(0)}$ and $\gamma$ to $g_{(0)}$ as $r_0 \to \infty$.

In performing this computation, we see that the presence of corners gives
rise to corner terms, which arise from the integration by parts that is necessary in varying the counterterm action \eqref{eq:sctscalar}. For example, for the variation of the second term in \eqref{eq:sctscalar} we obtain:
\[
\delta \Big( \frac{1}{2}\int_{\partial_r M}  \sqrt{|\gamma|} \frac{\Phi \square_\gamma \Phi}{2(1-k)} \Big) = \int_{\partial_r M}  \sqrt{|\gamma|} \frac{\delta \Phi \square_\gamma \Phi}{2(1-k)} + \frac{1}{2}\int_{C_1}\sqrt{|\sigma|} \frac{e^{-2r} (\partial_t \Phi \delta \Phi - \Phi \partial_t \delta \Phi)}{2(1-k)}\,.
\]
The second term on the right hand side is a corner contribution. However, a similar corner term arises in $S_{\text{ct},0}$, and in the total action \eqref{eq:stotalscalar} these two corner terms cancel each other precisely by the matching conditions.

The subsequent terms in the counterterm action are all of the form $\sqrt{\gamma}\Phi \square_\gamma^n \Phi$ for $n < k$, plus a log term of the form $\sqrt{\gamma}\Phi\square_\gamma^k \Phi \log e^{-r_0}$. After the integration by parts, these all give corner terms as well, which involve a higher number of derivatives of $\Phi$. More precisely, the corner expressions that one obtains from such terms are of the form
\begin{equation}
\label{eq:cornerterms}
\int_C \sqrt{\gamma} e^{-2r} \Phi\partial_t \square^{n-1}_\gamma \Phi\,,
\end{equation}
and equivalent terms with some of the derivatives shifted to the first $\Phi$.

Let us now systematically show that all such terms cancel against a matching solution, using the higher-order smoothness of $\phi_{(0)}$ that we derived before. First of all, recall that the matching conditions imply that $\phi_{(0)}$ should actually be $C^{2k+1}$-continuous. This in turn means that $\phi_{(2)}$ is $C^{2k-1}$ continuous, $\phi_{(4)}$ is $C^{2k-3}$-continuous, etc., up to $\tilde \phi_{(2k)}$ and $\phi_{(2k)}$, which are just $C^1$-continuous. Substituting this in the Fefferman-Graham expansion \eqref{eq:fgphi}, we see that $\Phi$ is not only $C^1$-continuous by the matching conditions, but also $C^3$-continuous up to terms of order $e^{-(k + d/2) r}$, and $C^5$ continuous up to terms of order $e^{(-k - d/2 + 2)r}$, etc.

We now rewrite the leading piece of \eqref{eq:cornerterms} as
\be
\int_C e^{(k + d/2 -2n)r}\sqrt{g_{(0)}}\phi_{(0)} \big(\partial_t \square^{n-1}_0 \Phi\big) + \ldots
\ee
A complete cancellation of this term between $M_1$ and $M_0$ takes place if $\Phi$ is $C^{2n-1}$-continuous up to and including terms of order $e^{-(k+ d/2 - 2n)r}$. However, the previous argument shows that $C^{2n-1}$-continuity for $\Phi$ holds up to terms of order $e^{-(k+ d/2 - 2n + 4)r}$, and the continuity condition is satisfied indeed, for all $n < k$. Therefore, as $r_0 \to \infty$, the terms coming from $M_0$ and $M_1$ cancel indeed and no corner contributions to the one-point functions arise. A similar argument shows that there is no problem with the log term with $n=k$ either.

Having shown the absence of corner contributions in \eqref{eq:vevscalarbare}, one finds that the expression for the one-point function becomes of the standard form, given for example for $k=2$ by:
\be
\vev{\op_{[1]}(x)} = \lim_{r_0 \to \infty} e^{k+ d/2 r_0} \Big[\partial_r \Phi(x) - (k - \frac{d}{2}) \Phi(x) - \frac{\square_\gamma \Phi(x)}{2(1-k)} - \frac{1}{2} \square_\gamma^2 \Phi(x)\log e^{-2r}\Big]_{r = r_0}\,.
\ee
Substitution of the expansion \eqref{eq:fgphi} yields the familiar result:
\be
\vev{\op_{[1]}(x)} = - 2k \phi_{(2k)[1]}(x)\,,
\ee
which is actually valid for all nonzero $k$, see for example \cite{Skenderis:2002wp}.

Finally, consider the one-point function on $M_2$, where we should use the Euclidean version of the source-operator coupling, $-\int \sqrt g_{(0)} \phi_{(0)} \op$.
Repeating the above procedure, we find again:
\be
\vev{\op_{[0]}(x)} = - 2k \phi_{(2k)[0]}(x)\,.
\ee
Since $\phi_{(2k)}(x)$ is continuous across the matching surface by the first matching condition, and since localized corner terms are absent, the one-point function is continuous across the corner as well.

\subsection{Gravity}
For gravity the procedure requires modification and becomes more
involved. We therefore begin with an outline of the steps taken below.

The first step in this procedure is to establish the variational 
principle for the Einstein-Hilbert action for a manifold whose
boundary has corners.
Recall that in the Euclidean setup a well-defined variational problem 
requires the addition 
of the boundary counterterms \cite{Papadimitriou:2005ii} 
and the variational derivatives w.r.t. boundary data lead to the boundary 
correlators. In the Lorentzian setup the variational derivatives w.r.t. 
initial and final data are also important and 
lead to matching conditions. The analysis of the variational
problem is done in subsection \ref{sec:finiteboundaries}.
We will find that there is a need for a special corner term.

The next step is to understand how to glue the various pieces together.
Given a corner in the boundary contour there should exist a 
corresponding bulk hypersurface across which the various bulk pieces
are matched. So we need to understand the possible bulk extensions of the 
boundary contour.  This is analyzed in subsection 
\ref{sec:fgcoordinates} where we show that the extensions are parametrized 
by a single function $f(r,x^a)$ with a certain asymptotic expansion.

Using these results we then derive the matching conditions in 
subsection \ref{sec:matchinggravity} and find their
implications for the radial expansion of the bulk fields near
the corner in \ref{sec:matchinggravityimposed}.
These are all the data we need to analyze whether there are any 
new contributions to the on-shell action and the 
one-point function from the matching surfaces.
This is done for the on-shell action 
in subsection \ref{sec:onshellactiongravity} and for the 1-point functions 
in subsection \ref{sec:contonepointgravity}. We find that 
there are possible contributions from 
each segment but the matching conditions imply complete cancellation
between the contributions of the two pieces that one glues to each other.

The upshot of the discussion is therefore very similar to
the scalar field: we will show that no localized corner terms arise
and that the one-point function of the stress energy tensor is
(appropriately) continuous across the corner.

\subsubsection{Setup}
\label{sec:volumelement}
As we mentioned earlier, we consider manifolds $M_C$ consisting of a number of segments $M_j$ where the metric is Lorentzian or Euclidean. To simplify the computation of the on-shell action for these spacetimes, we introduce a notation where the Einstein-Hilbert action $S_j$ for each separate segment $M_j$ is \emph{always} written as
\be
\label{eq:sehsegment}
S_j  = \frac{1}{2\kappa^2} \int_{M_j} d^{d+1}x \, \sqrt{-G}(R - 2\Lambda)\, ,
\ee
where $\kappa^2 = 8 \pi G_{d+1}$ and $\Lambda = -d(d-1)/(2 \ell)$ with
$\ell$ the AdS radius. Throughout this paper, we set $\ell = 1$.  
In (\ref{eq:sehsegment}) the square root is defined with a branch cut just
above the real axis. For example, for a Euclidean metric $\sqrt{-G} =
- i\sqrt{|G|}$, so that $i S = - S_{E}$ with $S_E =\int\sqrt{|G|}(-R +
2\Lambda)$ the correct Euclidean action. Similarly, for a Lorentzian
metric on a backward-going contour we obtain an extra minus sign since
we are on the other branch of the square root. (To see this, notice
that the time coordinate $t_c$ on this segment is given by $t = e^{i
  \pi}t_c$. If $G,G_c$ denote the metric determinant in the $t,t_c$
coordinate system, respectively, then $G_c = e^{2\pi i}G$, and we make
a full turn indeed.) The advantage of this formalism is that the total
Einstein-Hilbert action $S_{EH}$ for $M_C$ becomes \be
\label{eq:sehtot}
i S_{EH} = i S_0 + i S_1 + \ldots
\ee
for all vertical or horizontal segments $M_0,M_1,\ldots$ We see that all the signs are absorbed in the volume element. This action for $M_C$ needs to be supplemented with various surface terms which we define in due course.

Although we will not discuss this in detail, this prescription can be extended to general complex metrics, allowing for the `filling' of more general QFT contours that are not just built up from horizontal and vertical segments in the complex time plane. In such cases the bulk metric $G_{\mu \nu}$ may be complex, but it should always be non-degenerate for the scalar curvature to be well-defined. Allowing for a complex metric implies that one has to allow for complex diffeomorphisms as well, for example to bring the metric to a Fefferman-Graham form. Complex diffeomorphisms are discussed in some detail in \cite{Halliwell:1989dy}. For such cases, our choice for the branch cut in the volume element is then precisely consistent with the requirement that a QFT contour cannot go upward in the complex time plane.

\subsubsection{Finite boundaries}
\label{sec:finiteboundaries}
In equation \eqref{eq:sehtot}, we split the on-shell action for $M_C$ as a sum over the various segments $M_i$. Just as for the scalar field, we will find the matching conditions via a saddle-point approximation which involves taking functional derivatives of the on-shell action with respect to the initial and final data. This only works if we have a well-defined variational principle for each segment separately, which is what we investigate in this subsection.

Consider a single Asymptotically locally AdS (AlAdS) manifold $M$ with a (possibly complex) metric $G_{\mu \nu}$ and two `initial' and `final' boundaries which we denote here as $\partial_\pm M$. The manifold $M$ also has a radial conformal boundary, which we denote as $\partial_r M$, and the corners where $\partial_\pm M$ meets $\partial_r M$ are denoted as $C_\pm$. We pick coordinates $(r, x^i)$ on $M$, with $x^i = (t,x^a)$, and we will also use $x^A = (r,x^a)$. The conformal boundary is again at $r \to \infty$. We regulate the computation of the on-shell action by imposing $r < r_0$. In this subsection we consider the variational principle in the case where one keeps $r_0$ finite throughout.

A well-defined variational principle for Dirichlet boundary conditions in the presence of corners requires the Einstein-Hilbert action to be supplemented not only with the usual Gibbons-Hawking-York boundary terms on $\partial_r M$ and $\partial_\pm M$, but also with special corner terms defined on $C_\pm$ \cite{Hayward:1993my, Hawking:1996ww, Brown:2000dz}. To find these corner terms, we choose coordinates such that $\partial_r M$ is given by $r = r_0$ and $\partial_\pm  M$ by $t=t_\pm$. The metric near the corners can be put in the following two ADM-forms:
\begin{equation}
\label{eq:1stadm}
\begin{split}
G_{\mu \nu}dx^\mu dx^\nu &= (\hat H^2 + \hat H_i \hat H^i) dr^2 + 2 \hat H_i dx^i dr + \hat \gamma_{ij}dx^i dx^j\,,\\
\hat \gamma_{ij}dx^i dx^j &= (-\hat M^2 + \hat M_a \hat M^a) dt^2 + 2 \hat M_a dx^a dt + \sigma_{ab}dx^a dx^b\,,
\end{split}
\end{equation}
as well as
\be
\label{eq:2ndadm}
\begin{split}
G_{\mu \nu}dx^\mu dx^\nu &= (- M^2 + M_A M^A) dt^2 + 2 M_A dx^A dt + h_{AB} dx^A dx^B\,,\\
h_{AB} dx^A dx^B &= (\hh^2 + \hh_a \hh^a) dr^2 + 2 \hh_a dx^a dr + \sigma_{ab} dx^a dx^b.
\end{split}
\ee
Relating the two metrics, we find
\begin{align}
\label{eq:relationsadmexpansion}
\hh^2 &= \frac{\hat H^2 M^2}{M^2 + (M^r)^2 \hat H^2} & \hat H_t &= M_r \\ 
\hat M^2 &= M^2  - (M^r)^2 \hh^2 & -\frac{M^r}{M^2} &= \frac{\hat H^t}{\hat H^2}\,.\nonumber
\end{align}
For a real Lorentzian metric $M^2$ and $\hat M^2$ are positive, whereas they are negative for a Euclidean metric. We will henceforth assume that $\sigma$, the determinant of $\sigma_{ab}$, is real and positive. This will simplify the discussion and is sufficient for all the examples below.

The standard Gibbons-Hawking-York surface terms involve the extrinsic curvature $^\pm \cK_{AB}$ of $\partial_\pm M$ and $\hat K_{ij}$ of $\partial_r M$, which we will define using the (possibly complex) unit normals,
\bea
\label{eq:unitnormals}
\partial_r M:\
\hat n_\mu dx^\mu = \frac{\sqrt{-G}}{\sqrt {-\hat \gamma}} dr
& \rightarrow &
\hat K_{ij} {=} \frac{\sqrt{\hat H^2 \hat M^2}}{2 \sqrt{\hat M^2}\hat H^2} (\hat D_i \hat H_j + \hat D_j \hat H_i - \partial_r \hat \gamma_{ij}) \,,\\
\partial_\pm M: \ ^\pm n_\mu dx^\mu = \pm \frac{\sqrt{-G}}{\sqrt{h}}  dt
& \rightarrow &
^\pm\! \cK_{AB}
{=} \pm \frac{\sqrt{\hh^2 M^2}}{2 \sqrt{\hh^2} M^2}
(D_A M_B + D_B M_A - \partial_{t}h_{AB}) \,. \nonumber
\eea
Adding the Gibbons-Hawking-York terms, we define the bare action as:
\begin{equation}
\label{eq:S0}
S_{b} = \frac{1}{2\kappa^2}\Big[\int d^{d+1} x \sqrt{-G}(R - 2\Lambda) + 2\int_{\partial_\pm M}\!\!\!\!
d^d x \sqrt{h}\, {}^\pm\! \cK +2\int_{\partial_r M} \!\!\!\! d^d x \,\sqrt{-\hat \gamma}\,\hat K\Big]\,,
\end{equation}
where here and below the summation over $\partial_\pm M$ is implicit and we use the conventions of subsection \ref{sec:volumelement} for the square roots. For a real Lorentzian metric all the above terms are real, but for a real Euclidean metric all terms in \eqref{eq:S0} are purely imaginary (because $\sqrt{-G}$ and $\sqrt{-\hat \gamma}$ are then imaginary, from which it follows that $^\pm n_\mu dx^\mu$ and therefore $^\pm \cK$ are imaginary as well). As one may verify explicitly, in the latter case our choice of branch cut for the square roots in the volume elements implies that $iS_b = -S_E$ with $S_E$ the Euclidean action with the correct Gibbons-Hawking-York terms.

In the case of corners, \eqref{eq:S0} is not the correct action to use for Dirichlet boundary conditions. This is because we cannot perform a diffeomorphism at the corner mixing $t$ and $r$ without changing the definition of the two slices and therefore $\hat H_t$, $M_r$, $M^2$ and $\hat H^2$ are no longer pure gauge at the corner. With this in mind, the variation of the bare action \eqref{eq:S0} is given by the equations of motion, the conjugate momenta for all the various boundaries, plus a corner term
\be
\label{eq:deltasb}
\delta S_b = \frac{1}{2\kappa^2}\int_{C_\pm} d^{d-1}x \sqrt{\sigma}  \delta X_\pm + \ldots\,,
\ee
with $X$ given implicitly by
\begin{equation}
\label{eq:deltaX}
\delta X_\pm = \pm 2 \frac{\sqrt{\hh^2 M^2}}{M^2}\delta M^r\,.
\end{equation}
To find an explicit form of $X_\pm$, we have to integrate $\delta X$ for fixed $\hat M^2$ and $\hh^2$, using the relations \eqref{eq:relationsadmexpansion}. If the metric is completely real and $\hh^2$ and $\hat M^2$ are positive, then we find
\be
\delta X_\pm = \pm 2 \, \delta \,
\text{arcsinh}\Big( \frac{\hh M^r}{\hat M} \Big)\,,
\ee
whereas if $\hat M^2$ is negative and $\hh^2$ and $M^r$ are positive we get
\be
\delta X_\pm = \mp 2 i \, \delta \, \text{arccos}\Big( \frac{\hh M^r}{\sqrt{-\hat M^2}}\Big) \, .
\ee
We can rewrite these expressions in a covariant form using the unit normals defined in \eqref{eq:unitnormals}. Their inner product is given by:
\begin{equation}
\label{eq:nnpm}
^\pm n^\mu \hat n_\mu = \pm \frac{\sqrt{\hh^2}}{\sqrt{\hat M^2}} M^r\,.
\end{equation}
For real $M^2$, $\sqrt{H^2}$ and $\hat M^2$, we can therefore write without branch cut ambiguities:
\begin{equation}
\label{eq:Xpm}
X_\pm =
\begin{cases}
2 \text{ arcsinh}(\,\,^\pm n^\mu \hat n_\mu) & \hat{M}^2 > 0\\
- 2 i  \text{ arcsin}(\,i\, ^\pm n^\mu \hat n_\mu) & \hat{M}^2 < 0\,.
\end{cases}
\end{equation}
In the more general case, the required corner term has the same structure but one needs to be careful about the branch cuts. Notice that $X$ is defined up to a local piece, for example a constant.

Following \cite{Hayward:1993my, Hawking:1996ww, Brown:2000dz}, we aim for a variational
principle that is well-defined for a fixed induced metric on the boundaries,
\ie for fixed $\hat \gamma_{ij}$ and $h_{AB}$. In that case, we should add a
corner term to cancel the unwanted variation $\delta X$ in \eqref{eq:deltasb}. Such a corner term is given by
\be
S_{C_\pm} = - \frac{1}{2\kappa^2}\int_{C_\pm} d^{d-1}x \sqrt{\sigma} X_\pm\,.
\ee
Adding corner terms to the action \eqref{eq:S0} defines an improved (but still bare) action $S_I$,
\begin{eqnarray}
\label{eq:simproved}
S_I &=& S_b + S_{C_\pm} \nonumber \\
&=& \frac{1}{2\kappa^2}\Big[\int d^{d{+}1} x \sqrt{-G}(R - 2\Lambda) + 2\int_{\partial_\pm M}\! \! \! \! \! d^d x \sqrt{h}\, {}^\pm \!\cK  \\&&
\qquad \qquad \qquad + \,2 \int_{\partial_r M}\! \! \! \! \! d^d x \sqrt{-\hat \gamma} \hat K -  \int_{C_\pm}\! \! \! d^{d{-}1}x \sqrt{\sigma} X_\pm \Big]\,,\nonumber
\end{eqnarray}
whose variation is of the form
\begin{eqnarray}
\delta S_I &=& \frac{1}{2\kappa^2}\Big[\int_{\partial_r M} \sqrt{-\hat \gamma}(\hat \gamma^{ij} \hat K - \hat K^{ij})\delta \hat \gamma_{ij}
+ \int_{\partial_\pm M} \sqrt{h}(h^{AB}\, {}^\pm\! \cK
- \, {}^\pm\!\cK^{AB})\delta h_{AB} \nonumber \\ & &\qquad \qquad
- \int_{C_\pm} d^{d-1}x \delta(\sqrt \sigma) X_\pm\Big]\,,\label{eq:varprinciple}
\end{eqnarray}
which is the correct variation for Dirichlet boundary conditions indeed. We will henceforth use this improved action as the bare action and drop the subscript $I$.

\subsubsection{Fefferman-Graham coordinates}
\label{sec:fgcoordinates}
The above discussion was valid for a general spacetime whose boundary has corners. Since we are interested in AlAdS spacetimes where the metric diverges near the radial boundary, we will run into divergences as we let $r_0 \to \infty$. To investigate these divergences, we pick a coordinate system in which the metric is of the Fefferman-Graham form,
\be
ds^2 = dr^2 + \gamma_{ij}dx^i dx^j \,,
\ee
with the radial expansion
\be
\label{eq:fgmetric}
\gamma_{ij} = e^{2r}(g_{(0)ij} + e^{-2r}
g_{(2)ij} + \ldots + e^{-dr} [g_{(d)ij} + \tilde g_{(d)ij}\log
e^{-2r}] + \ldots) \,.
\ee
From the Einstein equations we find that all coefficients $g_{(2n)ij}$ with $2n <
d$, as well as $\tilde g_{(d)ij}$, are locally determined by
$g_{(0)ij}$, and involve up to $2n$ or $d$ derivatives of
$g_{(0)ij}$. The term $g_{(d)ij}$ is not locally determined (except
for its trace and its divergence) and this term directly enters in the
one-point function of the stress energy tensor \cite{deHaro:2000xn}.

The disadvantage of the Fefferman-Graham form of the metric is that one can generally no longer pick a coordinate $t$ such that the surfaces $\partial_\pm M$ are given by slices of constant $t$. On the other hand, one can use the leftover gauge freedom to make sure that $\partial_\pm M$ are asymptotically given by:
\begin{equation}
\label{eq:pmpm}
\partial_\pm M: t = f_\pm(r,x^a)\,,
\end{equation}
with
\begin{equation}
\label{eq:fasympt}
\lim_{r \to \infty} f_\pm(r,x^a) = t_\pm
\end{equation}
and $t_\pm$ constants. We will discuss the asymptotic behavior of $f_\pm$ more precisely below.

Let us consider a single initial or final boundary. Dropping for now the subscript $\pm$, we write an ADM-decomposition of $\gamma_{ij}$ near the corner:
\be
\label{eq:admgamma}
\gamma_{ij}dx^i dx^j = (-N^2 + N^a N_a) dt^2 + 2 N_a dt dx^a + \tau_{ab}dx^a dx^b\,.
\ee
We may pick boundary Gaussian normal coordinates centered at the
corner, so that $N_a \sim O(1)$. Furthermore,
$N^2= e^{2r} N^2_{(0)} + N_{(2)}^2 + \ldots$ and
$\tau_{ab} = e^{2r} \tau_{(0)ab} + \tau_{(2)ab} + \ldots$ We can relate this ADM-decomposition to the double ADM-decomposition \eqref{eq:2ndadm} of the previous subsection by introducing a new coordinate
\begin{equation}
\label{eq:diffeot}
t' = t - f(r,x^a)\,,
\end{equation}
after which the initial slice is given by $t'=0$. In the new coordinates, the metric is of the form \eqref{eq:2ndadm}, with $t$ replaced by $t'$, and with the components
\begin{equation}
\label{eq:metricrelations}
\begin{split}
-M^2 + M_A M^A &= -N^2 + N_a N^a\\
M_r &= (-N^2 + N_a N^a)\partial_r f\\
M_a &= N_a + (-N^2 + N_c N^c) \partial_a f\\
\hh^2 + \hh_a \hh^a &= 1 + (-N^2 + N_a N^a) (\partial_r f)^2\\
\hh_a &= (-N^2 + N_c N^c) \partial_a f \partial_r f + N_a \partial_r f\\
\sigma_{ab} &= \tau_{ab} + (-N^2 + N_c N^c) \partial_a f \partial_b f + N_a \partial_b f + N_b \partial_a f\,,\\
\end{split}
\end{equation}
where indices are raised with the appropriate metric. We use these equations below to write down a radial expansion of the components on the left-hand side in terms of the Fefferman-Graham expansion and a radial expansion of $f$.

For AlAdS spacetimes the Dirichlet boundary data are given by $g_{(0)ij}$ and $h_{AB}$. Asymptotically, $g_{(0)ij}$ determines a Fefferman-Graham radial coordinate as well as the subleading coefficients up to $g_{(d)ij}$ in the Fefferman-Graham expansion of the metric. Of course, the initial and final metric $h_{AB}$ should be such that $\partial_\pm M$ can be embedded in the asymptotic metric dictated by $g_{(0)ij}$ and this condition constrains the asymptotic form of $h_{AB}$. To be precise, $h_{AB}$ should have a radial expansion that is compatible with the last three equations in \eqref{eq:metricrelations} for a certain $f$. However, as long as $f$ is unspecified, $h_{AB}$ is not to any order determined in terms of $g_{(0)ij}$.

We remark that the last three lines in \eqref{eq:metricrelations} signify constraints on $h_{AB}$ only. Therefore, they are different from the usual constraints on the initial data in a Hamiltonian formalism of general relativity, which also involve the extrinsic curvature. These usual constraints are satisfied if the extrinsic curvature of the initial slice is computed using the embedding of the initial slice as a hypersurface in the solution. Therefore, they are automatically satisfied if we compute the extrinsic curvature using the first three lines of \eqref{eq:metricrelations}. Since this is how we compute the extrinsic curvature below, we will not worry about these constraints.

\subsubsection{Gluing and matching conditions}
\label{sec:matchinggravity}
In the previous subsections, we found an improved action \eqref{eq:simproved} and discussed the Fefferman-Graham expansion for a single AlAdS spacetime with corners. We now take two of such spacetimes and glue them together along the initial and final hypersurfaces $\partial_\pm M$.

We will denote the two segments by $M_0$ and $M_1$ and we glue $\partial_+ M_0$ to $\partial_- M_1$, which we from now on we denote as $\partial_t M$. The corner, \ie the intersection of $\partial_t M$ with $\partial_r M_0$ and $\partial_r M_1$, is denoted by $C$. As before, a subscript (sometimes in square brackets) indicates the manifold under consideration. We make no assumptions about the signature of the metric on $M_0,M_1$ and in fact the metric may even be complex. We write the total action as
\be
\label{eq:stotgravity}
i S_0 + i S_1\,,
\ee
with the individual actions given by \eqref{eq:simproved}. We recall that we use the conventions of subsection \ref{sec:volumelement}, so extra factors of $i$ might be included in the volume elements and extrinsic curvatures. As we did for the scalar field, we will henceforth ignore the contribution from other segments than $M_0$ and $M_1$ as well as the contribution to the on-shell actions of $M_0$ and $M_1$ that may arise from other matching surfaces.

Let us now find the precise matching conditions that the metrics on $M_0$ and $M_1$ have to satisfy near $\partial_t M$. The first matching condition is continuity of the initial and final Dirichlet data. For gravity, this becomes continuity of the induced metric:
\be
\label{eq:conth}
h_{[0]AB} = h_{[1]AB}\,.
\ee
The second matching condition is obtained from the variation of the on-shell regularized action with respect to the data on $\partial_t M$. Let us first suppose the variation vanishes at the corner $C$. In that case, we read off from \eqref{eq:varprinciple} that the second matching condition becomes:
\begin{equation}
\label{eq:contK}
\cK_{[0]AB} + \cK_{[1]AB} = 0\,.
\end{equation}
We can also consider a variation that does not vanish at $C$, for which \eqref{eq:varprinciple} shows that
\be
\label{eq:contXbare}
(X_{[1]} + X_{[0]}) \delta(\sqrt{\sigma})= 0\,,
\ee
where we included the $\delta(\sqrt{\sigma})$ because of the following
reason. Notice that this is a corner matching condition which is therefore not valid to all orders in $r$. However, since $\sqrt{\sigma} \sim e^{(d-1)r}$, \eqref{eq:contXbare} is actually divergent as $r_0 \to \infty$. Therefore, it only vanishes completely if the $X$'s match to high order in their radial expansion. If there are no log terms, then we find
\begin{equation}
\label{eq:contX}
X_{[1]}+ X_{[0]} = O(e^{-dr})\,.
\end{equation}
Equation \eqref{eq:contX} is the corner analogue of the bulk matching condition \eqref{eq:contK}. Notice that such a corner condition was absent when we discussed the scalar field discussed above. Its implications will be investigated in the next subsection.

We showed before that $\cK_{AB}$ and $X$ are imaginary for a Euclidean metric. Therefore, although it is not transparent in our notation, the matching conditions \eqref{eq:contK} and \eqref{eq:contX} do contain factors of $i$ when joining a Lorentzian to a Euclidean metric.

\subsubsection{Imposing the matching conditions}
\label{sec:matchinggravityimposed}
For the scalar field, the matching conditions were crucial in demonstrating the cancellation of corner divergences and the absence of localized corner contributions to the one-point function. A similar cancellation will occur for gravity, but imposing the three matching conditions \eqref{eq:conth}, \eqref{eq:contK} and \eqref{eq:contX} will not be as straightforward as for the scalar field.

In this subsection we shall impose the matching conditions order by order in a radial expansion of $h_{AB}$, $\cK_{AB}$ and $X$. We start with a detailed analysis of the leading-order terms in the matching conditions. We then discuss continuity in the complex time plane of the boundary metric. Just as for the scalar field, the higher-order continuity is related to the continuity of the subleading terms in the Fefferman-Graham expansion of the bulk metric. Afterwards, we show that our leading-order results extend to the higher-order terms as well.

\paragraph{Leading order matching conditions\\}
We will work in the Fefferman-Graham coordinates, with the matching surface $\partial_t M$ given by \eqref{eq:pmpm}. Without loss of generality, we assume that the corner is given by $t = 0$ on $\partial_r M_1$ and $\tau = 0$ on $\partial_r M_0$, so $\lim_{r \to \infty} f(r,x^a) =0$ on either side. We suppose that $f$ behaves asymptotically as
\begin{equation}
\label{eq:fasymptlin}
f = e^{-r}f_{(1)}(x^a) + \ldots
\end{equation}
This is the leading asymptotic behavior of $f$, since any
slower falloff near $r \to \infty$ would yield a non-spacelike induced
metric on $\partial_t M$ in a real Lorentzian spacetime.
Substituting (\ref{eq:fasymptlin}) and the leading-order terms in the ADM-decomposition \eqref{eq:admgamma} of $\gamma_{ij}$ in \eqref{eq:metricrelations}, we find the leading behavior of $\hh^2$, $M^2$ and $M^r$. The inner product between the
unit normals, given in \eqref{eq:nnpm}, becomes to leading order:
\be
\begin{split}
^\pm n^\mu \hat n_\mu = \mp \frac{\sqrt{N_{(0)}^2} f_{(1)}}{\sqrt{1 - N_{(0)}^2 f_{(1)}^2}}\,.
\end{split}
\ee
Since continuity of $X_\pm$ follows from continuity of
$^\pm n^\mu \hat n_\mu$, the corner matching condition \eqref{eq:contX} becomes to leading order:
\begin{equation}
\label{eq:contf}
\sqrt{N_{(0)[0]}^2} f_{(1)[0]} = \sqrt{N_{(0)[1]}^2} f_{(1)[1]}\,,
\end{equation}
where we reinstated the subscripts to indicate the manifold under consideration.

\begin{figure}
\centering
\psfrag{0}{$0$}
\psfrag{t}{$t_0$}
\psfrag{t2}{$t_1$}
\psfrag{tf0}{$t_0 = f_{[0]}$}
\psfrag{tf1}{$t_1 = f_{[1]}$}
\includegraphics[width=8cm]{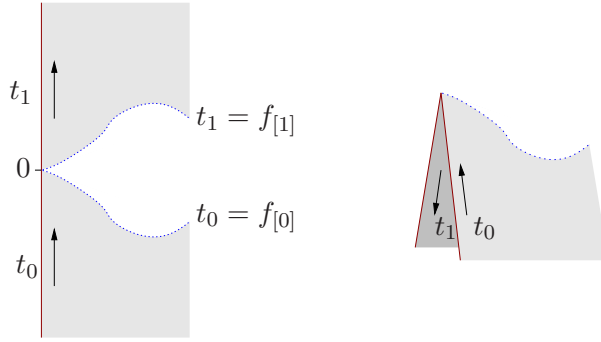}
\caption{\label{fig:fullturn}On the left, the dotted lines represent two bulk hypersurfaces given by $t = f$ in the vicinity of a corner in the boundary contour at $t=0$. On the right, we see that around a full turn in the boundary contour it is natural to expect that $f_{[1]} = -f_{[0]}$.}
\end{figure}

Let us work out the consequences of this condition. Recall that we absorbed factors of $i$ in the square roots of the metric determinant, and therefore \eqref{eq:contf} is not necessarily a relation between real quantities. For example, if we match a Lorentzian to a Euclidean solution, then $N_{(0)}^2$ changes sign across the corner and the square root on the Euclidean side of \eqref{eq:contf} becomes imaginary. On the other hand, the square root on the Lorentzian side is real, and so is $f_{(1)}$ since we use real coordinates. This means that in that case we must have:
\be
f_{(1)[0]} = f_{(1)[1]} = 0\,,
\ee
which more generally holds in all cases for which the phase of $N_{(0)}^2$ is discontinuous across the corner. Actually, this phase is only continuous when we match two solutions with the same signature (recall that we chose boundary coordinates in which $N_{a(0)} = 0$). This happens either if we have no corner at all, or if the corner makes a 180-degree turn. In the first case, we can pick boundary coordinate systems in which $N_{(0)}^2$ is continuous across the corner and \eqref{eq:contf} becomes simply
\begin{equation}
\label{eq:nocorner}
f_{(1)[0]} = f_{(1)[1]}\, .
\end{equation}
Since we just artificially split a spacetime in two parts, it is natural that there is no further constraint on $f$. The case in which the corner makes a full turn is slightly more involved. First of all, the two boundary segments ending on the corner must be straight horizontal lines in the complex time plane, since the boundary contour cannot go up in this plane. We may again assume that $N_{(0)}^2$ is continuous across the corner, but that does not mean that the square roots in \eqref{eq:contf} are. Namely, one of the segments is backward-going in the complex time plane and in subsection \ref{sec:volumelement} we already mentioned that the square root for a backward-going contour results in a minus sign. The matching condition for a full turn therefore becomes
\begin{equation}
\label{eq:fullturn}
f_{(1)[0]} = - f_{(1)[1]}\, .
\end{equation}
This implies that, at least at this order, we can freely move the hypersurface $\partial_t M$ up and down in the bulk, as long as we move it by the same amount on both components and keep the location of the corner fixed. We have sketched this in Fig.~\ref{fig:fullturn}.

We have worked out the leading order term in the corner matching condition in three cases, corresponding to three different corners. We emphasize that our formalism of subsection \ref{sec:volumelement} allowed us to summarize all three cases in the single equation \eqref{eq:contf}. We will see below that the subleading behavior of $f$ is constrained in an analogous way.

As a sidenote, let us also compute the leading order term in the radial expansion of the second matching condition \eqref{eq:contK}. If we use \eqref{eq:unitnormals} to expand the trace of the extrinsic curvature $^\pm \! \cK_{AB}$ the leading order term becomes:
\be
^\pm \cK
= \pm d \frac{\sqrt{N_{(0)}^2} f_{(1)}}{\sqrt{1 - N_{(0)}^2 f_{(1)}^2}}\, .
\ee
The trace part of \eqref{eq:contK} therefore results to leading order again
in \eqref{eq:contf}.
It is plausible that for AlAdS spacetimes the corner matching
condition \eqref{eq:contX} follows from \eqref{eq:contK} and does not need to be imposed separately. This would be related to the fact that the asymptotics of the bulk metric are
completely determined by the Fefferman-Graham data, but a more
complete analysis is required to settle this issue completely. This
will not be attempted here and we will instead continue to treat \eqref{eq:contX} as an additional condition.

\paragraph{Continuity in the complex time plane\\}
Just as for the scalar field, the Fefferman-Graham expansion relates subleading terms in the matching conditions to higher-order continuity in the complex time plane of the sources. Before proceeding with the subleading terms in the matching conditions, let us therefore first discuss the notion of smoothness in the complex time plane for the boundary metric.

Consider a contour in the complex time plane with a corner. We define $C^k$-smoothness for the boundary metric as the condition that the $k$'th order $t$-derivatives of the metric components exist at the corner of the contour. Although this is a natural definition, in our notation a complication arises because we do not work directly with a complex time coordinate on for example the vertical segments. Instead, we rather use a contour time like $t_c$ or $\tau$ which is real on a particular segment of the contour and for such parametrizations the continuity condition has a different form. We may find this new form by regarding these local parameters as related to $t$ via a \emph{complex} diffeomorphism, for example $t = -i\tau$ or $t = 2T -t_c$. If we use these parameters to express continuity of the metric, then we need to take care of the transformation properties under the diffeomorphism as well. For example, $C^0$ continuity of $g_{(0)ij}$ across the corner of Fig.~\ref{fig:onecorner}, where $t = - i \tau$, becomes the condition that at the corner
\be
g_{(0)[0]\tau \tau}  = - g_{(0)[1]tt} \,,\qquad g_{(0)[0]\tau a} = -i g_{(0)[1]t a}\,, \qquad g_{(0)[0]ab} = g_{(0)[1]ab}\, .
\ee
Similarly, $C^1$ continuity in the complex time plane becomes $\partial_\tau (g_{(0)[0]ab}) = -i \partial_t (g_{(0)[1]ab})$ and $\partial_\tau (g_{(0)[0]\tau \tau})  = i \partial_t (g_{(0)[1]tt})$. The extension to higher orders and other components is analogous. As an example, take $ds_{[0]}^2 = d\tau^2 + \delta_{ab}dx^a dx^b$ and $ds_{[1]}^2 = -dt^2 + \delta_{ab}dx^a dx^b$. Although there is an apparent discontinuity in the metric components, with our definitions the metric is $C^\infty$ at the corner.

We will from now on \emph{assume} that the boundary metric at the corner is $C^{d}$ continuous in the complex time plane. The reason for this smoothness condition is the same as that for the scalar field: it guarantees the existence of a Fefferman-Graham expansion of the metric at the corner, and the locally determined coefficients in this expansion are then automatically continuous across the corner as well. Since we continue to use real coordinates like $\tau$, we will always need to supplement the continuity condition with the transformation under the complex diffeomorphism.

\paragraph{Higher order matching conditions\\}
We showed above that the leading order matching conditions imply that $f_{(1)}$ usually vanishes, except in special cases when $N_{(0)}^2$ does not change across the corner. In this subsection, we show that the matching conditions and the $C^d$ continuity of the boundary metric fix the higher-order terms in $f$ to behave just as $f_{(1)}$, at least up to terms that vanish faster than $e^{-dr}$.

We first assume that the leading order term in $f$ is:
\begin{equation}
\label{eq:fasymptsubl}
f(r,x^a) = e^{-nr}f_{(n)}(x^a)\,.
\end{equation}
One may easily check that in this case
\be
^\pm n^\mu \hat n_\mu = \mp \sqrt{N_{(0)}^2}f_{(n)} e^{(1-n)r} + \ldots
\ee
and a repetition of the previous analysis shows that, for $n \leq d$, the leading order term in \eqref{eq:contX} becomes equivalent to
\be
\sqrt{N_{(0)[0]}^2} f_{(n)[0]} = \sqrt{N_{(0)[1]}^2} f_{(n)[1]}\, .
\ee
Therefore, if the phase of $\sqrt{N_{(0)}^2}$ is discontinuous across the corner, we find that not only $f_{(1)}$ but all terms up to and including order $e^{-dr}$ in $f(r,x^a)$ vanish as well.

If $N_{(0)}^2$ is continuous, then $f_{(1)}$ does not necessarily vanish, equation \eqref{eq:fasymptsubl} no longer holds, and the above derivation for the subleading terms is no longer valid. However, the $C^d$ continuity of the boundary metric implies that the locally determined terms in the Fefferman-Graham expansion \eqref{eq:fgmetric} are continuous across the matching surface as well and the metric is thus the same to high order on either side (up to the complex diffeomorphism discussed above). A discontinuity in \eqref{eq:fgmetric} may appear at the earliest for the nonlocally determined term $g_{(d)ij}$, which is at overall order $e^{(2-d)r}$ in the radial expansion of the bulk metric. By substituting this radial expansion in the fourth equation in \eqref{eq:metricrelations}, and using the continuity to all orders of $H^2 + H_a H^a$, we find that $f$ has to be continuous across the corner up to and including terms of order $e^{-dr}$. (Notice that the fourth equation in \eqref{eq:metricrelations} is invariant under $f \leftrightarrow -f$, but we fixed the overall sign already at leading order.)

This finishes our discussion about imposing the matching conditions: the previous two paragraphs show that $f$ `matches' up to and including terms of order $e^{-dr}$ for all three cases. Up to this order, we find that $f_{[0]} = -f_{[1]}$ for a full turn, that $f_{[0]}$ and $f_{[1]}$ both vanish for any other corner, and that  $f_{[0]} = f_{[1]}$ if there is no corner at all. In the next subsection, we will use these conditions to demonstrate the absence of localized (divergent) corner contributions to the on-shell action, in order to eventually show the continuity of the one-point function of the stress energy tensor around the corner.

\subsubsection{Computation of the on-shell action}
\label{sec:onshellactiongravity}
The bare on-shell action \eqref{eq:simproved} has the usual Gibbons-Hawking-York contribution from $\partial_t M$ as well as an extra corner contribution. However, the matching conditions directly imply that these terms cancel between the two spacetimes. The total action \eqref{eq:stotgravity} becomes:
\bea
iS_0 + iS_1 &=& \frac{i}{2\kappa^2} \int_{M_0}d^{d+1}x\sqrt{-G}(R - 2 \Lambda)
+ \frac{i}{\kappa^2}\int_{\partial_r M_0}d^{d}x \sqrt{-\gamma} K \nonumber \\
&+& \frac{i}{2\kappa^2} \int_{M_1}d^{d+1}x\sqrt{-G}(R - 2 \Lambda)
+ \frac{i}{\kappa^2}\int_{\partial_r M_1} d^{d}x\sqrt{-\gamma} K\,.
\label{eq:stotgravityexp}
\eea
This action can again be renormalized with the usual radial
counterterms, except for a subtlety involving the bulk
integrals in this action. Namely, the $t$-integrals
do not run between fixed endpoints, say $t=0$ and $t=T$, but now
rather end on $t= f_\pm(r,x^a)$. The usual radial
counterterms, however, assume an $r$-independent limit on the bulk
integral and the radial counterterms may not exactly cancel
all divergences.

We will now show that these extra divergences also cancel between the two matching
solutions. To first order, the cancellation can be shown very explicitly.
Namely, if $f$ is of the form \eqref{eq:fasymptlin},
then we can radially expand the volume element as:
\begin{align}
\int_{M_0}\sqrt{-G} \, d^{d+1}x &= \int^{r_0}dr \int dx^a \int_{f(r,x^a)}\!\!\!\!\!dt\sqrt{N^2 \sigma} \label{eq:radialexpansionbulk}
\\ &= \int^{r_0} dr \Big[ \int dx^a \int_{f(r_0,x^a)}\!\!\!\!\!dt \sqrt{N^2 \sigma} - e^{(d-1)r}\int dx^a f_{(1)} \sqrt{N_{(0)}^2 \sigma_{(0)}} + \ldots \Big] \,.\nonumber
\end{align}
The first term has an $r$-independent lower limit on the $t$-integral
and so all divergences in this term are dealt with by integrating the
usual radial counterterms also until $f(r_0,x^a)$. The second term
is not cancelled by counterterms and may lead to new divergences.
However, in \eqref{eq:stotgravityexp} a similar term comes
from the expansion of the action $S_1$ for $M_1$ and by
the corner matching condition
\eqref{eq:contf} the terms exactly cancel each other.
Notice that an extra sign on $M_1$ arises
because we expand the upper rather
than the lower limit of the $t$-integral.

For higher orders, we recall that $f$ is continuous or vanishing up to and including terms of order $e^{-dr}$. Using also the higher-order continuity of the bulk metric, a continuation of the expansion \eqref{eq:radialexpansionbulk} shows that the corrections cancel up to finite terms. This means that no extra divergences arise from the discrepancy between the limits of the $t$-integration.

Having eliminated all possible sources of corner divergences, we may conclude that the usual radial counterterms are sufficient to make the total on-shell action finite. For example, in $d=4$ the counterterm action is of the form:
\begin{equation}
\label{eq:sctgravity}
S_{\text{ct}} = \frac{1}{2\kappa^2} \int d^d x \sqrt{-\gamma} \Big( 3 + \frac{1}{4}R + \frac{1}{4}\log e^{-r_0} \Big[\frac{1}{4}R^{ij}R_{ij} - \frac{1}{6}R^2 \Big]\Big)\,,
\end{equation}
where the curvatures are those of the boundary metric $\gamma_{ij}$ at $r = r_0$. This counterterm action is valid for all signatures if we define $\sqrt{-\gamma}$ in the same way as $\sqrt{-G}$ above, \ie with the branch cut above the positive real axis.

\subsubsection{Continuity of the one-point function}
\label{sec:contonepointgravity}
We have shown that the on-shell action can be holographically renormalized with the usual counterterms in the presence of corners. It remains to show that the one-point function is appropriately continuous around the corners as well.

The renormalized one-point function of the stress energy tensor is obtained by varying the renormalized on-shell action with respect to radial boundary data. As for the scalar field, the integration by parts in the variation of a counterterm action like \eqref{eq:sctgravity} may result in localized corner contributions to the one-point function. However, a similar analysis as for the scalar field shows that the higher-order continuity of the boundary metric in the complex time plane ensures that such contributions again cancel between two matching solutions.

Let us explicitly show the cancellation of the first corner term that arises from the integration by parts of the radial counterterms, which originates from the second term in \eqref{eq:sctgravity}. This is just an Einstein-Hilbert like term and
it cancels against the matching solution if the extrinsic curvature of
the corner, which we denote $K_{(0)ab}$, is
continuous across the corner:
\be
K_{(0)[0]ab} + K_{(0)[1]ab} = 0\,.
\ee
Cancellation of the next term gives a higher-order continuity
condition. Explicitly, the variation of these terms gives
\begin{multline}
\label{eq:deltactgravity}
\delta\int_{\partial_r M} d^{d}x\, \sqrt{-\gamma} \Big[\frac{R_{ij}R^{ij}}{(d-2)^2} - \frac{dR^2}{4(d-1)(d-2)}\Big] = \int_{\partial_r M} d^d x \, \sqrt{-\gamma} (\ldots) \delta \gamma_{ij} \\ + \int_{C} d^{d-1}x\sqrt{\sigma} \Big[ n_i P^{ij}  (\na^l \delta \gamma_{lj} - \gamma^{kl}\na_j \delta \gamma_{kl}) + (\na_i P^{ij}) (n_j \gamma^{kl}\delta \gamma_{kl} -n^k \delta \gamma_{kj}) \Big]\,,
\end{multline}
where
\be
P^{ij} = - \frac{dR \gamma^{ij}}{4(d-1)(d-2)}  + \frac{R^{ij}}{(d-2)^2}
\ee
and $n^i$ is an appropriately defined unit normal for the corner as a
submanifold of $\partial_r M$. From \eqref{eq:deltactgravity} we explicitly see that the higher-order continuity condition involves up to three derivatives of the metric in $d=4$.

By the absence of initial or corner contributions, the holographic expression for the one-point function of the stress-energy tensor is completely analogous to the Euclidean case. In particular, it is expressed directly in terms of $g_{(d)ij}$ and terms that are determined locally by $g_{(0)ij}$. For example, in $d=4$ we find up to scheme-dependent terms that
\be
\vev{T_{ij}} = \frac{2}{\kappa^2}\Big( g_{(4)ij} - \frac{1}{8} [(\Tr\, g_{(2)})^2 - \Tr\, g_{(2)}^2] - \frac{1}{2}(g_{(2)}^2)_{ij} + \frac{1}{4}g_{(2)ij}\Tr\, g_{(2)} \Big)\,,
\ee
see \cite{deHaro:2000xn} for the exact expressions in other dimensions. Alternatively, we may use the `radial Hamiltonian' approach to holographic renormalization \cite{Papadimitriou:2004ap,Papadimitriou:2004rz}, which provides a more efficient way of obtaining renormalized correlators. In this approach, the one-point function can be more compactly written as
\be
\vev{T_{ij}} = \pi_{(d)ij}\,,
\ee
where $\pi_{(d)ij}$ is the term of dilatation weight $d$ in the expansion of the radial canonical momentum in eigenfunctions of the dilatation operator.

Since by assumption all locally determined terms in the Fefferman-Graham expansion of the metric are continuous, continuity of the one-point function will follow from continuity of $g_{(d)ij}$ across the corner. Fortunately, the continuity of $g_{(d)ij}$ follows directly if we substitute the expansion \eqref{eq:fgmetric} in the last equation of \eqref{eq:metricrelations}. The left-hand side in this equation is continuous to all orders by the first matching condition. On the other side, we know that $f$ is continuous up to and including terms of order $e^{-dr}$, and we know that all $g_{(2n)ij}$ with $2n < d$ as well as $\tilde g_{(d)ij}$ are continuous since they are locally determined by $g_{(0)ij}$. Collecting terms of overall order $e^{(2-d)r}$ then establishes that $g_{(d)ij}$ has to be continuous as well. (As shown in \cite{Papadimitriou:2005ii}, there is no diffeomorphism freedom at this order if we fix a boundary coordinate system and a boundary metric, so continuity of $g_{(d)ab}$ implies continuity of $g_{(d)ij}$ indeed.) We have thus established that the vev of the stress-energy tensor is continuous across the corner (in the sense discussed in subsection \ref{sec:matchinggravityimposed}).

We end this section with a remark about the function $f(r,x^a)$. Recall that we could in some cases freely specify this function at the corner, provided it was the same on both sides (possibly up to a sign). On the other hand, this function has no place in the QFT, and therefore holographically computed QFT correlators should be independent of $f$. Our prescription passes this test, since the one-point function we obtain is indeed independent of $f$.



\section{Examples} \label{examples}
In this section we apply the general prescription to several
concrete cases. The examples below are meant to illustrate the
applicability of the real-time gauge/gravity prescription for
computing time-ordered, retarded or Wightman correlation functions
in a variety of backgrounds directly from the bulk theory. Notice
that such correlation functions sometimes differ only by the form of
their $i\epsilon$ insertions (or other analyticity properties). Although the $i\epsilon$ insertions are often set by hand, in the QFT they can be obtained from a formal derivation which is briefly discussed in appendix \ref{sec:app:fieldthy}. Any first-principles real-time gauge/gravity prescription should therefore also be able to correctly determine these $i\epsilon$ insertions via bulk computations. The examples below show that our prescription indeed produces $i\epsilon$ insertions that are always in agreement with field theory expectations (as described in appendix \ref{sec:app:fieldthy}), which provides a nontrivial check of the prescription.

\subsection{Examples involving global AdS$_3$}
\label{sec:globalads3}
\begin{figure}
\centering
\psfrag{E}{$E$}
\psfrag{L}{$L$}
\psfrag{a}{(a)}
\psfrag{b}{(b)}
\psfrag{c}{(c)}
\psfrag{d}{}
\psfrag{e}{}
\psfrag{f}{}
\includegraphics[width=13cm]{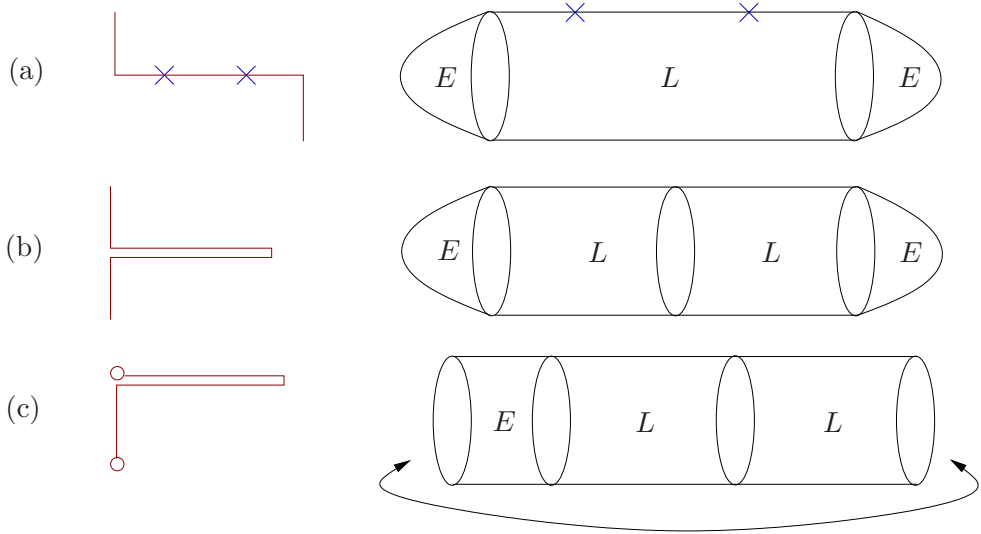}
\caption{\label{fig:contoursintplane}On the left, various contours in the complex time plane. The vertical segments in the first two contours should be thought of as extending to infinity, yielding a vacuum state on the corner. The circles in the third contour should be identified; it is then a thermal contour. The crosses represent an example of the operator insertions we consider. On the right, we sketch the spacetimes consisting of piecewise Euclidean and Lorentzian AdS$_3$ that fill the given contours. One should impose matching conditions on the hypersurfaces between the segments.}
\end{figure}

For the examples in this subsection, we will consider a two dimensional CFT with a holographic dual defined on a cylinder with metric
\be
ds^2 = -dt^2 + d\phi^2
\ee
and a contour for the CFT which consists of the $\phi$ circle times a
path $C$ in the complex time plane, with $C$ being piecewise
horizontal or vertical. The discussion can  be extended straightforwardly
to a CFT in $d$ dimensions, but we restrict ourselves to $d=2$ for now.
Various possibilities for $C$ are indicated on the
left of Fig.~\ref{fig:contoursintplane}. We will compute the two-point function for operators inserted on the last two contours drawn in Fig.~\ref{fig:contoursintplane}; the first contour, with the indicated operator insertions, was discussed in \cite{us}.

As we mentioned in the section \ref{prescription}, the general idea is to `fill in' the entire field theory contour with bulk spaces. In the case when all sources
vanish along $C$, one can fill each horizontal segment of $C$ with a
segment of empty Lorentzian AdS$_3$ and each vertical segment with a
segment of Euclidean AdS$_3$. The metric on the Lorentzian segments is
of the form:
\begin{equation}
\label{eq:metricemptyads3}
ds^2 = -(r^2 + 1) dt^2 + \frac{dr^2}{r^2 + 1} + r^2 d\phi^2
\end{equation}
and the Euclidean metric can be obtained by the replacement $t =
-i\tau$. In this metric, surfaces of constant $t$ or $\tau$ have
vanishing extrinsic curvature and the induced metric is independent of
the signature of the spacetime metric. Therefore, the matching
conditions for gravity are satisfied if we glue the Euclidean and Lorentzian
segments together along such surfaces. The complete bulk solution $M_C$ consisting of
Lorentzian and Euclidean segments glued together along these constant
$t$ or $\tau$ surfaces therefore satisfies all the conditions stated
above and can be taken as a filling for the given contour. We have
drawn such fillings schematically on the right of
Fig.~\ref{fig:contoursintplane}. Note that these `piecewise
AdS' spacetimes may not be the only bulk solutions for the given class of
contours, a point which we will come back to when we discuss black
holes.

By switching on boundary sources, we can perturb such backgrounds, with the provision that the matching conditions are satisfied for the perturbations as well. In the two examples below, we will add a massive scalar field in the bulk and compute a contour-time ordered two-point function of the dual operator. We will work in the approximation in which the scalar field is free and propagates without backreaction.

\subsubsection{Generalities}
\label{sec:generalities}
Before considering specific contours, we first discuss some generalities regarding the solutions to the Klein-Gordon equation that are valid for each Lorentzian and Euclidean segment separately.

We start from the action
\be
\label{eq:Sfreescalar}
S= \frac{1}{2}\int d^{d+1}x\, \sqrt{-G} (-\partial_\mu\Phi \partial^\mu \Phi - m^2 \Phi^2)\,,
\ee
with $d=2$ and the metric $G_{\mu \nu}$ given by \eqref{eq:metricemptyads3}. As usual, we have $m^2 = \Delta(\Delta -2)$ with $\Delta - 1 = l \in \{0,1,2,\ldots\}$. As we already discussed in \cite{us}, the regular mode solutions to the Klein-Gordon equation are of the form
\be
e^{-i\omega t + ik\phi} f(\omega,|k|,r)\,,
\ee
with
\be
\label{eq:radialmodesads}
f(\omega,k,r) = C_{\omega k l} (1+r^2)^{\omega/2} r^{k}
F((\omega + k + 1 + l)/2,(\omega + k + 1 - l)/2;k+1;-r^2) \,,\\
\ee
where $F$ is a hypergeometric function and $C_{\omega k l}$ is a normalization factor chosen such that the coefficient of the leading term equals 1. For large $r$, this solution behaves as
\begin{equation}
\label{eq:expansionf}
f(\omega,k,r) = r^{l-1}+ \ldots + r^{-l-1}\alpha(\omega,k,l) [\ln(r^2) + \beta(\omega,k,l)] + \ldots
\end{equation}
with
\begin{eqnarray}
\alpha(\omega,k,l)&=&
\frac{((\omega + k + 1 - l)/2)_l ((\omega - k + 1 - l)/2)_l}{l! (l-1)!}\,,
\nonumber \\
\beta(\omega,k,l) &=& - \psi((\omega + k + 1 + l)/2) -
\psi((-\omega + k + 1 - l)/2)\,,
\end{eqnarray}
where $(a)_n = \Gamma(a+n)/\Gamma(a)$ is the Pochhammer symbol and $\psi(x)=d \ln \Gamma(x)/dx$ is the digamma function. In the expansion \eqref{eq:expansionf} we omitted terms of lower powers of $r$ and some terms polynomial in $\omega$ and $k$ (which would lead to contact terms in the 2-point function).

\begin{figure}
\centering
\psfrag{w}{$\omega$}
\psfrag{0}{$0$}
\includegraphics[width=9cm]{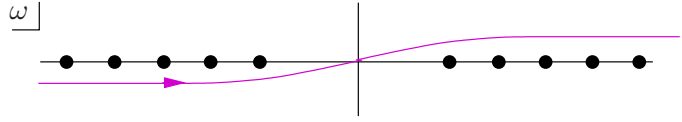}
\caption{\label{fig:freqcontour}The dots represent poles in complex frequency space and the curved line a Feynman contour that avoids them.}
\end{figure}

A bulk-boundary propagator can be obtained by an integral over $\omega$ and a sum over $k$ of these modes. However, if the frequency equals
\be
\omega = \omega_{nk}^\pm \equiv \pm(2n + k + 1 + l)\,,  \quad n \in \{0,1,2,\ldots\}\,,
\ee
then the term $\alpha(\omega,k,l)\beta(\omega,k,l)$ in the radial expansion of the modes has a pole and therefore the modes become singular. To obtain a well-defined bulk-boundary propagator, one needs to specify a contour in $\omega$-space around these poles, for example the contour sketched in Fig.~\ref{fig:freqcontour}. Furthermore, the residues of these poles form exactly the normalizable modes, since the poles only occur at normalizable order in the radial expansion of the modes (indeed, the first terms in this expansion are always local so they cannot contain poles in frequency space). Let us denote them by $g(\omega_{nk},k,r)$:
\be
\begin{split}
g(\omega_{nk},k,r) &= \oint_{\omega_{nk}} d\omega f(\omega_{nk},k,r)\\
&= r^{-l-1} \alpha(\omega_{nk},k,l) \Big( \oint_{\omega_{nk}} d\omega \beta(\omega,k,l)  \Big) + \ldots\\&= r^{-l-1} 4 \pi i \, \alpha(\omega_{nk},|k|,l)  + \ldots\,,
\end{split}
\ee
where the contour is defined as counterclockwise for $\omega_{nk}^-$ and clockwise for $\omega_{nk}^+$, so that $g(\omega_{nk}^+,|k|,r) = g(\omega_{nk}^-,|k|,r)$ and $\alpha(\omega_{nk}^+,|k|,l) = \alpha(\omega_{nk}^-,|k|,l)$.

These normalizable modes can be added at will to any solution without changing the asymptotics for large $r$. The most general solution (without specifying any initial or final data) therefore involves an arbitrary sum over these modes and is thus of the form
\begin{equation}
\label{eq:mostgeneralphi}
\begin{split}
\Phi(t,\phi,r) &= \frac{1}{4\pi^2}\sum_{k \in \mathbb Z} \int_C d\omega
\int d\hat t \int d \hat \phi e^{-i\omega (t - \hat t)
+ ik(\phi - \hat \phi)} \phi_{(0)}(\hat t, \hat \phi) f(\omega,|k|,r)
\\&\qquad + \sum_{\pm} \sum_{k \in \mathbb Z}
\sum_{n = 0}^{\infty} c_{nk}^\pm e^{-i\omega_{nk}^\pm t + ik\phi}g(\omega_{nk},|k|,r)\,,
\end{split}
\end{equation}
with so far arbitrary coefficients $c_{nk}^\pm$ (provided the sum converges). For convenience, let us fix the contour $C$ to be of the Feynman form sketched in Fig.~\ref{fig:freqcontour}. Any different contour can then be implemented by changing the normalizable modes. As we show in detail below, the initial and final data, so the other segments and matching conditions, will eventually completely fix the $c_{nk}^\pm$.

Below, we will often make use of the following observation. To the past of all the sources, the contour of the $\omega$-integral can be closed in the upper half of the complex frequency plane. The choice for a Feynman contour implies that we pick up the poles at the negative frequencies only, which we repeat are just the normalizable modes. The solution can then be fully written as a sum over normalizable modes,
\be
\begin{split}
\Phi &= \frac{1}{4\pi^2}\sum_{n=0}^\infty
\sum_{k \in \mathbb Z} e^{-i \omega_{nk}^- t  + ik \phi}
\phi_{(0)} (\omega_{nk}^-,k) g(\omega_{nk},|k|,r)
\\&\qquad + \sum_{\pm} \sum_{k \in \mathbb Z}
\sum_{n = 0}^{\infty} c_{nk}^\pm
e^{-i\omega_{nk}^\pm t + ik\phi}g(\omega_{nk},|k|,r)\,,
\end{split}
\ee
which is to be expected by completeness of the normalizable modes. Similarly, to the future of all the sources, we can deform the contour in the lower half plane and pick up the residues at the positive frequencies.

Next, consider the solution on the Euclidean segments. One can obtain the mode solutions by a replacement of the form $t = -i \tau$. We will set all sources to zero along the Euclidean segments, so the solutions there will always consist of normalizable modes only,
\be
\Phi_E(\tau,\phi,r) = \sum_{\pm} \sum_{k \in \mathbb Z} \sum_{n = 0}^{\infty} d_{nk}^\pm e^{\omega_{nk}^\pm \tau + ik\phi}g(\omega_{nk},|k|,r)\,,
\ee
with to be determined coefficients $d_{nk}^\pm$. Note that, if a contour extends all the way to $\tau \to \infty$, then we also require finiteness of the solution in this limit. This corresponds to the absence of any sources at this point. Such a condition directly implies that all the $d_{nk}^+$ are zero, whereas the $d_{nk}^-$ are still unconstrained. The converse statement holds for a contour extending to $\tau \to - \infty$.

This finishes the introduction of the solutions on the various segments; we can now consider specific contours and see how the matching conditions specify the coefficients of the normalizable modes for us.

\subsubsection{Wightman functions}
\label{sec:wmfunction}
Our first example is the computation of a vacuum-to-vacuum two-point function using an in-in formalism. As explained in appendix \ref{sec:app:fieldthy}, the in-in formalism in particular allows for the computation of Wightman functions directly from a path integral. In our case, we can do the same holographically.

\begin{figure}
\centering
\psfrag{tE0}{$\tau_{0}$}
\psfrag{tL1}{$t_{1}$}
\psfrag{tL2}{$t_{2}$}
\psfrag{tE2}{$\tau_{3}$}
\includegraphics[width=7cm]{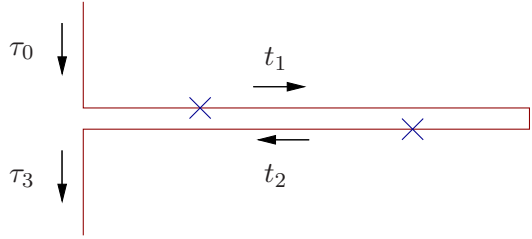}
\caption{\label{fig:wmcontour}The in-in contour we use to compute a Wightman function. We choose time coordinates that increase in the direction of the arrows.}
\end{figure}

Let us therefore consider the contour sketched in
Fig.~\ref{fig:contoursintplane}b, given again in Fig.~\ref{fig:wmcontour}. It runs from $i\infty$ to $0$, then to $T$ (with $T$ real and positive), then back to the origin and then to $-i\infty$. As we outlined above and sketched on the right of Fig.~\ref{fig:contoursintplane}b, for such a contour we consider a filling that consists of two Lorentzian AdS$_3$ spacetimes between two Euclidean AdS$_3$ caps. These four space(time)s will be denoted as $M_i$, with $i$ running from $0$ to $3$. We will use a subscript $i$ also on other quantities to distinguish on which of the $M_i$ they are defined. Sometimes, in order to avoid confusion with other subscripts, we will put this subscript in square brackets, writing for example $c_{[i]}$.

We can again split the contour-integrated action into the following combination:
\be
- \int_{-\infty}^0 d\tau_0 L_E(\Phi_0) + i \int_0^T dt_1 L_L(\Phi_1) - i \int_T^{2T} dt_2 L_L(\Phi_2) - \int_{0}^\infty d\tau_3 L_E(\Phi_3)\,,
\ee
with the Lagrangians
\be
\begin{split}
L_L(\Phi) &= \frac{1}{2} \int d^2 x \sqrt {|G|} (- \partial_\mu \Phi \partial^\mu \Phi - m^2 \Phi^2)\,,\\
L_E(\Phi) &= \frac{1}{2} \int d^2 x \sqrt {|G|} (\partial_\mu \Phi \partial^\mu \Phi + m^2 \Phi^2)\,.
\end{split}
\ee
We use a real contour time coordinate on every segment $M_i$ whose direction is indicated in Fig.~\ref{fig:wmcontour}. We glue the surface given by $\tau_3 = 0$ to that given by $t_2 = 2T$, and similarly the surfaces $t_1 = T$ to $t_2 = T$ and $t_1 = 0$ to $\tau_0 = 0$.

A full list of matching conditions is now given by continuity of the fields, plus continuity of their derivatives with appropriate signs. These signs are easily found by equating the conjugate momenta obtained from functional differentiation of the on-shell actions. One obtains:
\begin{equation}
\label{eq:2ndconditions}
\begin{split}
- \partial_{\tau_0} \Phi_0(\tau_0 = 0) - i \partial_{t_1} \Phi_1(t_1 = 0) &= 0\\
+ i \partial_{t_1} \Phi_0(t_2 = T) + i \partial_{t_2} \Phi_2(t_2 = T) &= 0\\
- i \partial_{t_2} \Phi_2(\tau_2 = 2T) + \partial_{\tau_3} \Phi_3(\tau_3 = 0) &= 0\,.
\end{split}
\end{equation}
Consider the case of a nonzero source $\phi_{(0)[1]}$ only on the conformal boundary of $M_1$, so $\Phi_1$ is given by (cf. \eqref{eq:mostgeneralphi}):
\bea
\Phi_1(t_1,\phi,r) &=& \frac{1}{4\pi^2}\sum_{k \in \mathbb Z} \int_C d\omega
\int_{M_1} d\hat t \int d \hat \phi e^{-i\omega (t_1 - \hat t)
+ ik(\phi - \hat \phi)} \phi_{(0)[1]}(\hat t, \hat \phi) f(\omega,|k|,r)
\nonumber \\ & &\qquad + \sum_{\pm} \sum_{k \in \mathbb Z}
\sum_{n = 0}^{\infty}
c_{[1]nk}^\pm e^{-i\omega_{nk}^\pm t_1 + ik\phi}g(\omega_{nk},|k|,r)\,,
\label{eq:phi1general}
\eea
with to be determined coefficients $c_{[1]nk}^\pm$. As in \cite{us}, we take the source to vanish near $t_1 = 0$ and $t_1 = T$. By performing
the $\omega$-integral, we write the solution as a sum over normalizable
modes in the vicinity of these hypersurfaces:
\bea
\Phi_1(t_1 \sim 0, \phi,r) &=& \frac{1}{4\pi^2}\sum_{n=0}^\infty
\sum_{k \in \mathbb Z} e^{-i \omega_{nk}^- t_1  + ik \phi}
\phi_{(0)[1]} (\omega_{nk}^-,k) g(\omega_{nk},|k|,r)
\nonumber \\ &&\qquad+ \sum_{\pm} \sum_{k \in \mathbb Z}
\sum_{n = 0}^{\infty} c_{[1]nk}^\pm
e^{-i\omega_{nk}^\pm t_1 + ik\phi}g(\omega_{nk},|k|,r) \,,\label{eq:phi1nearboundaries}
\\
\Phi_1(t_1 \sim T, \phi,r) &=& \frac{1}{4\pi^2}\sum_{n=0}^\infty
\sum_{k \in \mathbb Z} e^{-i \omega_{nk}^+ t_1  + ik \phi}
\phi_{(0)[1]} (\omega_{nk}^+,k) g(\omega_{nk},|k|,r)
\nonumber \\ && \qquad+ \sum_{\pm} \sum_{k \in \mathbb Z}
\sum_{n = 0}^{\infty} c_{[1]nk}^\pm
e^{-i\omega_{nk}^\pm t_1 + ik\phi}g(\omega_{nk},|k|,r)\,. \nonumber
\eea
Since there is no source on the other segments, the solutions
$\Phi_0$, $\Phi_2$ and $\Phi_3$ are just sums over normalizable modes.
For $\Phi_2$ we obtain:
\begin{equation}
\label{eq:phi2general}
\Phi_2(t_2,\phi,r) = \sum_{\pm} \sum_{k \in \mathbb Z}
\sum_{n = 0}^{\infty} c_{[2]nk}^\pm
e^{-i\omega_{nk}^\pm t_2 + ik\phi}g(\omega_{nk},|k|,r)\,.
\end{equation}
For $M_0$ we can only allow for modes of the form $e^{+|\omega|\tau_0}$, since $\tau_0$ extends to $-\infty$. Similarly, since $\tau_3 \to \infty$ on $M_3$, the modes there are of the form $e^{-|\omega|\tau_3}$. We thus find that
\begin{equation}
\begin{split}
\Phi_0(\tau_0,\phi,r) = \sum_{k \in \mathbb Z}
\sum_{n = 0}^{\infty} c_{[0]nk}
e^{\omega_{nk}^+ \tau_0 + ik\phi}g(\omega_{nk},|k|,r)\,,\\
\Phi_3(\tau_3,\phi,r) = \sum_{k \in \mathbb Z}
\sum_{n = 0}^{\infty} c_{[3]nk}
e^{-\omega_{nk}^+ \tau_3 + ik\phi}g(\omega_{nk},|k|,r)\,.\\
\end{split}
\end{equation}
The matching conditions will now determine the $c_{[i]nk}^\pm$ for us. Since the different modes $g(\omega_{nk},|k|,r)$ are orthogonal (up to the symmetry $g(\omega^+_{nk},k,r) = g(\omega^-_{nk},k,r)$), we can do the matching `mode-wise', \ie we can compare the coefficients of the various modes. For example, the first matching between $M_1$ and $M_0$, which is $\Phi_1(t_1 = 0,\phi,r) = \Phi_0(\tau_0 = 0,\phi,r)$, yields
\be
c_{[0]nk} = \frac{1}{4\pi^2} \phi_{(0)[1]}(\omega_{nk}^-,k)
+ c_{[1]nk}^+ + c_{[1]nk}^-\,,
\ee
and the second matching condition, which is the first
equation in \eqref{eq:2ndconditions}, becomes
\be
- \omega_{nk}^+ c_{[0]nk} - \frac{1}{4\pi^2}\omega_{nk}^- \phi_{(0)[1]}(\omega_{nk}^-,k) - \omega_{nk}^+ c_{[1]nk}^+ - \omega_{nk}^- c_{[1]nk} = 0\, .
\ee
Recalling that $\omega_{nk}^+ = - \omega_{nk}^-$ and combining the two matching conditions, we find that
\begin{equation}
\label{eq:nopositivefreq}
c_{[1]nk}^+ = 0 \,,
\end{equation}
which is the statement that there are no positive frequencies to the past of the sources.

Similarly, from the matching conditions between $M_2$ and $M_3$, one deduces
\begin{equation}
\label{eq:c2nk+}
c_{[2]nk}^+ = 0\,,
\end{equation}
so on $M_2$ we can only allow for negative frequencies with respect to $t_2$. Then, from the matching condition between $M_1$ and $M_2$, we see that frequencies should be inverted on $M_2$: positive frequencies on $M_1$ become negative frequencies on $M_2$ (with respect to $t_2$)  and vice versa. Therefore, on $M_1$ there can only be positive frequencies close to $t_1 = T$. Indeed, working out the details results in
\be
c_{[1]nk}^- = 0\,,
\ee
which, combined with \eqref{eq:nopositivefreq}, completely fixes the $c_{[1]nk}^\pm$ on $M_1$ to be zero. We thus obtain the usual Feynman prescription for the bulk-boundary propagator on $M_1$, which is reassuring: using the in-in instead of the in-out formalism should not have changed our result of \cite{us} and indeed we found that it did not.

The solution is now completely fixed and one may compute all of the $c_{[2]nk}^\pm$ and the $c_{[0]nk}$ and $c_{[3]nk}$ using the matching conditions. For the Wightman function, we will only be interested on the solution on $M_2$ for a source on $M_1$, so we will only need the $c_{[2]nk}^\pm$. Equation \eqref{eq:c2nk+} already fixed half of them, and the first matching condition between $M_1$ and $M_2$ yields
\be
c_{[2]nk}^- = \frac{1}{4\pi^2} \phi_{(0)[1]}(\omega_{nk}^+,k) e^{- 2 i\omega_{nk}^+ T}\,.
\ee

With the solutions determined, consider the one-point functions. As we mentioned in section \ref{ren}, the gluing of different solutions does not affect the usual prescription that the renormalized one-point function in the presence of sources is given by the renormalized radial conjugate momentum. For the case under consideration, we thus obtain (up to contact terms)
\be
\label{eq:vevscalar}
\vev{\op_{[i]}(x)} =\frac{i}{\sqrt{-g_{(0)}}}\fdel{\phi_{(0)[i]}}(- S_{0}+ i S_{1} - i S_{2} - S_{3} + S_{ct}) = -2l\phi_{(2l)[i]} (x)\,,
\ee
with $\phi_{(2l)[i]}$ the term of order $\sim r^{-l-1}$ in the radial expansion of $\Phi_{i}$.

We are in particular interested in the Wightman function:
\be
\begin{split}
\vev{\op(x) \op(x')} &= \vev{T_C \op_{[2]}(x) \op_{[1]}(x')} \\&= \frac{i}{\sqrt{-g_{(0)}}}\fdel{\phi_{(0)[1]}(x')} \vev{\op_{[2]}(x)} \\&= - 2li \fdel[\phi_{(2l)[2]}(x)]{\phi_{(0)[1]}(x')}\,.
\end{split}
\ee
With the solution $\Phi_2$ we found above, we obtain
\be
\phi_{(2l)[2]}(x) = \frac{i}{\pi} \sum_{k \in \mathbb Z}
\sum_{n = 0}^{\infty}  \alpha(\omega_{nk},|k|,l)e^{-i\omega_{nk}^+ (2T -t_2) + ik\phi} \int_{M_1} d\hat t d\hat \phi e^{i\omega_{nk}^+ \hat t -ik \hat \phi} \phi_{(0)[1]}(\hat t,\hat \phi)\,.
\ee
Using $t = 2T - t_2$ and taking the functional derivative, we get:
\be
\begin{split}
\langle \op(x) \op(x') \rangle &=
\frac{2l}{\pi}\sum_{n\geq 0} \sum_{k \in \mathbb Z} e^{-i\omega_{nk}^+ (t-t') + ik(\phi - \phi')}\alpha(\omega_{nk}^+,|k|,l)\,.
\end{split}
\ee
This expression satisfies some standard checks that are expected for a Wightman function, namely it vanishes for $\omega < 0$ and the coefficients are real and positive definite, see appendix \ref{sec:app:fieldthy}. Evaluating the summations, we find
\be
\langle \op(x) \op(x') \rangle = \frac{l^2/(2^l \pi)}{[\cos(t -i\epsilon) - \cos(\phi)]^{l+1}}\,,
\ee
which has poles when $t - i\epsilon$ is real, so it is analytic in the lower half plane, also as expected.

\subsubsection{Thermal AdS}
\label{sec:thermalads}
Let us now consider the thermal contour indicated in
Fig.~\ref{fig:contoursintplane}c. We again take $t_1$ to run from $0$ to $T$ on $M_1$, $t_2$ to run from $T$ to $2T$ on $M_2$, and $\tau$ to run from $0$ to $\beta$ on $M_3$. The novelty here is that we glue the part with $\tau = \beta$ to the surface with $t_1 = 0$ in order to obtain a thermal state. The action splits into:
\be
i \int_0^T dt_1 L_L[\Phi_1] - i \int_T^{2T} dt_2 L_L[\Phi_2] - \int_0^\beta d\tau L_E[\Phi_3]\,.
\ee
Again, we consider a source living only on $M_1$ and solve the Klein-Gordon equation. The general expressions for $\Phi_1$ and $\Phi_2$ (without specification of initial and final data) are exactly the same as before and are given by the equations \eqref{eq:phi1general} and \eqref{eq:phi2general}. We can also use the expressions \eqref{eq:phi1nearboundaries} for $\Phi_1$ when $t \sim 0$ or $t \sim T$. For $\Phi_3$, we may allow for both positive and negative frequencies and the most general purely normalizable solution is:
\be
\Phi_3(\tau,\phi,r) =\sum_\pm \sum_{k \in \mathbb Z}
\sum_{n = 0}^{\infty} c_{[3]nk}^\pm
e^{\omega_{nk}^\pm \tau + ik\phi}g(\omega_{nk},|k|,r)\,.
\ee

The full list of matching conditions is now
\begin{align*}
\Phi_1(t_1 = T) &= \Phi_2(t_2 = T) &  \partial_{t}\Phi_1(t_1 = T) &= - \partial_{t} \Phi_2(t_2 = T)\\
\Phi_2(t_2 = 2T) &= \Phi_3(\tau = 0) & i \partial_{t}\Phi_2(t_2 = 2T) &= \partial_{\tau} \Phi_3(\tau = 0)\\
\Phi_1(t_1 = 0) &= \Phi_3(\tau = \beta) & -i \partial_{t}\Phi_1(t_1 = 0) &= \partial_{\tau}\Phi_3(\tau = \beta)\,,
\end{align*}
which, after some algebraic manipulations, results in
\be
c_{[1]nk}^\pm = \frac{1}{4\pi^2} \phi_{(0)[1]}(\omega_{nk}^\pm,k) \frac{1}{e^{\beta \omega_{nk}^+ } - 1}\,.
\ee
The nonzero $c_{[1]nk}^\pm$ directly enter into the two-point function and we get
\be
\begin{split}
\vev{T\op_{[1]}(x')\op_{[1]}(x)}_\beta &= \frac{l}{2\pi^2 i}\sum_{k \in \mathbb Z} \int_C d\omega
e^{-i\omega (t - t')
+ ik(\phi - \phi')} \alpha(\omega,|k|,l)\beta(\omega,|k|,l)
\\ &\qquad + \frac{2l}{\pi} \sum_{\pm} \sum_{k \in \mathbb Z}
\sum_{n = 0}^{\infty}   \frac{\alpha(\omega_{nk},|k|,l)}{e^{\beta |\omega_{nk}^\pm|}- 1}  e^{-i\omega_{nk}^\pm (t - t')
+ ik(\phi - \phi')}\,,
\end{split}
\ee
with the subscript $\beta$ indicating the temperature.
As in the `free-field' approximation, we find the sum of the
zero-temperature Feynman propagator and a heat-bath
contribution. Also, notice the symmetry $x \leftrightarrow x'$. After
rewriting the thermal contributions as geometric series, one readily
finds that this expression becomes
\be \label{2pt_EAdS}
\langle T \op(x) \op(x') \rangle_\beta = \sum_{n\in \mathbb Z} \frac{l^2/(2^l \pi)}{[\cos(t -i\epsilon t + i n\beta) - \cos(\phi)]^{l+1}}\,,
\ee
which satisfies the KMS condition, so it corresponds to a thermal two-point function indeed. It is a sum over images of the zero temperature result,
reflecting the fact that Euclidean thermal AdS is obtained by identifications in the
time direction of Euclidean global AdS.

One can actually arrive more directly at
(\ref{2pt_EAdS}) by using the relation between
thermal AdS and global AdS to first
obtain the Euclidean correlator by a sum over images and then analytically continue to real-time.
This was the way
(\ref{2pt_EAdS}) was obtained earlier in \cite{Birmingham:2002ph}. Of course,
the $i\epsilon$ insertions then have to be fixed by hand.
The emphasis here is on the fact that we can unambiguously arrive at
(\ref{2pt_EAdS}), including the correct $i\epsilon$ insertions, by employing a Lorentzian signature gauge/gravity dictionary and without assuming any special properties of the background under consideration.

\subsection{Poincar\'e coordinates}
\label{sec:poincare}
For our next example we consider a CFT in
$d$-dimensional Minkowski ($Mink_d$) spacetime.
As is well known, Minkowski spacetime is conformally isometric to an
open region of the Einstein static universe, $\Real \times S^{d-1}$.
Thus the correlators for the CFT in Minkowski spacetime can
be obtained from those of the Einstein universe (as we
demonstrate for $d=2$ in subsection \ref{2pt_poinc}).
Since boundary Weyl transformations are a specific class of bulk
diffeomorphisms \cite{Imbimbo:1999bj,Skenderis:2000in},
a similar procedure can be done holographically.

Nevertheless, it is still interesting to directly compute the
correlators in Minkowski spacetime, not least because this is the typical background for most QFT computations.
Furthermore, for a CFT on $Mink_d$ to be  exactly equivalent
to the theory on $\Real \times S^{d-1}$
the boundary conditions of all QFT fields
at infinity of  $Mink_d$ must be the ones dictated by the theory
on $\Real \times S^{d-1}$. One may however wish to study the QFT on
$Mink_d$ with fields satisfying different boundary conditions at infinity.
For example, the ground state
of a conformally coupled scalar $\phi$ on $\Real \times S^{d-1}$ has
necessarily $\vev{\phi}=0$ because of the curvature coupling of the scalar.
The same theory on $Mink_d$ however allows for ground states with
non-vanishing $\vev{\phi}$, since in this case the curvature coupling vanishes.
In such cases the nonzero scalar vev spontaneously breaks conformal
invariance. This is described in the bulk by domain wall
spacetimes containing additional bulk
fields capturing the vevs of gauge invariant operators.
One can extend the methods described here to apply to the
computation of real-time correlators along holographic RG flows,
extending the Euclidean computations in \cite{Bianchi:2001de,Bianchi:2001kw},
but we shall not discuss this in detail here.

Instead we will compute vacuum-to-vacuum amplitudes for
the CFT without vevs. To this end, we consider the field
theory path of Fig.~\ref{fig:contour}a, but with $\Real^{d-1}$ as the
spacelike part of the boundary manifold. We can compactify the entire
contour by adding a single point, resulting in the boundary geometry
shown in Fig.~\ref{fig:pacman}a. The Lorentzian segment is cut off at
finite initial and final times $t = \pm T$. Below
we holographically compute a time-ordered two-point function for a
CFT in this background.

\subsubsection{Bulk spacetime}
As before, the first step is to find a suitable bulk manifold that fills in the contour.

\begin{figure}
\centering
\psfrag{a}{(a)}
\psfrag{b}{(b)}
\psfrag{t}{$t$}
\psfrag{t = -i}{$t\!=\!-\infty$}
\psfrag{t = i}{$t\!=\!\infty$}
\psfrag{t=-T}{$t\!=\!-T$}
\psfrag{t=T}{$t\!=\!T$}
\includegraphics[width=9cm]{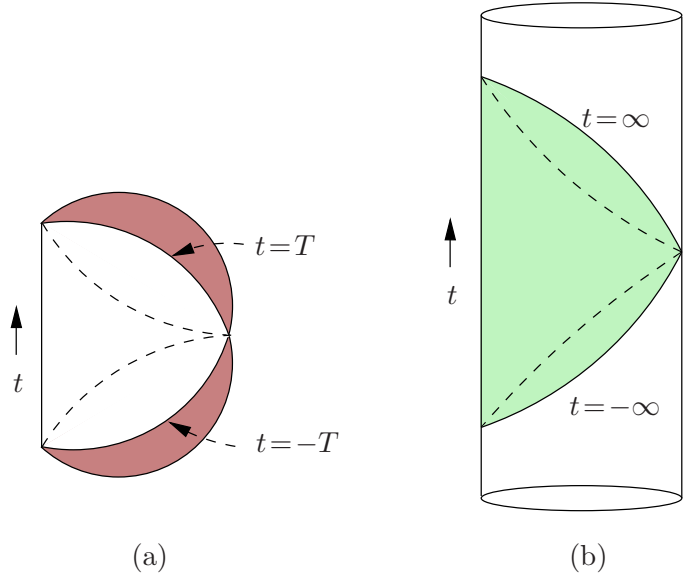}
\caption{\label{fig:pacman}(a) The geometry used for the computation of the two-point function in $Mink_d$. The Lorentzian manifold is cut off at slices given by $t = \pm T$, to which the darker shaded Euclidean caps are glued. (b) The Poincar\'e coordinate system covers only a part of AdS$_{d+1}$. Both the global AdS time and the Poincar\'e time run upward. The planes $t = \pm \infty$ bound the coordinate system.}
\end{figure}

We begin with the Lorentzian segment of the contour. In the absence of any sources and vevs, it is filled in with a segment of empty AdS$_{d+1}$ in Poincar\'e coordinates:
\begin{equation}
\label{eq:adspoinc}
ds^2 = \frac{dz^2 - dt^2 + d\vec x^2}{z^2}\,.
\end{equation}
The Poincar\'e coordinate system covers only a part of all of
AdS$_{d+1}$, as indicated in Fig.~\ref{fig:pacman}b. We will however cut
off the bulk manifold along the hypersurfaces $t = \pm T$ and therefore we will not
need the rest of the AdS$_{d+1}$ spacetime anyway. The Lorentzian
segment with the above metric and $-T < t <T$ will be referred to as
$M_1$ below.

The two Euclidean segments can be filled with Euclidean AdS$_{d+1}$, whose metric can be obtained from \eqref{eq:adspoinc} by the replacement $t = - i\tau$. We again need only a part of these spaces and cut off the Euclidean solutions along hypersurfaces of constant $\tau$. More precisely, we call $M_0$ the Euclidean manifold with the metric
\begin{equation}
\label{eq:eadspoinc}
ds^2 = \frac{dz^2 + d\tau_0^2 + d\vec x^2}{z^2}
\end{equation}
and $\tau_0 < 0$. Similarly, we take $M_2$ the Euclidean manifold with the $\tau_2 > 0$ and the same metric \eqref{eq:eadspoinc} with the replacement $\tau_0 \to \tau_2$.

Next, we glue the three components together by gluing the surface given by $\tau_0 = 0$ on $M_0$ to the surface $t = - T$ on $M_1$, and the surface $\tau_2 = 0$ to the surface $t = T$ on $M_1$. One may easily verify that the matching conditions for gravity are satisfied, since the induced metric on surfaces of constant $t$ or $\tau$ is the same and these surfaces are totally geodesic. We conclude that the combination of $M_0$, $M_1$ and $M_2$ satisfies all the holographic boundary data as well as all the matching conditions, and so it can serve as the background around which we study perturbations below.

\subsubsection{Solutions}
We will again obtain a time-ordered two-point function of a scalar operator of conformal weight $\Delta = \frac{d}{2}+ l$, with $l \in \{0,1,2,\ldots\}$, which is dual to a bulk scalar field of mass $m^2 = \Delta(\Delta -d)$. As we did in the previous examples, we take the scalar field to propagate freely and without backreaction.

On $M_1$ the action for the scalar field is again \eqref{eq:Sfreescalar}, this time with the metric \eqref{eq:adspoinc}. Solutions to the equations of motion satisfy the Klein-Gordon equation:
\be
z^{d+1} \partial_z (z^{-d+1} \partial_z \Phi) + z^2 \square_0 \Phi - m^2 \Phi = 0\,.
\ee
After separation of variables we find modes labeled by $(\omega,\vec k)$:
\be
\label{eq:modespoinc}
e^{-i\omega t + i\vec k \cdot \vec x} z^{d/2} K_{l}(q z) \,,\qquad \qquad e^{-i\omega t + i\vec k\cdot \vec x} z^{d/2} I_l (qz)\,.
\ee
For spacelike momenta $q^2 = -\omega^2 + \vec k^2 > 0$, these modes are unambiguously defined. For timelike momenta $q^2 < 0$, we have to consider possible branch cuts. First of all, we put the square root in defining $q = \sqrt{q^2}$ just above the negative real axis. We indicate this by using
\be
q_\epsilon = \sqrt{-\omega^2 + \vec k^2 -i\epsilon}\,.
\ee
Second, $K_l$ has a branch cut along the negative real axis, which is however unimportant since $|\arg (q_\epsilon z)| \leq \pi/2$. Finally, $I_l$ has no branch cut since $l$ is an integer.

To select the right solution on $M_1$, we should look at the asymptotics:
\be
\begin{split}
z^{d/2}K_l(qz \to 0) &= \Gamma(l)\frac{z^{d/2-l}}{2^{l+1}q^l}+\ldots\\
z^{d/2}I_l(qz \to 0) &= \frac{1}{\Gamma(l+1) }\frac{z^{d/2 + l}}{2^lq^{-l}} + \ldots\\
z^{d/2}K_l(qz \to \infty) &= \sqrt{\frac{\pi z^{d-1}}{2 q}} e^{- qz} + \ldots\\
z^{d/2}I_l(qz \to \infty) &= \sqrt{\frac{z^{d-1}}{2\pi q}} [e^{qz} + e^{-qz - (l+ \frac{1}{2})\pi i}] +\ldots
\end{split}
\ee
For spacelike momenta, finiteness as $z\to \infty$ selects
$z^{d/2}K_l(qz)$ as the only correct solution. On the other hand, for
timelike momenta no linear combination of the solutions remains finite
as $z \to \infty$, which means that any solution that does remain
finite as $z \to \infty$ should be obtained as an infinite sum over
the modes. Furthermore, from the asymptotics as $z \to 0$, we
find that the modes $z^{d/2}K_l(q_\epsilon z) \sim z^{d/2 -l}$
correspond to sources on the conformal boundary, whereas the
$z^{d/2}I_l(q_\epsilon z) \sim z^{d/2 + l}$ are the normalizable
modes.

For timelike momenta $q_\epsilon z = - i |q| z$ and we will henceforth rewrite the modified Bessel function of the first kind using $I_l(-i|q|z) = e^{-i\pi l/2} J_l(|q|z)$. Although we could also have rewritten $K_l(-i|q|z) = (i\pi e^{il\pi/2}/2)H_n^{(1)}(|q|z)$, we do not do so below, since $z^{d/2}K_l(q_\epsilon z)$ is needed for both spacelike and timelike momenta. We emphasize that $K_l(q_\epsilon z)$ is unambiguously defined for all real $q^2$.

Next, consider the manifolds $M_0$ and $M_2$, both with the Euclidean
metric \eqref{eq:eadspoinc} and $0 < \tau < \infty$ and $-\infty <
\tau < 0$, respectively. Although we will mainly work in position
space below, we will for completeness present the mode solutions here
as well. First of all, the mode solutions on $M_0$ and $M_2$ are
obtained by the usual substitution $t \to - i\tau$ in the Lorentzian
modes \eqref{eq:modespoinc}. Since we will not switch on any
sources on these segments, the solutions on $M_0$ and $M_2$ need to be
purely normalizable. As we just showed, this implies that only the
modes $z^{d/2}J_l(|q| z)$ with $q^2 < 0$ are allowed. Furthermore,
since no operators are inserted at the points $\tau \to \pm \infty$,
we will also request finiteness of the solution in this limit. This
implies a restriction to negative frequencies on $M_0$ and to
positive frequencies on $M_2$. More explicitly, the solutions on these
segments are built up from the modes
\be
\begin{split}
e^{|\omega| \tau_0 + i\vec k \cdot \vec x} z^{d/2}J_l(|q|z) \qquad \text{on } M_0\,,\\
e^{-|\omega| \tau_2 + i\vec k \cdot \vec x} z^{d/2}J_l(|q|z) \qquad \text{on } M_2\,,\\
\end{split}
\ee
with $- \omega^2 + \vec{k}^2 < 0$. Since the individual modes diverge as $z \to \infty$, we should again sum an infinite number of these modes in order to get a solution that vanishes also at this point.

\subsubsection{Bulk-boundary propagator}
\begin{figure}
\centering
\psfrag{w}{$\omega$}
\psfrag{-k}{$-k$}
\psfrag{k}{$k$}
\includegraphics{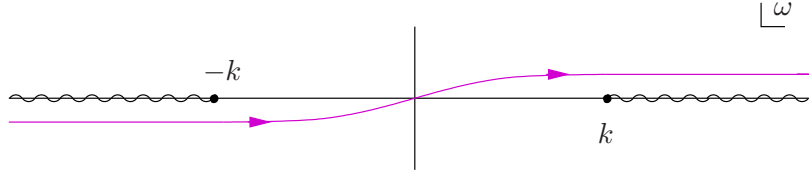}
\caption{\label{fig:wintegral} The contour around the branch cuts (wavy lines) in the complex frequency plane used to define a bulk-boundary propagator.}
\end{figure}

The next step is to compute a bulk-boundary propagator, which we denote by $X(t,\vec x , z)$. We will consider the propagator for a source on the conformal boundary of $M_1$ only. Let us first investigate the solution on $M_1$. Inspired by the Euclidean bulk-boundary propagator, we may try:
\begin{equation}
\label{eq:bbpoincfourier}
X_1(t,\vec x,z) = \frac{1}{(2\pi)^d}
\int_C d\omega \int d\vec k \, e^{-i\omega t + i\vec k \cdot \vec x} \frac{2^{l+1}q_\epsilon^l}{\Gamma(l)} z^{d/2}K_l(q_\epsilon z)\,.
\end{equation}
The $i\epsilon$-prescription is equivalent to a Feynman contour
$C$ in the $\omega$-plane around the branch cuts which we show in Fig.~\ref{fig:wintegral}. The expression \eqref{eq:bbpoincfourier} is not obviously
convergent as $z\to \infty$. However, we can perform the Fourier transform by closing and deforming the contour. The $i \epsilon$-prescription tells us which branch cuts we pick up and the corresponding position-space expression is equal to
\begin{equation}
\label{eq:bbpoinc}
X_1(t,\vec x,z) = i\Gamma(l)\Gamma(l+\frac{d}{2}) \pi^{-\frac{d}{2}} \frac{z^{l + \frac{d}{2}}}{(-t^2 + \vec x^2 + z^2 + i\epsilon)^{l+ d/2}}\,,
\end{equation}
which clearly converges for large $z$.

As in the previous section, this bulk-boundary propagator is not unique without imposing initial and final conditions. Indeed, one may always add a normalizable solution, which we will denote as $Y(t,\vec x, z)$. Notice that we know that normalizable solutions on $M_1$ exist from our discussion of the previous section,
where we used global coordinates. In the previous subsection we found that
$Y(t,\vec x, z)$ must be a linear combination of
the modes $z^{d/2} I_l(q_\epsilon z)$ with $q^2 < 0$, which we write as
\begin{equation}
\label{eq:ypoinc}
Y_{1}(t,\vec x,z) = \frac{1}{(2\pi)^d} \int d\omega \int dk\, e^{-i\omega t + i\vec k\cdot \vec x} \theta(-q^2) c_{[1]}(\omega,\vec k) z^{d/2}J_l(|q| z)\,,
\end{equation}
with further constrains on $c_{[1]}(\omega,\vec k)$ by requesting finiteness for $z \to \infty$ that we will not work out here. To reiterate, without initial or final conditions such normalizable solutions can be added at will to our suggested bulk-boundary propagator \eqref{eq:bbpoincfourier}, so the normalizable solutions parametrize the ambiguity in the bulk-boundary propagator. In particular, any different $i\epsilon$-prescription than the one we fixed above can be implemented by changing these $c_{[1]}(\omega,\vec k)$.

\subsubsection{Matching}
With the solutions on $M_1$ specified, let us now discuss the matching. We will show that the matching conditions imply that $X_1(t,\vec x, z)$ is the right bulk-boundary propagator and that no normalizable solution can be added since $Y_1(t,\vec x,z)$ can never be matched to a regular and normalizable solution on $M_0$ and $M_2$.

We begin with the matching conditions between $M_0$, $M_1$ and $M_2$:
\begin{align}
\Phi_1(t_1=T,\vec x, z) &= \Phi_2(\tau = 0,\vec x, z) & i\partial_{t_1}\Phi_1(t_1=T,\vec x, z) + \partial_\tau \Phi_2(\tau = 0,\vec x, z) &= 0 \nonumber \\
\Phi_1(t_1= - T,\vec x, z) &= \Phi_2(\tau = 0,\vec x, z) & -i\partial_{t_1}\Phi_1(t_1=T,\vec x, z) - \partial_\tau \Phi_2(\tau = 0,\vec x, z) &= 0\,.
\label{eq:matchingpoinc}
\end{align}
Let us now show that we can find solutions $X_0$ and $X_2$ on $M_0$ and $M_2$ that can be matched to $X_1$. This is straightforward in position space, where we can verify that the position-space expressions
\begin{equation}
\label{eq:bbpoinceucl}
\begin{split}
X_0(\tau_0,\vec x,z) = i\Gamma(l)\Gamma(l+\frac{d}{2}) \pi^{-\frac{d}{2}} \frac{z^{\frac{d}{2}+ l}}{(-(-T - i\tau_0)^2 + \vec x^2 + z^2 + i \epsilon)^{l+ d/2}}\,,\\
X_2(\tau_2,\vec x,z) = i\Gamma(l)\Gamma(l+\frac{d}{2}) \pi^{-\frac{d}{2}} \frac{z^{\frac{d}{2}+ l}}{(-(T - i\tau_2)^2 + \vec x^2 + z^2 + i\epsilon)^{l+ d/2}}\,,
\end{split}
\end{equation}
satisfy the equations of motion on all of $M_0,M_2$ and are normalizable. Furthermore, they actually satisfy the matching conditions as well. To see this, notice that the $(+i\epsilon)$-insertions in the denominators of \eqref{eq:bbpoinceucl} are not necessary for nonzero $\tau$, but they are necessary to ensure that \eqref{eq:bbpoinceucl} are well-defined (distributions) on the initial and final hypersurfaces given by $\tau_0 = 0$ and $\tau_2 = 0$. With the given $i\epsilon$-insertions, we can compare \eqref{eq:bbpoinceucl} to \eqref{eq:bbpoinc} and one readily verifies that the matching conditions are satisfied.

Let us now show that one could not have picked any other
$i\epsilon$-insertions ($-i\epsilon$, $+i\epsilon t$, etc.) on the
Lorentzian side. If we would have done so, the matching conditions
would directly dictate a corresponding change in
\eqref{eq:bbpoinceucl}. However, such a change in the Euclidean
solutions is not allowed, because any other $i\epsilon$-insertion in
\eqref{eq:bbpoinceucl} would give a singularity in either $X_0$ or in
$X_2$. For example, if we would replace the $+i\epsilon$ with
$-i\epsilon$ on $M_2$, then $X_2$ would be singular at $\tau_2 =
\epsilon/2T$, around the point given by $\vec x^2 + z ^2 = T^2$
and thus  this solution should be discarded. We conclude that
the  $i\epsilon$-insertion in
\eqref{eq:bbpoinc} is the only one that
moves the singularity everywhere away from the contour.

It remains to show that \eqref{eq:bbpoinc} is the indeed the unique
bulk-boundary propagator by demonstrating that there are no matching
Euclidean solutions for the normalizable solution
\eqref{eq:ypoinc}. Using the normalizable modes we found above, the
solution on $M_0$ should necessarily be of the form
\be
Y_0(\tau_0,\vec x,z) = \int d\omega \int d\vec k \, e^{|\omega| \tau_0 + i\vec k\cdot \vec x} \theta(-q^2) c_{[0]} (|\omega|,\vec k) z^{d/2}J_l(|q|z)\,,
\ee
for some coefficients $c_{[0]}(|\omega|,\vec k)$. A similar expression holds for the solution on $M_2$:
\be
Y_2(\tau_2,\vec x,z) = \int d\omega \int d\vec k\, e^{-|\omega| \tau_2 + i\vec k\cdot \vec x} \theta(-q^2) c_{[2]} (|\omega|,\vec k) z^{d/2}J_l(|q|z)\,.
\ee
Consider now the matching conditions, for example the continuity condition between $M_1$ and $M_2$:
\be
Y_1(T,\vec x,z) = Y_2(\tau_0 = 0)\,.
\ee
Although this is an equality between two integrals, the modes $z^{d/2}J_l(|q|z)$ are orthogonal,
\be
\int_0^\infty dz \,z^{-1} J_l(|q|z) J_l(|q'|z) = c \delta(|q|-|q'|)\,,
\ee
with $c$ a constant. We can therefore equate the integrands (up to $\omega \leftrightarrow -\omega$), which results in
\be
c_{[1]}(\omega,\vec k) + c_{[1]}(-\omega,\vec k) = c_{[0]}(|\omega|,\vec k)\,.
\ee
The other matching conditions can be imposed in a similar way and they ultimately determine $c_{[1]}(\omega,\vec k) = c_{[0]}(\omega,\vec k) = c_{[2]}(\omega,\vec k) = 0$. There is thus no normalizable solution and the bulk-boundary propagator $X(t,\vec x, z)$ is unique.

\subsubsection{Two-point function} \label{2pt_poinc}
As for the computation of the time-ordered two-point function, the only difference with the Euclidean case are the $i\epsilon$-insertions in the bulk-boundary propagator, which in Fourier space corresponds to the replacement $q \to q_\epsilon$. Just as for AdS$_{d+1}$ in global coordinates, these enter directly in the two-point function which, up to contact terms, is then given by:
\be
\vev{T\op(q)\op(-q)} = \frac{i (-1)^{l}}{2^{2l-1} \Gamma(l)^2} q_\epsilon^{2l} \log q_\epsilon\,.
\ee
The $i \epsilon$-insertion corresponds again to the Feynman contour of Fig.~\ref{fig:wintegral} around the branch cuts, signifying time-ordering indeed. In position space, we find
\bea
\vev{T\op(x)\op(0)} &=& \frac{1}{(2\pi)^d} \frac{i (-1)^{l}}{2^{2l-1} \Gamma(l)^2} \int e^{-i\omega t + i\vec k \cdot \vec x} q_\epsilon^{2l} \log q_\epsilon
\nonumber \\
&=& \frac{2l \Gamma(l+ d/2)}{\pi^{d/2}\Gamma(l)}\frac{1}{(-t^2 + \vec x^2 + i\epsilon)^{l + \frac{d}{2}}} \,,\label{eq:twopointpoinc}
\eea
and the $i\epsilon$-insertion agrees with \cite{Mack:1973mq}.

\paragraph{Normalization\\}
Let us compare the normalization of the time-ordered two-point function on the cylinder with that on two-dimensional Minkowski space. We start from the time-ordered two-point function on the cylinder:
\begin{align*}
\vev{T\op(x) \op(0)} &= \frac{l^2/(2^l \pi)}{[\cos(t - i\epsilon t) - \cos(\phi)]^{l+1}}\,, & ds^2 &= -dt^2 + d\phi^2\,,\\
\intertext{and apply the coordinate transformation $t = u-v$, $\phi = u +v$, after which we find}
\vev{T\op(x) \op(0)} &= \frac{l^2/(2^{2l+1} \pi)}{[\sin(u - i\eta)\sin(v+ i\eta)]^{l+1}}\,, & ds^2 &= 4 du dv\,,\\
\intertext{where now $\eta = \epsilon(u-v)$. We then Weyl transform and use covariance of the two-point function:}
\vev{T\op(x) \op(0)} &= \frac{2 l^2/\pi}{[\tan(u- i\eta)\tan(v+ i\eta)]^{l+1}}\,, & ds^2 &= \frac{du dv}{\cos^2(u)\cos^2(v)}\,,\\
\intertext{where we should remember that the two-point function is multiplied by two Weyl factors; one evaluated at $x$ and one at $0$. For $-\pi/2< u,v < \pi/2$, we can rewrite the denominator using
\[
\tan(u - i\eta) \tan(v + i\eta) = \tan(u)\tan(v) + i \epsilon (u-v)[\tan(u) - \tan(v)] = \tan(u)\tan(v) + i \epsilon'
\]
with $\epsilon'$ positive and constant. Finally, using $x+ y = \tan(u)$ and $x-y = \tan(v)$, we obtain}
\vev{T\op(x) \op(0)} &= \frac{2 l^2/\pi}{[-y^2 + x^2 + i\epsilon']^{l+1}}\,, & ds^2 &= -dy^2 + dx^2\,,
\end{align*}
and the normalization is indeed the same as in \eqref{eq:twopointpoinc} evaluated at $d=2$.

\subsection{Higher-point correlation functions}
\label{sec:higherpoint}
In this subsection, we briefly discuss how real-time higher-point correlation functions can be computed with our prescription. We take an interacting scalar field with potential
\be
V(\Phi) = \frac{1}{2}m^2 \Phi^2 + \frac{\lambda}{3} \Phi^3 + \ldots
\ee
so that the equation of motion becomes:
\be
\square \Phi - m^2 \Phi - \lambda \Phi^2 = 0\,.
\ee
This equation can be solved perturbatively. We first compute the sequence:
\be
\begin{split}
&\square \Phi_{\{0\}} - m^2 \Phi_{\{0\}}= 0\,,\\
&\square \Phi_{\{1\}} - m^2 \Phi_{\{1\}} = \lambda \Phi_{\{0\}}^2\,,\\
&\square \Phi_{\{2\}} - m^2 \Phi_{\{2\}} = \lambda \Phi_{\{1\}}^2\,,\\
&\ldots
\end{split}
\ee
where $\Phi_{\{0\}}$ satisfies the radial boundary data and $\Phi_{\{i\}}$ with $i \geq 1$ vanish asymptotically. The full solution is then obtained as:
\be
\Phi = \Phi_{\{0\}} + \Phi_{\{1\}} + \Phi_{\{2\}} + \ldots
\ee
To compute the series $\Phi_{\{i\}}$, we need to compute the bulk-bulk propagator $Z$. This propagator satisfies
\begin{equation}
\label{eq:bubu}
\square_G Z(x,x') = \frac{-1}{\sqrt{-G}}\delta^{d+1}(x - x')
\end{equation}
and vanishes asymptotically. In terms of $Z(x,x')$, we find:
\begin{equation}
\label{eq:bubuinaction}
\Phi_{\{i+1\}}(x) = \int_M d^{d+1}x' \sqrt{-G}  Z(x,x') \Phi_{\{i\}}(x')\,.
\end{equation}
In our case, the bulk manifold $M$ splits into multiple parts and we need to integrate the bulk-bulk propagator against $\Phi_{\{i\}}$ on the various segments. The bulk-bulk propagator therefore also splits in multiple components depending the segment that $x$ and $x'$ lie on. We will indicate this by a subscript. For example, $Z_{[12]}(x,x')$ denotes the bulk-bulk propagator with $x$ on $M_1$ and $x'$ on $M_2$. Equation \eqref{eq:bubuinaction} then becomes:
\begin{equation}
\label{eq:phijiplus1}
\Phi_{[j]\{i+1\}}(x) = \sum_k \int_{M_k} d^{d+1}x' \sqrt{-G}  Z_{[jk]}(x,x') \Phi_{[k]\{i\}}(x')\,,
\end{equation}
with the sum over all of the components $M_k$. Of course, $Z_{[jk]}$ is homogeneous on $M_j$ if $j\neq k$. Also, we will explicitly see below that $Z_{[jk]}(x,x') = Z_{[kj]}(x',x)$.

Let us now find this matrix of bulk-bulk propagators. These bulk-bulk propagators need to satisfy the matching conditions, since then so will all the $\Phi_{\{i\}}$ and consequently also $\Phi$. (Our derivation of the matching conditions for a scalar field in section \ref{sec:globalads3} was independent of the potential $V[\Phi]$, so the matching conditions are unchanged by the interaction terms.) For concreteness, consider the bulk spacetime of the previous subsection, with a Lorentzian segment $M_1$ sandwiched between two Euclidean segments $M_0$ and $M_2$. The matching conditions become important when we move $x$ from, say $M_1$ to $M_0$ while keeping $x'$ fixed. For example, we get
\be
\begin{split}
Z_{[11]}(t_1 = -T,\vec x,z;t'_1,\vec x',z') = Z_{[01]}(\tau_0 = 0,\vec x,z;t'_1,\vec x',z')\,,\\
-i\partial_t Z_{[11]}(t_1 = -T,\vec x,z;t'_1,\vec x',z') -\partial_\tau Z_{[01]}(\tau_0 = 0,\vec x,z;t'_1,\vec x',z') = 0\,,
\end{split}
\ee
just as in \eqref{eq:matchingpoinc}, and all the other matching conditions are similar.

The uniqueness of the bulk-bulk propagator is clear from the previous section, where we showed that there is no normalizable homogeneous solution that satisfies the matching conditions. As for existence, the bulk-bulk propagator for Lorentzian AdS in Poincar\'e coordinates is already known, see for example \cite{D'Hoker:2002aw} and the references therein, where one may find that
\be
Z_{[11]}(x,x') = Z[\xi_{11}]\,,
\ee
with $\xi_{11}$ an AdS-invariant function,
\begin{equation}
\label{eq:xi11}
\xi_{11} = \frac{(z - z')^2 - (t-t')^2 + (\vec x-\vec x')^2 + i\epsilon}{z^2 + z'^2 - (t-t')^2 + (\vec x - \vec x')^2}
\end{equation}
and $Z$ given by
\be
Z[\xi_{11}] = \frac{2^{-\Delta}\Gamma(\Delta)}{\pi^{d/2}\Gamma(\Delta - \frac{d}{2}) \Gamma(2\Delta - d)} \Big(1 - \xi_{11}\Big)^\Delta F(\frac{\Delta}{2},\frac{\Delta+1}{2};\Delta - \frac{d}{2}+1;[1 - \xi_{11}]^2)\,,
\ee
with $F(a,b;c;z)$ the hypergeometric function. This solution is regular except when $\xi_{11} \to 0$. Analytically continuing $t$, $t'$ to $M_0$ or $M_2$ by the replacement $t  = -T -i\tau_0$ or $t = T -i\tau_2$ yields other $\xi_{ij}$, and the $i\epsilon$-insertions again ensure that these $\xi_{ij}$ satisfy the matching conditions when either $x$ or $x'$ moves from one segment to the other. So if we define
\be
Z_{[ij]}(x,x') = Z[\xi_{ij}]\,,
\ee
then the various $Z_{[ij]}$ satisfy \eqref{eq:bubu}, the matching conditions, and vanish asymptotically. Therefore, the full matrix of bulk-bulk propagators can be obtained by this analytic continuation. Just as for the bulk-boundary propagator, the matching conditions uniquely fix the $i\epsilon$-insertions to be those in equation \eqref{eq:xi11}.

Again, these $i\epsilon$-insertions enter directly into the higher-point correlation functions. These are obtained as usual by further functional differentiation of the renormalized one-point function. For example, for a time-ordered vacuum-to-vacuum three-point function with all three arguments on $M_1$ we obtain
\be
\vev{T\op(x_1) \op(x_2) \op(x_3)} = (2\Delta -d) \frac{\delta^2 \phi_{(2\Delta-d)} (x_1)}{\delta \phi_{(0)}(x_2)\delta \phi_{(0)}(x_3)}\Big|_{\phi_{(0)}=0}\,,
\ee
with $\phi_{(2\Delta-d)}$ the coefficient of the normalizable mode (of
order $z^{\Delta}$) in the $z$-expansion of $\Phi$, and the source
$\phi_{(0)}$ should be set to zero after the functional
differentiation. Given the bulk solution, the procedure to obtain
these correlation functions is therefore just as for Euclidean
metrics, except for the replacement $t \to t - i\epsilon t$ (so $-t^2
\to -t^2 + i\epsilon$).

\subsection{Stationary black holes}
\label{sec:stationarybh}
The thermal contour drawn in Fig.~\ref{fig:contour}c admits another possible bulk solution, which corresponds to an eternal black hole. In this section, we will use this filling to compute the time-ordered two-point function for an operator dual to a free scalar field moving in the black hole background. We will again work in $d=2$, so the bulk spacetime is the static three-dimensional BTZ black hole. The rotating black hole will be discussed in the next subsection.

\begin{figure}
\centering
\psfrag{tL1}{$t_{L1}$}
\psfrag{tL2}{$t_{L2}$}
\psfrag{tR1}{$t_{R1}$}
\psfrag{tR2}{$t_{R2}$}
\psfrag{tE1}{$\tau_0$}
\psfrag{tE2}{$\tau_3$}
\includegraphics[width=8cm]{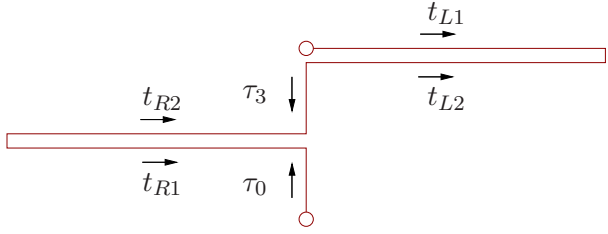}
\caption{\label{fig:bhcontour}The contour we use for the black hole. The circles should be identified.}
\end{figure}

Below, we will actually use the deformed contour of Fig.~\ref{fig:bhcontour} rather than the contour of Fig.~\ref{fig:contour}c. As we will see shortly, this has the advantage of `opening up' the second boundary of the black hole spacetime as well. In the next subsection, we describe a bulk manifold that fills in this deformed contour. Afterwards, we proceed by switching on a scalar field and holographically compute correlation functions.

\subsubsection{Bulk spacetime}
\begin{figure}
\centering
\psfrag{L}{$L$}
\psfrag{R}{$R$}
\psfrag{F}{$F$}
\psfrag{P}{$P$}
\psfrag{a}{(a)}
\psfrag{b}{(b)}
\psfrag{tl1}{$t_{L}$}
\psfrag{tr1}{$t_{R}$}
\includegraphics[width=10cm]{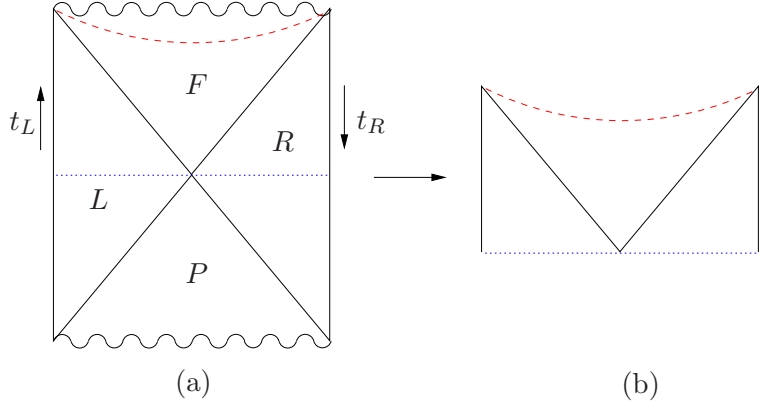}
\caption{\label{fig:btzpenrose}(a)
The Penrose diagram for the eternal BTZ black hole. The arrows indicate
the  direction of time. In the diagram,
  every point represents a circle. The horizons, which are the solid
  diagonal lines, separate the spacetime in four regions labelled by
  L, R, F and P. (b) We cut off the spacetime along the dotted lines
  and keep the part in between them.}
\end{figure}

Consider the eternal Lorentzian massive non-rotating BTZ black hole,
whose Penrose diagram is given in Fig.~\ref{fig:btzpenrose}a. The black hole splits into four parts, which we denote by L, R, F and P. On either part the
metric is
\be
ds^2 = -(r^2 -r_+^2)dt^2 + \frac{dr^2}{(r^2-r_+^2)} + r^2 d\phi^2\, .
\ee
If necessary, we will use a subscript like L or R to
indicate the corresponding part of the spacetime.
Notice that time runs backward on R. The mass and temperature
of the black hole are given by
\be
\label{eq:mt}
M = \frac{r_+^2}{8 G_3}\,, \qquad T = \frac{r_+}{2 \pi}\, .
\ee
(Recall that we set the AdS radius to one, $\ell^2=1$.)
To simplify the notation, we make the coordinate transformation
\be \label{r+}
t = \frac{t'}{r_+}\,, \qquad r = r' r_+\,, \qquad \phi = \frac{\phi'}{r_+}\,,
\ee
after which the metric reads
\be
\label{eq:lorbtz}
ds^2 = -(r^2 -1)dt^2 + \frac{dr^2}{(r^2-1)} + r^2 d\phi^2\,,
\ee
where we have dropped the primes. Note that the periodicity of $\phi$
has now changed to
\be \label{per_BTZ}
\phi \sim \phi + 2 \pi r_+\,.
\ee
At the very end of the computation we will return to standard
conventions.

To use this spacetime as a filling for (a part of) the contour of
Fig.~\ref{fig:bhcontour}, we first have to cut it off along an initial
slice, which we take to be the $t_L = t_R = 0$ slice, as well as a final slice,
which we choose to be the
$r_F = \hat r$ slice, with $\hat r < 1$ a constant. These segments are the
(blue) dotted and the (red) dashed lines of Fig.~\ref{fig:btzpenrose}a,
respectively. As is shown in Fig.~\ref{fig:btzpenrose}b, we keep the
segment in between these surfaces. Notice that $t_L > 0$ but $t_R < 0$
on this segment. We will need two copies of the segment, which we
denote by $M_1$ and $M_2$.

\begin{figure}
\centering
\psfrag{a}{(a)}
\psfrag{b}{(b)}
\psfrag{t}{$\tau$}
\psfrag{t=0}{$\tau=0$}
\psfrag{t=pi}{$\tau=\pi$}
\includegraphics[width=10cm]{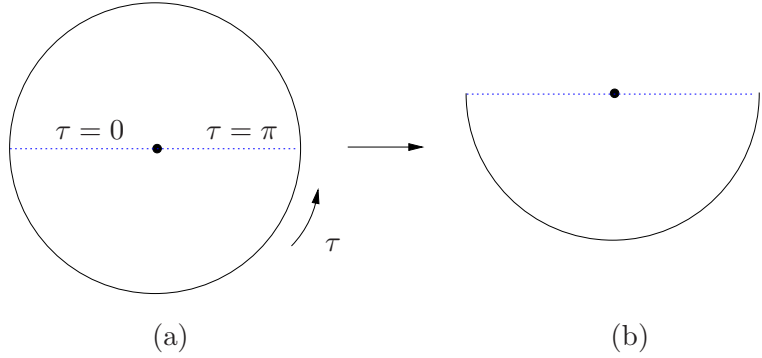}
\caption{\label{fig:euclbtz}(a) The Euclidean BTZ black hole, where again every point represents a circle. Euclidean time $\tau$ runs as indicated. (b) We cut off the spacetime along the dotted line given by $\tau = 0$ and $\tau = \pi$ and keep the lower part.}
\end{figure}

Next, consider the Euclidean solution with the metric
\be
\label{eq:eucbtz}
ds^2 = (r^2 -1)d\tau^2 + \frac{dr^2}{(r^2-1)} + r^2 d\phi^2
\ee
and with periodicities
\be
\begin{split}
\tau \sim \tau + 2\pi \,,\qquad \qquad \qquad
\phi \sim \phi + 2\pi r_+\,.
\end{split}
\ee
Topologically, this solution is $D_2 \times S^1$, with $D_2$ a two-dimensional disk and the $S^1$ is parametrized by $\phi$. As shown in Fig.~\ref{fig:euclbtz}, we will cut it in half along the hypersurface given by $\tau = 0$ and $\tau = \pi$, and keep the part given by $0 < \tau <\pi$. We will again need two copies of this part, which we denote as $M_0$ and $M_3$. In Fig.~\ref{fig:euclbtz}b, we have drawn these spacetimes as half a disk.

\begin{figure}
\centering
\psfrag{m0}{$M_0$}
\psfrag{m1}{$M_1$}
\psfrag{m2}{$M_2$}
\psfrag{m3}{$M_3$}
\psfrag{tl1}{$t_{L1}$}
\psfrag{tl2}{$t_{L2}$}
\psfrag{tr1}{$t_{R1}$}
\psfrag{tr2}{$t_{R2}$}
\psfrag{t0}{$\tau_0$}
\psfrag{t3}{$\tau_3$}
\includegraphics[width=10cm]{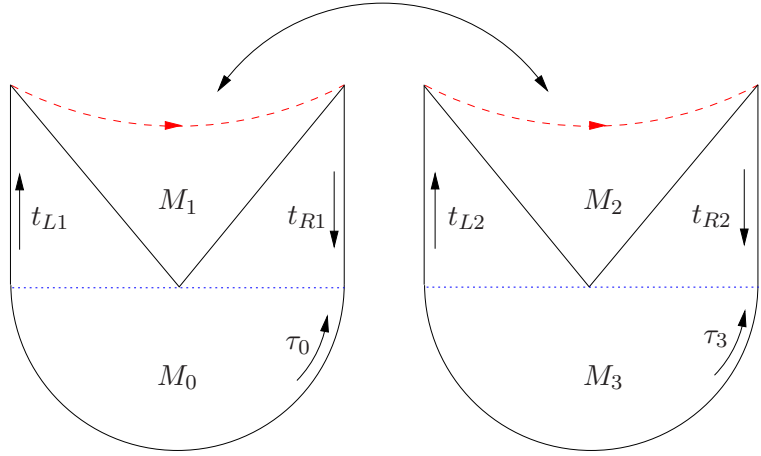}
\caption{\label{fig:filledbhcontour}The four components $M_0$, $M_1$, $M_2$ and $M_3$ are glued together to create a manifold that fills the contour of Fig.~\ref{fig:bhcontour}. The direction of the various time coordinates is the same as in Fig.~\ref{fig:bhcontour}.}
\end{figure}

We now glue the four manifolds together as shown in Fig.~\ref{fig:filledbhcontour}. Notice that $M_0$ is glued to $M_1$ such that the part with $\tau = 0$ is glued to the part with $t_L =0$, and the part with $\tau =\pi$ is glued to the part with $t_R = 0$. The same holds for the gluing between $M_2$ and $M_3$.

Let us verify that the matching conditions for gravity are
satisfied. First of all, the fact that $M_1$ and $M_2$ are identical
means that the matching conditions for gravity are trivially satisfied
along their gluing surface, which is the (red) dashed line in
Fig.~\ref{fig:filledbhcontour}. In fact, we could have glued $M_1$ and
$M_2$ along any spacelike bulk hypersurface extending all the way to
the two radial boundaries (and disjoint from the surfaces $t_L = t_R =
0$), and the matching conditions would still be satisfied.

For the matching between the Euclidean and the Lorentzian segments, one may directly see from the metrics \eqref{eq:lorbtz} and \eqref{eq:eucbtz} that any surface of constant $t$ or $\tau$ has the same induced metric. One may also use reflection and translation symmetry to find that the extrinsic curvature of such slices must vanish. Therefore, the matching conditions for gravity are satisfied for this gluing, too. Finally, by passing to a coordinate system that is regular everywhere at the gluing surface, one may verify that there are no problems at the coordinate singularity at $r = 1$, either.

Let us now turn to the matching conditions for a scalar field. The overall action \eqref{eq:Sfreescalar} can be split into a separate piece for each segment:
\begin{equation}
\label{eq:actionsbh}
iS_1 - iS_2 - S_0 -S_3\,.
\end{equation}
Continuity and the saddle-point approximation for the combination of actions \eqref{eq:actionsbh} determines the matching conditions to be:
\begin{align}
\Phi_1(r = \hat r) &= \Phi_2(r = \hat r) & i \partial_r \Phi_1(r=\hat r) - i \partial_r \Phi_2(r=\hat r) &= 0  \nonumber \\
\Phi_1(t_L = 0) &= \Phi_0(\tau = 0) & -i \partial_t \Phi_1(t_L=0) + \partial_\tau \Phi_0(\tau = 0) &= 0 \nonumber \\
\Phi_1(t_R = 0) &= \Phi_0(\tau = \pi) & i \partial_t \Phi_1(t_R= 0) - \partial_\tau \Phi_0(\tau = \pi) &= 0 \label{eq:matchingcondbh} \\
\Phi_2(t_L = 0) &= \Phi_3(\tau = 0) & i \partial_t \Phi_2(t_L = 0) + \partial_\tau \Phi_3(\tau = 0) &= 0 \nonumber \\
\Phi_2(t_R = 0) &= \Phi_3(\tau = \pi) & -i \partial_t \Phi_2(t_R = 0) - \partial_\tau \Phi_3(\tau = \pi) &= 0 \,.\nonumber
\end{align}
Incidentally, one may have wondered why the second set of horizontal
line segments in Fig.~\ref{fig:bhcontour} points to the left rather
than to the right. This can be seen from the matching conditions \eqref{eq:matchingcondbh}:
as one may verify they correspond to $C^1$
continuity in the complex time plane only if the contour has the shape
of Fig.~\ref{fig:bhcontour}. One may also verify that a
replacement $t_{R2} \to -t_{R2}$ has no effect on the \emph{shape} of
the contour.

\subsubsection{Mode solutions}
We can now turn to the computation of two-point functions. We start by finding mode solutions to the Klein-Gordon equation,
\be
\square_G \Phi - m^2 \Phi = 0\,,
\ee
on the various components. As usual, $m^2 = \Delta (\Delta -2)$ and we assume $\Delta = 1+l$ with $l \in \{0,1,2,\ldots\}$. In Lorentzian signature, we find two possible solutions, which we denote by $\psi_\pm$,
\be
\psi_\pm = e^{-i\omega t + ik\phi} f(\pm \omega,k,r)\,,
\ee
with a radial part given by
\begin{multline}
\label{eq:radialmodes}
f(\omega,k,r) = C_{\omega k l}\Big(1 -\frac{1}{r^2}\Big)^{i\omega/2} r^{-l-1} \\ \times F(\frac{i}{2}(\omega - k) + \frac{1}{2}(1+l), \frac{i}{2}(\omega + k) + \frac{1}{2}(1+l);i\omega + 1; 1 - \frac{1}{r^2})\,,
\end{multline}
with $F(a,b;c;z)$ a hypergeometric function and
 \be
C_{\omega k l} = \frac{\Gamma(\frac{i}{2}(\omega + k) + \frac{1}{2}(1+l))\Gamma(\frac{i}{2}(\omega - k) + \frac{1}{2}(1+l))}{\Gamma(i \omega + 1)\Gamma(l)}
 \ee
chosen such that the coefficient
of the leading behavior of $f(\pm \omega,k,r)$ as $r \to \infty$ equals one.
The asymptotic expansion of the modes is given by
\be
\label{eq:btzmodesasympt}
\psi_\pm = e^{-i \omega t + ik\phi} \Big(r^{l - 1} + \ldots + \alpha(\pm \omega,k,l) r^{-l-1}[\ln(r^2) + \beta(\pm \omega,k,l)] + \ldots \Big)\,,
\ee
with
\be
\begin{split}
\alpha(\omega,k,l) &= (-1)^l \frac{(\frac{i}{2}(\omega + k)+ \frac{1}{2}(1-l))_l(\frac{i}{2}(\omega - k)+ \frac{1}{2}(1-l))_l}{l!(l-1)!}\,,\\
\beta(\omega,k,l) &= - \psi(\frac{i}{2}(\omega + k) + \frac{1}{2}(1+l)) - \psi(\frac{i}{2}(\omega - k)+\frac{1}{2}(1+l)) + \text{local}\,,
\end{split}
\ee
where the local terms we omitted from $\beta(\omega,k,l)$
originate from the expansion of the prefactor $(1 -
1/r^2)^{i\omega/2}$ up to the relevant order.
Such local terms lead to contact terms in the two-point function
and will be omitted. The similarity between the modes
\eqref{eq:radialmodes} and \eqref{eq:radialmodesads} is not
accidental: one may verify that the backgrounds with the metrics
\eqref{eq:lorbtz} and \eqref{eq:metricemptyads3} are related by
analytic continuation in complex $(t,r,\phi)$-space, and so are the
corresponding mode solutions in these backgrounds. Since the behavior
of the modes in the interior of the spacetime is different, we will
not use this fact here.

Near the horizon both modes oscillate infinitely rapidly. To see this, we transform to Poincar\'e coordinates, given by
\be
\tanh (t) = -\frac{y}{x}\,, \qquad
r^2 = \frac{x^2 - y^2 + z^2}{z^2}\,, \qquad
e^{2\phi} = x^2 -y^2 + z^2\,,
\ee
which brings the metric to the form
\be
ds^2 = \frac{1}{z^2}(dx^2 - dy^2 + dz^2)\,.
\ee
With this definition the future and past horizons
on the L quadrant are mapped to
$x=-y$ and $x = y$, respectively.
Taking the near-horizon limit $x \pm y \to 0$, we find
\be
\label{eq:btzmodeshor}
\psi_{\pm} = C_{\omega k l}\exp\Big(\mp \frac{i\omega}{2}\ln(x \pm y)^2 + i (k \mp \omega)\phi\Big)(1 + \ldots)\,.
\ee
We can create modes that are well-defined almost everywhere on the
Lorentzian segments via analytic continuation across the horizons, in
the way specified by Unruh \cite{Unruh:1976db,birrell}, see also
\cite{Herzog:2002pc}. Depending on whether we analytically continue
from L to R via the lower or the upper half of the complex $y$
plane, an extra factor of $e^{\pi \omega}$ or $e^{-\pi \omega}$
should be added to the L mode to produce an R mode.
Since this is the
case for both $\psi_+$ and $\psi_-$, we find four different
combinations:
\be
\label{eq:lrmodes}
\begin{split}
\phi_{+ +} &=
\begin{cases}
e^{-i\omega t + ik\phi} f(\omega,k,r)& \text {on L}\\
e^{-i\omega t + ik\phi + \pi \omega} f(\omega,k,r)& \text{on R}
\end{cases}\\
\phi_{+ -} &=
\begin{cases}
e^{-i\omega t + ik\phi} f(\omega,k,r)& \text {on L}\\
e^{-i\omega t + ik\phi - \pi \omega} f(\omega,k,r)& \text{on R}
\end{cases}\\
\phi_{- +} &=
\begin{cases}
e^{-i\omega t + ik\phi} f(- \omega,k,r)& \text {on L}\\
e^{-i\omega t + ik\phi + \pi \omega} f(- \omega,k,r)& \text{on R}
\end{cases}\\
\phi_{- -} &=
\begin{cases}
e^{-i\omega t + ik\phi} f(- \omega,k,r)& \text {on L}\\
e^{-i\omega t + ik\phi - \pi \omega} f(- \omega,k,r)& \text{on R}\,.
\end{cases}\\
\end{split}
\ee
These modes form a complete set both on L and on R, and can thus be used to decompose any solution. In particular, solutions that are regular at the horizons can be obtained as an infinite sum over these modes.

Finally, on the Euclidean solutions $M_0$ and $M_3$, the mode solutions are as usual obtained by the replacement $t \to -i\tau$ in the $\psi_\pm$. In this case, there is no need for an analytic continuation, and we find two rather than four solutions, which we denote by $\phi_\pm$:
\be
\label{eq:btzeuclmodes}
\phi_\pm = e^{\omega \tau + ik\phi} f(\pm \omega,k,r)\,.
\ee
Going through the same arguments as before, one finds that these modes also oscillate infinitely fast near the horizon.

\subsubsection{No normalizable solution}
Let us now show the absence of a normalizable solution satisfying
the matching conditions. This would imply uniqueness of any solution
satisfying given radial boundary data.

We begin on $M_1$ where we write
\be
Y_1 = \sum_k \int d\omega (c_{[1]++}\phi_{++} + c_{[1]+-}\phi_{+-} + c_{[1]-+}\phi_{-+} + c_{[1]--}\phi_{--})\,,
\ee
with the $c_{[1]\pm \pm}$ some functions of $\omega$ and $k$. Notice that the sum is over $r_+ k \in \mathbb Z$ to comply with the periodicity \eqref{per_BTZ}. The solution looks different in the various regions. Using \eqref{eq:lrmodes} we obtain
\bea
Y_{1,L} &=& \sum_k \int d\omega e^{-i\omega t + ik \phi}[(c_{[1]++}+c_{[1]+-})f(\omega) +(c_{[1]-+}+c_{[1]--})f(-\omega)] \,,\nonumber
\\
Y_{1,R} &=& \sum_k \int d\omega e^{-i\omega t + ik\phi}[(c_{[1]++}e^{\pi \omega}+c_{[1]+-}e^{-\pi \omega})f(\omega)
\nonumber \\
&& \qquad \qquad
+(c_{[1]-+}e^{\pi \omega}+c_{[1]--}e^{-\pi \omega})f(-\omega)] \,,
\eea
where here and below we suppress the $k,r$ arguments from
$f(\omega,k,r)$  for notational simplicity.
By substituting the asymptotic behavior \eqref{eq:btzmodesasympt} of the modes, we find that $Y_{1}$ is normalizable on both L and R if
\be
\label{eq:btznormcond}
\begin{split}
c_{[1]++} + c_{[1]--} + c_{[1]+-} + c_{[1]-+} &= 0\,,\\
(c_{[1]++} + c_{[1]-+} )e^{\pi \omega} + (c_{[1]+-} + c_{[1]--}) e^{-\pi \omega}  &= 0\,.
\end{split}
\ee
Similarly, on $M_2$ we consider
\be
Y_2 = \sum_k \int d\omega (c_{[2]++}\phi_{++} + c_{[2]+-}\phi_{+-} + c_{[2]-+}\phi_{-+} + c_{[2]--}\phi_{--})
\ee
and the same argument as above leads to the the same conditions
\eqref{eq:btznormcond} but with $c_{[1]\pm\pm}$ replaced by  $c_{[2]\pm\pm}$.
Besides satisfying the
same radial boundary data, the matching conditions between $M_1$ and
$M_2$ imply that $Y_1$ and $Y_2$ also have the same initial
data on the matching surface. Since the
solution on either $M_1$ or $M_2$ is uniquely specified by boundary and
initial data, we find that $c_{[2]\pm \pm} =
c_{[1]\pm \pm}$.

On the Euclidean parts $M_0$ and $M_3$ the solution
should be a
linear combination of the Euclidean modes \eqref{eq:btzeuclmodes}.
We write it as
\be
\begin{split}
Y_0 &= \sum_k \int d\omega e^{\omega \tau_0 - ik \phi}[c_{[0]+}f(\omega) + c_{[0]-}f(-\omega)]\,,\\
Y_3 &= \sum_k \int d\omega e^{\omega \tau_3 - ik \phi}[c_{[3]+}f(\omega) + c_{[3]-}f(-\omega)]\,.
\end{split}
\ee
As for $Y_1$ and $Y_2$, the demand for normalizability implies
\be
c_{[0]+} + c_{[0]-}=0, \qquad c_{[3]+} + c_{[3]-}=0\,.
\ee

We now impose the matching conditions \eqref{eq:matchingcondbh}
between the Euclidean and the Lorentzian solution. Using the orthogonality
of the normalizable modes, this leads to algebraic relations between
the individual coefficients $c_{[i]\pm}$ and $c_{[j]\pm\pm}$.
In particular, the matching conditions between $M_0$ and $M_1$ determine
\be
c_{[1]+-} = 0\,,
\ee
while those between $M_2$ and $M_3$ fix
\be
c_{[1]-+} = 0\, .
\ee
Using \eqref{eq:btznormcond} we conclude that all the $c_{[1]\pm \pm} = 0$
and thus no normalizable solution exists.

\subsubsection{Bulk-boundary propagator}
We will now find the bulk-boundary propagator for a delta-function source at $(\hat t, \hat \phi)$ on the L part of $M_1$. Since we have just shown the absence of any normalizable solution, any bulk-boundary propagator that satisfies the matching conditions is guaranteed to be unique. Let us therefore make an educated guess and consider a solution $X_1$ on $M_1$ that contains only the modes $\phi_{++}$ and $\phi_{--}$:
\be
X_1 = \frac{1}{4\pi^2r_+}\sum_k \int d\omega e^{i\omega \hat t - ik\hat \phi}(a_{[1]++}\phi_{++} + a_{[1]--}\phi_{--})\,,
\ee
with new coefficients $a_{[1]\pm\pm}$ which are to-be determined functions of $\omega$ and $k$. Again, to comply with the periodicity of $\phi$ given in \eqref{per_BTZ}, we need $r_+ k \in \mathbb Z$ as well as the extra prefactor of $1/r_+$ to normalize the boundary delta function. Notice that we already split off a factor $e^{i\omega \hat t - ik\hat \phi}$ from $a_{[1]++}$ and $a_{[1]--}$. On the two regions R and L, our ansatz takes the following form:
\be
\label{eq:X1LR}
\begin{split}
X_{1,L} &= \frac{1}{4\pi^2 r_+}\sum_k \int d\omega e^{-i\omega (t-\hat t) + ik(\phi -\hat \phi)} [a_{[1]++}f(\omega) + a_{[1]--}f(-\omega)]\,,\\
X_{1,R} &= \frac{1}{4\pi^2 r_+}\sum_k \int d\omega e^{-i\omega (t-\hat t) + ik(\phi - \hat \phi)} [a_{[1]++}e^{\pi\omega}f(\omega) - a_{[1]--}e^{-\pi \omega}
f(-\omega)]\,.
\end{split}
\ee
As we mentioned above, we put a delta-function source on $(\hat t, \hat \phi)$ on the conformal boundary of L and no sources on the conformal boundary of R. Substituting the asymptotics \eqref{eq:btzmodesasympt}, such boundary conditions for $X_1$ lead to
\be
\begin{split}
a_{[1]++} + a_{[1]--} &= 1\,,\\
a_{[1]++}e^{\pi \omega} + a_{[1]--}e^{-\pi \omega} &= 0\,.\\
\end{split}
\ee
Notice that these conditions already fix the solution on $M_1$ to be:
\be
a_{[1]++} = \frac{-1}{e^{2\pi \omega}-1} \qquad \qquad a_{[1]--} = \frac{e^{2\pi \omega}}{e^{2\pi \omega}-1}\,.
\ee
In passing, we mention that it is not manifest that $X_1$ is finite at the horizons. To check this, one substitutes the near-horizon expansion \eqref{eq:btzmodeshor} of the modes and then computes the $\omega$-integral by contour deformation. One finds that an $i\epsilon$-insertion is necessary to ensure convergence and to regulate the lightcone singularity. (A subtle point is that the $a_{[1]++}$ and $a_{[1]--}$ both have a pole at $\omega = 0$, but the residues cancel each other so the contour can be freely deformed around this singularity.) The sum over $k$ can be computed using similar methods as we employ for the two-point function below and the computation then shows that after the $i\epsilon$ insertion $X_1$ is regular at all the horizons indeed. Notice that the light-cone singularity is expected; we found a similar singularity when we wrote down the position-space expression \eqref{eq:bbpoinc} in Poincar\'e coordinates. It can be removed by integrating the delta function on the boundary against a smooth source.

Let us now verify that we can find normalizable solutions on $M_0$, $M_2$ and $M_3$ such that the matching conditions are satisfied, so that $X_1$ is indeed
the bulk-boundary propagator on $M_1$.
We start with the matching solution $X_0$ on $M_0$.
It should be a linear combination of the modes $\phi_\pm$,
\be
X_0 = \frac{1}{4\pi^2 r_+}\sum_k \int d\omega e^{\omega \tau + ik \phi} e^{i\omega \hat t - ik\hat \phi} (a_{[0]+}f(\omega) + a_{[0]-} f(-\omega))\,.
\ee
Let us consider the following coefficients:
\be
a_{[0]+} = a_{[1]++} \,,\qquad \qquad a_{[0]-} = - a_{[1]++} = a_{[1]--} - 1\,,
\ee
with $a_{[1]\pm\pm}$ as given above. As one may directly verify by substituting
the asymptotic behavior \eqref{eq:btzmodesasympt} (now with $t = -i\tau$), the solution $X_0$ is normalizable since $a_{[0]+}+ a_{[0]-} = 0$. Notice furthermore that $0 < \tau < \pi$ on $M_0$. Therefore, despite the factor $e^{\omega \tau}$, the $\omega$-integral is still convergent along the real axis on $M_0$, because $a_{[0]\pm} \sim e^{-2\pi \omega}$ for large positive $\omega$.

To verify that the matching conditions are satisfied between $M_0$ and $M_1$, notice that the difference between the Euclidean and the Lorentzian solution on L,
\be
X_{1,L}(t=0) - X_0(\tau = 0) = \frac{1}{4\pi^2 r_+}\sum_k \int d\omega e^{i\omega \hat t - ik\hat \phi} \phi_{--} = 0
\ee
since $\hat t > 0$, so one can deform the contour in the upper half of the complex $\omega$-plane where $\phi_{--}$ has no poles even at normalizable order. (Actually, near the horizon, it is the oscillating behavior of radial part of the modes that determines where to deform the contour to. Since this is still the upper half plane, the difference vanishes there as well.) A similar argument
shows that the second matching condition on L as well as the both matching conditions on R are also satisfied.

Next we consider the solution on $M_2$,
\be
X_2 = \frac{1}{4\pi^2 r_+}\sum_k \!\int d\omega \,e^{-i\omega t_2 + ik \phi} e^{i\omega \hat t - ik\hat \phi} (a_{[2]++}\phi_{++} + a_{[2]+-}\phi_{+-} + a_{[2]-+} \phi_{-+} + a_{[2]--} \phi_{--})\,.
\ee
Since the radial boundary data on $M_1$ and $M_2$ are now different,
we cannot use the argument used earlier for the normalizable solution $Y$
to argue that $a_{[2]\pm\pm}$ is the same as $a_{[1]\pm\pm}$ and we
have to compute $a_{[2]\pm\pm}$.

To begin with, notice that the matching between $M_1$ and $M_2$ takes
places on the F component of the black hole as indicated in
Fig.~\ref{fig:btzpenrose}. Starting from the L quadrant one must
cross the future horizon but not the past horizon to arrive at the
F quadrant. Therefore, the modes $\phi_{++}$ and $\phi_{+-}$,
which become singular at the future horizon, acquire an additional
factor  of $e^{\pm \pi \omega}$ as we move from L to F. However, the
modes $\phi_{-\pm}$ become singular only at the past horizon and do
not get such a factor. These factors should be included both on $M_1$
and $M_2$ and show up in the matching conditions:
\be
\begin{split}
a_{[2]+-}e^{-\pi \omega} + a_{[2]++}e^{\pi \omega} &= a_{[1]+-}e^{-\pi \omega} + a_{[1]++}e^{\pi \omega} = \frac{-e^{\pi \omega}}{e^{2\pi \omega} -1}\,,\\
a_{[2]-+} + a_{[2]--} &= a_{[1]-+} + a_{[1]--} = \frac{e^{2\pi \omega}}{e^{2\pi \omega} -1}\,.
\end{split}
\ee
These two equations, together with those arising from normalizability on both sides of $M_2$, completely fix the $a_{[2]\pm \pm}$ to be:
\be
a_{[2]++} = 0\,, \qquad a_{[2]+-} = \frac{-e^{2\pi \omega}}{e^{2\pi \omega}-1}\,, \qquad a_{[2]-+} = 0\,, \qquad a_{[2]--} = \frac{e^{2\pi \omega}}{e^{2\pi \omega}-1}\,,
\ee
and we have found a normalizable solution $X_2$ on $M_2$ that matches to the solution $X_1$ on $M_1$.

Finally, we need to verify that we can obtain a normalizable solution $X_3$ on $M_3$ that matches to $X_2$. Since $X_2$, in contrast with $X_1$, is already fully normalizable, $X_3$ can be easily obtained by a simple analytic continuation of the solution on $X_2$. Just as for $M_0$, one may again verify that the $\omega$-integral in $X_3$ is convergent along the real axis, that $X_3$ is normalizable and that the matching conditions are satisfied.

Thus, the bulk-boundary propagator $X_1$ can be matched to
normalizable solutions on all segments. Since there are no
solutions that are everywhere normalizable, we have obtained \emph{the} bulk-boundary
propagator for the black
hole filling of the contour of Fig.~\ref{fig:bhcontour}. The same bulk-boundary propagator was actually written down in \cite{Herzog:2002pc}, where it was obtained by imposing boundary conditions at the horizon which are natural from considerations of quantum field theory in curved space \cite{birrell}. We have now derived that this is indeed the correct bulk-boundary propagator for the real-time gauge/gravity dictionary.

\subsubsection{Two-point functions}
We are mostly interested in the time-ordered and Wightman function for
real times. By looking at Fig.~\ref{fig:bhcontour}, we find that we
need operator insertions on the L component of either $M_1$ or
$M_2$, because these segments lie along the real time axis. To simplify the
notation we omit the subscript L, which should be understood in all
formulas in this subsection.

The one-point function in the presence of sources is again just the normalizable component $\phi_{(2l)}$ of the bulk-boundary propagator, times a factor $-2l$ which is fixed by holographic renormalization, see section \ref{ren}. Completely analogous to the analysis in section \ref{sec:globalads3}, this normalizable component $\phi_{(2l)}$ can be read off by substituting \eqref{eq:btzmodesasympt} in the solution $X_{1,L}$ or $X_{2,L}$. The two-point function computation is again completely analogous, and we find
\bea
\label{eq:timeorderedthermaltwopointfunction}
\vev{T \op(x)\op(x')} &=& \vev{T_C \op_{[1]}(x)\op_{[1]}(x')}
= \frac{li}{2\pi^2 r_+} \sum_k \int d\omega e^{-i\omega (t-t') + ik(\phi-\phi')}
\times \nonumber  \\ && [a _{[1]++} \alpha(\omega,k,l)\beta(\omega,k,l)
+ a_{[1]--}\alpha(-\omega,k,l)\beta(-\omega,k,l)] \,.
\eea
We recognize the structure of a time ordered propagator at finite temperature \cite{Landsman:1986uw}. Such a propagator is of the form
\be
\Delta(\omega,k) = - n(\omega) \Delta_A(\omega,k) + (1 + n(\omega)) \Delta_R(\omega,k)\,,
\ee
with $n(\omega)$ the Bose-Einstein distribution,
\be
n(\omega) = \frac{1}{e^{\beta \omega} - 1}\,,
\ee
and $\Delta_R$ and $\Delta_A$ are the retarded and advanced thermal propagators, which should be analytic functions in the respectively upper and the lower half of the complex $\omega$ plane, see appendix \ref{sec:app:fieldthy}. Since $\beta = 2\pi$ in our coordinates, we find $a_{[1]++} = - n(\omega)$. The structure of \eqref{eq:timeorderedthermaltwopointfunction} thus agrees with expectations.

To obtain a position-space expression, choose $t'=0$ and $t >0$. This allows us to perform the $\omega$-integral by deforming the contour to the lower half plane and picking up the poles. These poles come from $\beta(-\omega,k,l)$ and from the $a_{++}$ and the $a_{--}$. The former have poles at the quasinormal frequencies,
\be
\omega =\omega^\pm_{nk} \equiv -i(2n + l + 1) \pm k\,,
\ee
and the latter have poles at $\omega = -im$ with $m\in \{1,2,\ldots\}$ (the apparent pole at $\omega = 0$ in $\alpha_{++}$ and $\alpha_{--}$ has zero residue).

Afterwards, we compute the sum over $k$ as follows. We first use Poisson resummation to replace the sum by an integral and a sum over images $\phi \sim \phi + 2 \pi r_+ p$ with $p \in \mathbb Z$. The integral can again be done via contour deformation, replacing it by an infinite sum over residues as well. One then finds that the sum over the poles at $\omega = -im$ vanishes and we are left with the sum involving the quasinormal frequencies only,
\[
\frac{(-1)^{l+1} 2l}{\pi \Gamma(l)\Gamma(l+1) r_+}\sum_{\pm}\sum_{n=0}^\infty \sum_{m=1}^\infty
(\pm 1) e^{-i(2n+l+1 \pm m)(t-i\epsilon t) + i m \phi} \frac{\Gamma(1+n+l\pm m)\Gamma(1+n+l)}{\Gamma(1+n\pm m)\Gamma(1+n)}\,,
\]
where the $i\epsilon$ factor is uniquely fixed by requesting convergence away from contact points and
we suppressed the aforementioned sum over images. This expression can be evaluated without too much difficulty and adding the sum over images (remembering that the Poisson resummation yields an extra factor of $r_+$), we finally get
\be
\label{eq:btz2ptunn}
\vev{T \op(x)\op(0)} =  \sum_{m \in \mathbb Z} \frac{l^2/(2^l\pi)}{[-\cosh(t -i\epsilon t) + \cosh(\phi + 2m\pi r_+ )]^{l+1}}\,.
\ee

This computation was done using the metric in  (\ref{eq:lorbtz}) where the
mass of the BTZ entered through the periodicity of the angular
coordinate (\ref{per_BTZ}). To restore standard conventions,
we now perform the diffeomorphism $t \to r_+ t$ and
$\phi \to r_+ \phi$ followed by a Weyl transformation so that the boundary
background metric is $ds^2 = -dt^2 + d\phi^2$ with $\phi \sim \phi + 2\pi$.
Implementing these transformations in the two-point function
we obtain
\be
\vev{T \op(x)\op(0)} =  \sum_{m \in \mathbb Z} \frac{(2\pi T)^{4 l + 4}l^2 /(2^l \pi)}{[-\cosh(2 \pi T t -i\epsilon t)
+ \cosh(2 \pi T (\phi + 2m \pi))]^{l+1}}\,,
\ee
where we reinstated the temperature $T$ given in \eqref{eq:mt}.
This correlator satisfies the KMS condition and is a sum over images
in the $\phi$ direction. It was obtained earlier via an analytic continuation of the Euclidean correlator in \cite{KeskiVakkuri:1998nw}. As discussed in more detail in \cite{Birmingham:2002ph}, it is related to the thermal AdS two-point function by a double analytic continuation. This can directly seen from \eqref{eq:btz2ptunn}, where the substitution $t \to i\hat \phi$ and $\phi \to i\hat t$ yields precisely \eqref{2pt_EAdS} (up to $i\epsilon$ insertions which then have to be inserted by hand). This is the real-time manifestion of the fact that Euclidean thermal AdS$_3$ and Euclidean BTZ, which are both filled tori, are related by an S transformation of the boundary torus.

Let us also write down the Wightman function,
which can be obtained following the steps in section \ref{sec:wmfunction}:
\bea
\vev{\op(x)\op(x')} &=& \vev{T_C \op_{[2]}(x)\op_{[1]}(x')} = \frac{-li}{2\pi^2 r_+} \sum_k \int d\omega e^{-i\omega (t-t') + ik(\phi-\phi')}\times \nonumber \\ && [ a_{[2]+-} \alpha(\omega,k,l)\beta(\omega,k,l) + a_{[2]--}\alpha(-\omega,k,l)\beta(-\omega,k,l)]\,.
\eea
We can again obtain a position-space expression by closing the contour and picking up the poles, which results in
\be
\vev{\op(x)\op(0)} = \sum_{m \in \mathbb Z} \frac{(2\pi T)^{4 l + 4}l^2 /(2^l \pi)}{[-\cosh(2 \pi T t -i\epsilon) + \cosh(2\pi T(\phi + 2m \pi))]^{l+1}}\,.
\ee
Finally, the retarded two-point function is of the form
\be
\label{eq:retardedbh} 
i \Delta_R(x,0) \equiv \theta(x)\vev{[\op(x),\op(0)]} = \vev{T \op(x)\op(0)} - \vev{\op(0)\op(x)}\,.
\ee
From the above expressions, we find that it is analytic in the upper
half of the complex $\omega$-plane and vanishes for $t <
0$. Actually, it has support only on the forward lightcone, which
agrees with QFT expectations. Notice also that there is no need to insert
$i\epsilon$'s in the frequency-space expressions, since the poles in
the complex frequency plane all have non-zero imaginary part.
Such behavior however cannot arise in a CFT with a discrete energy
spectrum, at least at finite $N$, where one expects that retarded
correlators have  poles on the real axis.
Reconciling this behavior with expectations
from the AdS/CFT correspondence is still an open issue;
we refer to \cite{Maldacena:2001kr,Birmingham:2002ph}
for discussions of this point.

Let us finally remark that the retarded two-point function
\eqref{eq:retardedbh} can also be shown to be related to purely ingoing boundary
conditions at the horizon \cite{Herzog:2002pc}, leading eventually
to the recipe of \cite{Son:2002sd}. The derivation of this result 
from the current perspective is
presented in more detail in \cite{vanRees:2009rw}.

\subsection{Rotating black holes}
\label{sec:rotatingbh}

In the previous examples, we started with a CFT contour and obtained a corresponding bulk solution by the condition that it `filled' this contour.
In this section we will do the converse. We will start from a
Lorentzian solution and look for Euclidean solutions that can be
matched to it. This then leads to a specific CFT contour corresponding to
the combined solution.

Let us discuss the practical use of this procedure. As discussed in section
\ref{prescription}, the parts of the solution associated with vertical
segments of the contour
are directly related to the initial and final state or density matrix
of the field theory. The same information is also encoded in the
asymptotics of the Lorentzian solution, since from those one can compute the
holographic 1-point functions and from them in principle one can
reconstruct the dual state. The continuity of 1-point function
across the matching surfaces guarantees that the information
encoded in the Euclidean parts and the asymptotics of the Lorentzian
solution is indeed the same. Typically, it is not very easy
to extract the dual state starting from the vevs. The real-time
methods discussed here
present a new tool, namely given a Lorentzian solution
one looks for Euclidean solutions that can be matched to it. One
then uses this information to infer the holographic
interpretation of the solution.

In this subsection we illustrate how this is done
using the rotating BTZ black
hole \cite{btz1992,bhtz1993}. This
discussion readily generalizes to higher dimensional
rotating AdS-Kerr
black holes \cite{Carter:1968ks,Hawking:1998kw,Gibbons:2004js}.
As one may expect, the contour turns
out to be a thermal contour with a chemical
potential for angular momentum. Furthermore, this example
illustrates a number of additional issues as it provides
a concrete example of the use of a complex metric.

\subsubsection{Lorentzian solution}
The metric for the three-dimensional rotating BTZ black hole
\cite{btz1992,bhtz1993} is given by
\be
ds^2 = - (r^2 - r_+^2 - r_-^2) dt^2 + r^2 d\phi^2 + 2 r_+ r_- dt d\phi + \frac{r^2 dr^2}{(r^2 - r_+^2)(r^2 - r_-^2)}\,,
\ee
with $\phi$ periodic,
\be
\phi \sim \phi+ 2\pi\, .
\ee
The mass, angular momentum and temperature
of the black hole are related to $r_+$ and $r_-$ via
\be
M = \frac{r_+^2 + r_-^2}{8 G_3} \,,\qquad \qquad J = \frac{r_+ r_-}{4 G_3}\,,
\qquad \qquad T = \frac{r_+^2-r_-^2}{2 \pi r_+}\, .
\ee

It is convenient to use $(\hat t, \hat \phi, \hat r)$ coordinates:
\be
\begin{split}
\hat t &= \rpp t + \rmm \phi\,,\\
\hat \phi &= \rmm t + \rpp \phi\,,\\
\hat r^2 &= \frac{r^2 - r_-^2}{r_+^2 - r_-^2}\,.
\end{split}
\ee
Then the metric becomes
\begin{equation}
\label{eq:metricunitmassbtz}
ds^2 = -(\hat r^2 -1) d\hat t^2 + \frac{d\hat r^2}{\hat r^2-1} + \hat r^2 d\hat \phi^2\,,
\end{equation}
with the periodicity condition
\begin{equation}
\label{eq:periodicityunitmassbtz}
(\hat t,\hat \phi) \sim (\hat t + 2\pi \rmm, \hat \phi + 2\pi \rpp)\,,
\end{equation}
with $\rmm$ and $\rpp$ real, and we consider $0 \leq |\rmm| < \rpp$
but not the extremal case where $|\rmm| = \rpp$.

We consider an \emph{eternal} rotating BTZ black hole with two
radial boundaries. The rotating BTZ black
hole has a Penrose diagram that can be extended indefinitely to the
future and the past, across the various horizons \cite{bhtz1993}. We
will however cut off the spacetime along a spacelike hypersurface
extending from one radial boundary to another, just as for the static
BTZ example of the previous subsection. We thus explicitly avoid these
extra regions and the singularities.

\subsubsection{Euclidean solution}
To find a boundary contour corresponding to this spacetime, we will first look for a Euclidean solution that is to be matched to the Lorentzian solution across some initial hypersurface. Usually, in passing to the Euclidean version of a rotating black hole we not only make the replacement $t = -i\tau$, but also analytically continue the angular momentum parameter (which is $J$ or $\rmm$ in our case) to imaginary values. This way, the Euclidean metric one obtains is real. We will however show that the matching conditions are only satisfied for a complex Euclidean metric, given in coordinates $(\tau,r,\varphi)$ by
\begin{equation}
\label{eq:euclrotbtz}
ds^2 = (r^2 - r_+^2 - r_-^2) d\tau^2 + r^2 d\varphi^2 - 2 i r_+ r_- d\tau d\varphi + \frac{r^2 dr^2}{(r^2 - r_+^2)(r^2 - r_-^2)}\,,
\end{equation}
with coordinate ranges we make precise below. We now discuss this metric in more
detail. First of all, the Einstein equations are still satisfied for
the complex metric, since they are satisfied for any real or complex
$\rmm$. Second, a coordinate singularity arises at the
horizons. Insisting on a nondegenerate metric, a (complex) coordinate
transformation near the horizon shows that the necessary
periodicity in Euclidean time that avoids such a singularity is
\begin{equation}
\label{eq:complexperiodicity}
(\tau, \varphi) \sim (\tau + \frac{2\pi \rpp}{\rpp^2 - \rmm^2}, \varphi + \frac{2\pi i \rmm}{\rpp^2 - \rmm^2})\,,
\end{equation}
which notably involves a translation in the imaginary $\varphi$ direction. To comply with this periodicity, we will take the Euclidean manifold $M_0$ to be defined as follows. We first introduce $M_\Complex$ by extending the coordinates $(\tau,r,\varphi)$ to complex values, with the periodicities as above. The metric \eqref{eq:euclrotbtz} should then be seen as a nondegenerate holomorphic $(2,0)$-tensor on $M_\Complex$. Within $M_\Complex$, we take $M_0$ to be the submanifold given by real $\tau$ and $r$, but $\text{Im }\varphi = \tau (\rmm/\rpp)$. Notice that $M_0$ has three real dimensions. The metric restricts to $M_0$ as a complex tensor and the volume element is a three-form which we can integrate along $M_0$. If we introduce
\begin{equation}
\label{eq:torealeuclcoords}
\hat \varphi = \phi - \frac{i r_-}{r_+} \tau\,,
\end{equation}
then $\tau$, $r$ and $\hat \varphi$ are real on $M_0$ and therefore constitute an ordinary real coordinate system on $M_0$. In these coordinates the periodicity becomes
\be
(\tau, \hat \varphi) \sim (\tau + \frac{2\pi \rpp}{\rpp^2 - \rmm^2}, \hat \varphi)
\ee
and $\hat \varphi \sim \hat \varphi + 2\pi$ as well. However, the \emph{boundary} metric in $(\tau, \hat \varphi)$ coordinates is no longer diagonal. Since this will complicate the analysis below, we continue to use the complex $\varphi$ coordinate instead.

\subsubsection{Matching}
Let us now glue the `Euclidean' and the Lorentzian manifolds
together. We will first match the manifolds along a slice of
constant $t$ or $\tau$ away from the horizon. Afterwards, we will deal
with the subtleties introduced by the horizon.

We begin with the first matching condition. On the Lorentzian side, we find that the induced metric on a slice of constant $t$ is given by:
\be \label{ind}
\begin{split}
h_{AB}dx^A dx^B = r^2 d\phi^2 + \frac{r^2 dr^2}{(r^2 - r_+^2)(r^2 - r_-^2)}\,.
\end{split}
\ee
On the Euclidean side, we find exactly the same metric on a slice of constant $\tau$, with the replacement $\phi \to \varphi$, and therefore the first matching condition is satisfied. Let us now discuss the matching of the canonical
momenta.
On the Lorentzian side the extrinsic curvature (defined as usual with
a real outward pointing unit normal) is
\be
\begin{split}
^L\cK_{AB}dx^A dx^B &= \frac{2 \rpp \rmm}{\sqrt{(r^2 - r_+^2)(r^2 - r_-^2)}} dr d\phi
\end{split}
\ee
and on the Euclidean side we obtain
\be
^E\cK_{AB}dx^A dx^B = \frac{2 i \rpp \rmm}{\sqrt{(r^2 - r_+^2)(r^2 - r_-^2)}} dr d\phi\,.
\ee
A short computation then shows that the second matching condition is
satisfied as well, including the factor of $i$.

Notice that a
complex metric is already needed in the first matching condition,
\ie the continuity equation for the induced metric $h_{AB}$. Had we
continued $r_-$ to imaginary value
on the Euclidean side so that the bulk metric is real,
the induced metrics on the matching surface would not be the same
(because the factor $1/(r^2- r_-^2)$ in (\ref{ind}) would become
$1/(r^2+ r_-^2)$). The fact that the metric is complex is therefore not
directly related to the factor of $i$ appearing in the matching
conditions for the conjugate momenta.

The matching conditions should also be satisfied at the horizons.
Based on the example of the static BTZ black hole, one may
consider using  half a period of the Euclidean solution and
matching the surface given by $\tau = 0$ to $t_L = 0$ and the surface
given by $\tau = \pi r_+/(r_+^2 - r_-^2)$ to $t_R =
0$. However, moving from $\tau =0$ to $\tau =\pi r_+/(r_+^2 - r_-^2)$
on $M_0$ also involves an extra shift in the complex $\varphi$
direction. Therefore, a correct matching can be obtained by setting
$\phi_L = \varphi$ on the matching surface at L, and $\phi_R = \varphi
+ \pi i r_-/(r_+^2 - r_-^2)$ at R. One may then verify that the
matching condition are satisfied at the horizon by transforming to a
coordinate system that is nonsingular at the horizon. Since
$\partial_\varphi$ is a Killing vector, the matching conditions for
gravity are insensitive to the extra twist in $\varphi$. However, for
a scalar field the first matching condition becomes
\begin{equation}
\label{eq:matchingshiftphi}
\begin{split}
\Phi_0(\tau = 0, \varphi = 0,r) &=  \Phi_{1,L}(t_L = 0, \phi = 0,r)\,,\\
\Phi_0(\tau = \frac{\pi \rpp}{\rpp^2 - \rmm^2}, \varphi = \frac{\pi i \rmm}{\rpp^2 - \rmm^2},r) &=  \Phi_{1,R}(t_R = 0, \phi = 0,r)\,.
\end{split}
\end{equation}
which is clearly sensitive to the twist in $\varphi$.

\begin{figure}
\centering
\psfrag{It}{Im(t)}
\psfrag{Iph}{Im($\phi$)}
\psfrag{Rt}{Re(t)}
\includegraphics[width=8cm]{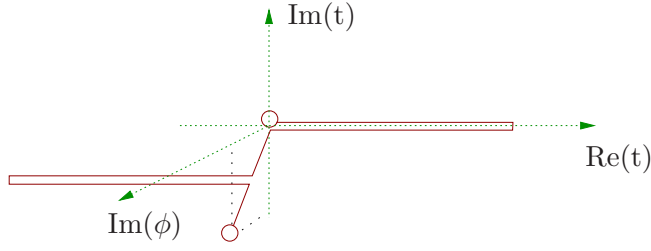}
\caption{\label{fig:rotbhcontour}The contour for the boundary CFT that corresponds to a rotating black hole does not only lie in the complex $t$ plane, but also extends into the complex $\phi$ plane. The circles should be identified.}
\end{figure}

We have shown that (half of) $M_0$ with the metric
\eqref{eq:euclrotbtz} can be matched to the Lorentzian rotating BTZ
black hole. We can therefore also glue two Euclidean and two
Lorentzian spacetimes in the same manner as shown in
Fig.~\ref{fig:filledbhcontour} for the BTZ black hole. Analyzing then
the boundary of this combination of spacetimes, we can finally read
off the boundary contour corresponding to the rotating BTZ black hole:
it is the contour of Fig.~\ref{fig:rotbhcontour}. This is the same
contour of Fig.~\ref{fig:bhcontour}, except that the vertical segments
involve a shift in the imaginary $\phi$ (or $\varphi$) direction of
total magnitude $2 \pi r_-/ (r_+^2 - r_-^2)$. Let us now interpret
this result in field theory.

First of all, notice that the \emph{boundary} metric on the vertical
segments is already complex, as can be verified by using real
coordinates for the boundary of $M_0$ via the coordinate
transformation \eqref{eq:torealeuclcoords}. This is in fact
consistent with the anticipated result that this contour corresponds
to a CFT at finite temperature and with non-zero chemical
potential for angular momentum.
Namely, for such an ensemble the density matrix is
\be
\rho = \exp(-\beta(H + \mu P_\phi))\,,
\ee
where $H$ is the Hamiltonian and $P_\phi$ is a translation in $\phi$. At the level of the path integral, such an ensemble corresponds to a contour that not only evolves in the imaginary time but also in the imaginary $\phi$ direction. From the periodicity \eqref{eq:complexperiodicity}, we immediately read off:
\be
\beta = \frac{2 \pi r_+}{r_+^2 - r_-^2}\qquad \qquad \mu = \frac{r_-}{r_+}\, .
\ee
Of course, if one works purely in Euclidean time, one may
also analytically continue $\mu$, $r_-$ and $J$ and then
both the boundary and the bulk metric would be real.
Our aim here was to develop a real-time formalism and this
led to complex metrics both in the boundary theory and in the bulk spacetime.

Let us finish this section with a brief comment on the use of complex
(but non-degenerate) metrics
in quantum gravity. First, it has been argued in the past
(see for example \cite{Halliwell:1989dy}) that
use of complex metrics after Wick rotation
might be essential for a path integral over
metrics. Second, saddle-point approximations often involve a deformation
of the integration contour to a point in the complex plane, even if
the integral originally is along the real axis. One particularly
elementary example where this happens is a discretized
vacuum-to-vacuum path integral for the harmonic oscillator, where the
initial and final vacuum wave functions require such a contour
deformation. Finally, complete reality of the bulk metric can no
longer be maintained when one studies perturbations, as the
$i\epsilon$ insertions that follow from our prescription necessarily
yield a complex graviton propagator.

\section{Summary of results and outlook}

We have presented a general prescription to holographically compute
real-time correlation functions within the
supergravity approximation. The main challenge in developing
such a real-time prescription, relative to Euclidean methods,
was to understand in detail how to deal with initial data. Our prescription
is a direct `holographic lift' of QFT real-time techniques to
the gravitational setting, namely there is a gravitational
counterpart of all QFT steps involved in such computations.
In more detail, in QFT one typically chooses
a contour in the complex time plane which usually
consists of a sequence
of horizontal (real) and vertical (imaginary) segments, the latter
being related to the choice of density matrix or initial/final state.
On the gravitational side, we construct solutions that directly correspond
to such QFT contours. Typically, real segments are associated with
Lorentzian solutions and vertical segments with Euclidean solutions, with appropriate matching conditions imposed on the joining surface.
The Euclidean parts
encode the initial and final state in the field theory and this is reflected in the bulk, where they can be thought of as Hartle-Hawking wave functions. These wave functions also provide the necessary initial and final data for the perturbations around a given supergravity background.

For the prescription to well defined, one must
establish that one can remove all infinities through a process of
(holographic) renormalization. Relative to the Euclidean discussion,
new infinities can appear at timelike infinity. In our setup
the analysis boils down to analyzing possible new contributions
from the joining surfaces. We show that no new counterterms are needed
and the holographic 1-point functions are continuous across the
matching surface. The continuity of the 1-point functions
is an important consistency condition of the entire setup:
as mentioned above, the Euclidean parts of the solution
are directly related to the
initial/final state but as is also well known the 1-point functions
encode the same information, too.

As a sidenote, the holographic
nature of the prescription also nicely shows up in the following issue that we encountered when demonstrating the renormalization.
Starting from a boundary state defined
at a boundary Cauchy surface, say the surface $t= t_0$, one
can extend this surface to the bulk, $t=f(r,\vec{x})$ with $f(r,\vec{x}) \to t_0$
as $r \to \infty$,
but clearly there is a certain amount of freedom of how this is done,
parametrized by the subleading behavior of $f(r,\vec{x})$.
These extensions are not part of the boundary theory, so the renormalized
theory should be independent of them. We explicitly find that possible
dependence on  $f(r,\vec{x})$ drops out indeed.

Having set up the prescription, we then moved on to demonstrate how to
apply it in a variety of examples that each illustrate different
points. The first example involved  the holographic computation of a vacuum
Wightman function. Although its
functional form was already known, we were able to compute it
completely holographically, without analytic continuation or insertion
of $i\epsilon$ by hand; instead, our computation provided for all the right signs and $i \epsilon$ insertions. We then computed a two-point function in real-time thermal
AdS. In this computation, we used the same Lorentzian background but different
initial data, which highlights the importance of properly defining the initial
and final boundary conditions.

The prescription can be used  to compute higher-point
functions as well and we explicitly demonstrated how to do 
such computations in an AdS background.
This discussion straightforwardly extends
to any other Asymptotically (locally) AdS bulk spacetime. 
It is worth mentioning that our prescription
resulted in a bulk-bulk propagator which is already of
quantum-mechanical nature (\ie Feynman rather than retarded). This
shows that the bulk fields are path-integral quantized and the
Euclidean caps provide the proper initial and final states. The
prescription thus naturally incorporates QFT in curved space and it is
not necessary to quantize perturbations by hand again.

A real-time thermal contour can also be `filled' with an eternal black
hole spacetime. Despite the presence of singularities and horizons, we
demonstrated how the initial and final conditions could again be
unambiguously specified via Euclidean caps. This procedure extends to
rotating black holes, where the analytic continuation is more subtle.
In our case, the reality conditions of the bulk
fields and factors of $i$ that arise in passing from real to imaginary
time agree with QFT arguments, where the situation is well-understood. In
particular, this procedure led to a
complex bulk (and boundary) metric in the case of a rotating black
hole.


The correlators we computed in the various examples were largely known
from earlier work, where they were obtained using special properties
of the backgrounds  and analytic continuation.
The emphasis here was on the coherent derivation of these
results using the new real-time prescription:
statistical factors and appropriate
$i \epsilon$ insertions in 2-point functions
all follow uniquely from solving the
matching conditions.

The true power of the new method however
should be in the applications that lie ahead of us.
Current and future applications of holography to RHIC and LHC physics
or to condensed matter systems require holographic modeling of
non-equilibrium phenomena and for such applications previous
methods are just not applicable.
On a more fundamental level, the new prescription may help us
addressing global issues and questions regarding the
holographic encoding of the bulk causal structure, including bulk
horizons, and the parts of spacetime beyond the horizon.
It would also be particularly interesting to extend the black hole analysis
of this paper to a collapsing shell of matter aiming at a
holographic description of the process of black hole formation.
Work about some of these issues is under way.

\section*{Acknowledgments}
We would like to thank Jan Smit for very illuminating conversations about
the subtleties of real-time quantum field theory.

\appendix
\section{Real-time quantum field theory}
\label{sec:app:fieldthy}
In this appendix, we discuss some aspects of real-time
quantum field theory relevant for our discussion.
The material presented here is not new and it is included
to make this paper self-contained.


\subsection{Vacuum wave function insertions}
In this section, we will analyze how the vacuum wave function insertions in the path integral lead to $i\epsilon$ insertions. In the main text, we mentioned how the wave functions can be obtained as path integrals along vertical segments in the complex time plane, leading ultimately to a contour as in Fig.~\ref{fig:contour}a. Let us begin by an explicit computation of these wave functions in a relatively simple case.

\subsubsection{Computation of the wave functions}
We will take a real free massive boson on flat Minkowski space $\Real^{1,d-1}$. As explained in the main text, the initial wave function $\inprod{\phi_-,-T}{\Omega}$ is computed via the projection:
\be
\lim_{\beta \to \infty} e^{\beta E_{vac}} \bra{\phi_-,-T} e^{-\beta \hat H} \ket{\Psi} = \inprod{\phi_-,-T}{\Omega} \inprod{\Omega}{\Psi}\,.
\ee
For simplicity, we shift the time coordinate such that $-T \to 0$ and we will take $\ket \Psi = \ket{\phi_\b,i \b}$ for some spatial field configuration $\phi_\b(x)$.

Since we take the field to be free, the path integral is Gaussian and
can be computed exactly. Let $\hat \phi(t,x)$ be the solution to the
equation of motion satisfying $\hat \phi(i\beta , x) = \phi_\b(x)$ and
$\hat \phi(0,x) = \phi_-(x)$.
We then obtain
\be
\inprod{\phi_-,0}{\Omega} = \lim_{\beta \to \infty} \cN e^{-S_E[\hat \phi]}\,,
\ee
with $\cN$ a normalization that does not depend on $\phi_-$ and $S_E$ the Euclidean on-shell action for the boson. Introducing a Euclidean time coordinate $\tau = it$, this action is given by:
\be
S_E[\phi] = \frac{1}{2}\int_{-\beta}^0 d\tau \int d^{d-1} x\, (\partial_\mu \phi \partial^\mu \phi + m^2 \phi^2)\,.
\ee
On-shell, it reduces to a surface integral,
\be
\inprod{\phi_-,0}{\Omega} = \lim_{\beta \to \infty} \cN \exp\Big( - \frac{1}{2}\int d^{d-1} x \, [\hat \phi(\tau,x) \partial_\tau \hat \phi(\tau,x)]^0_{\tau = -\beta}  \Big)\,.
\ee
Finding $\hat \phi$ is not hard and in the limit $\beta \to \infty$ we find that all dependence on $\phi_\beta$ can be absorbed in a shift of $\cN$
and  we recover the usual Gaussian wave function \cite{Weinberg:1995mt},
written in Fourier space as
\begin{equation}
\label{eq:wavefunction}
\inprod{\phi_-,0}{\Omega} = \cN' \exp\Big(-\frac{1}{2}\int \frac{d^{d-1}k}{(2\pi)^{d-1}} \, \phi_-(k) \omega_k \phi_-(-k) \Big)\,,
\end{equation}
with $\omega_k = \sqrt{k^2 + m^2}$. The conjugate final wave function $\inprod{\Omega}{\phi_+,T}$ can be computed using the same procedure,
leading to exactly the same result.

If interactions are switched on, the wave functions receive
corrections. However, as long as these interactions can be
switched off adiabatically for large times, the corrections
can also be ignored in the limit $t_i,t_f \to \infty$.
The analogous case in thermal field
theory, which we discuss below, is briefly discussed in \cite[section
2.4.1]{Landsman:1986uw}. For massless field theories there are
subtleties, but these considerations are not directly relevant for us
and they will not be discussed here. A computation of the
ground state wave function for electromagnetism and linearized
gravity can be found in \cite{Hartle:1984ke}.

\subsubsection{Effect of the wave function insertions}
Let us now show how the wave function insertions determine $i\epsilon$-insertions in the propagator. To this end, we introduce a source $J$ and compute
\be
Z[J] = \vev{\Omega|e^{-i \int J \phi}|\Omega}\,.
\ee
We suppose that the source vanishes smoothly at the endpoints $t = \pm T$ of the Lorentzian segment. Again via the usual slicing arguments, the path-integral representation one obtains is
\be
\begin{split}
Z[J] &= \int [\cD\phi] \exp \Big( iS[\phi] - i \int dt d^{d-1} x \, J \phi - \frac{1}{2} \int \frac{d^{d-1}k}{(2\pi)^{d-1}} \phi_-(k) \omega_k \phi_-(-k) \\ & \qquad - \frac{1}{2} \int \frac{d^{d-1}k}{(2\pi)^{d-1}} \phi_+(k) \omega_k \phi_+(-k)  \Big)\,,
\end{split}
\ee
where $\phi_\pm(k)$ is the Fourier transform of $\phi(\pm T,x)$ with respect to the spatial coordinates. Notice that the boundary values for the path integral $\int [\cD\phi]$ are not fixed.

To compute the path integral, we shift the integrand $\phi = \chi + \psi$,
where $\chi$ satisfies $\square \chi - m^2 \chi = J$ and $\psi$ is
the new integration variable. Notice that $\chi$ is not uniquely defined unless we specify some boundary conditions. To find these, notice that the aim of this shift is to get all the factors involving $J$ and $\chi$ to come out in front of the path integral, resulting in
\be
Z[J] = \cN \exp\Big( -\frac{i}{2} \int d^d x \, \chi J \Big)\,,
\ee
from which we would directly obtain the propagator as is shown below. However, an analysis of the boundary terms shows that such a factorization only occurs if one imposes additionally the two extra constraints:
\be
\begin{split}
- i\int d^{d-1}x \, \psi_- (x)\partial_t \chi (-T,x)  - \int \frac{d^{d-1}k}{(2\pi)^{d-1}} \psi_- (-k) \omega_k \chi_- (k) &= 0\,,\\
+ i\int d^{d-1}x \, \psi_+ (x)\partial_t \chi (T,x)  - \int \frac{d^{d-1}k}{(2\pi)^{d-1}} \psi_+ (-k) \omega_k \chi_+ (k) &= 0\,,\\
\end{split}
\ee
which should hold for all values of $\psi_\pm$. These conditions provide the boundary conditions for $\chi$. Since the source vanishes at the endpoints, $\chi$ is homogeneous for $t = \pm T$, and therefore has a Fourier expansion involving only modes of the form $e^{\mp i \omega_k t + i k x }$. The boundary conditions that one derives from these constraints are then simply that $\chi(-T,x)$ should contain only negative frequencies (\ie modes of the form $e^{-i\omega t + ikx}$ with $\omega < 0$) and $\chi(T,x)$ should contain only positive frequencies. But this uniquely fixes $\chi$ to be of the form
\be
\chi = \int dt' d^{d-1} x' \Delta_F(t-t',x-x') J(x')\,,
\ee
with $\Delta_F$ the Feynman propagator,
\begin{equation}
\label{eq:feynmanprop}
\Delta_F(t,x) = \int \frac{dt d^{d-1} x}{(2\pi)^d}\frac{e^{-i\omega t + i k x}}{-\omega^2 + k^2 - m^2 - i\epsilon}\,.
\end{equation}
As one may verify by contour deformation, one indeed obtains only positive/negative frequencies to the future/past of the source.

We can now take the limit $T \to \infty$. Assuming that the source
and any perturbatively added interactions vanish slowly at late
times, the propagator and the wave functions are unmodified and
all that is left are the $i\epsilon$-insertions which enter in the
perturbative expansion, which is precisely what we wanted to show.

Different (equivalent) arguments that translate wave functions
to $i\epsilon$ insertions can be found in the textbooks
\cite{Weinberg:1995mt} and \cite{Peskin:1995ev}. In particular
in \cite{Peskin:1995ev}, the
contour of Fig.~\ref{fig:contour}a is deformed to a straight line that
runs almost parallel to the real time axis, from $-T(1 - i\epsilon)$
to $T(1-i\epsilon)$, with $T \to \infty$. The projection property is
left unchanged and this way one still obtains vacuum-to-vacuum
amplitudes. The contour should always go downward or horizontal in the complex time plane so
that the operator $\exp(-i \hat H \Delta t)$ remains finite.

Finally, notice that the saddle-point $\chi$ is actually a
\emph{complex} solution, although we started with a real scalar field
and a real source $J(x)$. This can be viewed as a contour deformation
in field space before taking the saddle-point approximation.
Such a deformation is very explicit when one discretizes the path
integral. Nevertheless, the usual hermiticity constraints of $n$-point
functions are still satisfied. The fact that a saddle-point
approximation may involve complex fields holds for
gravity as well. In the context of holography, it is
the hermiticity of the boundary
stress energy tensor and its correlators that restricts
the allowed complex metrics.

\subsection{Contour time ordering}
The prescription above showed that time-ordered vacuum-to-vacuum correlation functions can be obtained via a path integral along a specific contour in the complex time plane. If the initial state is not an energy eigenstate then a corresponding path-integral formulation involves a so-called \emph{in-in contour} which runs from, say $t = 0$ to $t = T$ and back to $t=0$. At the endpoints of the contour, one may impose initial and final conditions on the fields corresponding to the state or ensemble under consideration.

As an example, suppose one wants to compute
\be
\vev{\Psi|\op(t')|\Psi} = \vev{\Psi|e^{i\hat H t'}\op e^{-i\hat H t'}|\Psi}\,.
\ee
We see that we first have to evolve the state $\ket{\Psi}$ for a time $t'$ before we insert the operator, but afterwards we also have to evolve \emph{back} in time before we insert the final wave function. This is the in-in or `closed time path' formalism of \cite{Schwinger:1960qe,Bakshi:1962dv,Bakshi:1963bn,Keldysh:1964ud}. Extending the contour to go beyond the point $t'$, say to some point $t''$, and then back again amounts to an insertion of the identity in the form $\exp[i\hat H (t''- t')] \exp[-i H (t'' -t')]$. Such an extension of the contour will not change the overall amplitude, which is something that is reflected in the dual gravity theory as well.

Another example where the in-in formalism is useful is real-time quantum field theory at finite temperature. In that case, the ensemble is described by a thermal density matrix,
\be
\hat \rho = \exp(-\beta \hat H)\,,
\ee
with $\beta = 1/T$ and $\hat H$ the Hamiltonian. Expectation values in such an ensemble are traces:
\be
\vev{\op}_\beta = \text{Tr}(e^{-\beta H} \op )\,.
\ee
One-point functions can be computed using the path-integral formalism
by taking a contour that runs straight down along the imaginary time
axis from $0$ to $-i\beta$, with (anti-)periodic boundary conditions
for the fields. However, a convenient way to obtain dynamical
information, \ie real-time correlation functions, is again via the
in-in formalism. In that case, one still rewrites the density matrix
as a Euclidean path integral, but the operator insertions at different
times force one to also path integrate along an in-in contour running
along the real time axis. The resulting contour is drawn in
Fig.~\ref{fig:contour}c on page \pageref{fig:contour}.

\subsubsection{Two-point functions}
For an in-in contour, one may insert operators along both the forward- and the backward-going segments of the contour. Via the usual slicing arguments, the correlation functions from a path integral along this contour are \emph{contour-time-ordered}. That is, if we pick a real `time' parameter $t_c$ that increases monotonically along the contour, then the correlation functions obtained from a path integral along this contour are ordered from small to large $t_c$. For example, let us take an in-in contour as in Fig.~\ref{fig:contour}b. The Lorentzian part runs up and then down the real $t$ axis, from $0$ to $T$ and then back to $0$. We denote the first segment by $C_1$ and the second by $C_2$. One may then choose $t_c = t$ on $C_1$ and $t_c = 2T - t$ on $C_2$.

Let us now introduce a source $J$ along the contour to compute the two-point functions. For an in-in contour, the source-operator coupling $-i \int_C dt J \op$ can be split in two parts and the partition function is defined as:
\be
Z[J_{[1]},J_{[2]}] = \vev{\exp\Big(-i \int_0^T dt_1 d^{d-1}x \, J_{[1]} \op_{[1]} + i \int_0^T dt_2 d^{d-1}x\, J_{[2]} \op_{[2]}\Big)}\,,
\ee
where a subscript in square brackets denotes the segment on which the field lives. The expectation values of course depend on the ensemble or state that is specified at $t=0$, but we will not write this explicitly.

Via functional differentiation one obtains four possible two-point functions,
\be
\vev{\cT_c \op_{[i]}(x)\op_{[j]}(x')} = (-1)^{\delta_{ij}}\frac{\delta^2 Z}{\delta J_{[i]}(x) \delta J_{[j]}(x')} \Big|_{J_{[1]} = J_{[2]} = 0}\,,
\ee
with $\cT_c$ denoting contour-time-ordering. Along the first segment contour-time-ordering coincides with normal time-ordering,
\be
\vev{\cT_c \op_{[1]}(x)\op_{[1]}(x')}=\vev{\cT \op(x)\op(x')}\,.
\ee
Along the second, backward-running segment, contour-time-ordering coincides with anti-time ordering, denoted by $\bar \cT$,
\be
\vev{\cT_c \op_{[2]}(x)\op_{[2]}(x')}=\vev{\bar \cT \op(x)\op(x')}\,.
\ee
If one puts one argument on the forward contour and the other on the backward contour, the latter one will always be later in contour time than the former and we get the Wightman functions:
\be
\begin{split}
\vev{\cT_c \op_{[1]}(x)\op_{[2]}(x')} &=  \vev{\op_{[2]}(x')\op_{[1]}(x)}  = \vev{\op(x')\op(x)}\,,\\
\vev{\cT_c \op_{[2]}(x)\op_{[1]}(x')} &=  \vev{\op_{[2]}(x)\op_{[1]}(x')}  = \vev{\op(x)\op(x')}\,.
\end{split}
\ee
Notice that the in-in path is also suitable to obtain vacuum-to-vacuum Wightman functions from a path integral. In the main text, we perform the corresponding holographic computation.

\subsubsection{Linear response}
Finally, we discuss the important role of the retarded two-point function, which describes the reaction of the system to an external perturbation. The perturbation can be described as a deformation of the theory such that the action $S$ changes to $S - \int J \op$. In the in-in formalism, the deformation should be present on both contours, so $J_{[1]} = J_{[2]} =  J$. Expanding then in $J$, we obtain the first-order response to the one-point function on $C_1$:
\be
\begin{split}
\delta \vev{\op(x')}_J &= \int_C dx J(x) \fdel{J(x)} \vev{\op_{[1]}(x')}_J\\
&= -i \int_0^T dt_1 \, J(x) \vev{\cT_c \op_{[1]}(x) \op_{[1]}(x')}
+ i \int_0^T dt_2\, J(x) \vev{\cT_c \op_{[2]}(x) \op_{[1]}(x')}\\
&= -i \int_0^T dt \, J(x) \vev{\cT \op(x) \op(x')} + i \int_{T}^{0} dt \, \vev{\op(x) \op(x')}\\
&= \int_0^T dt \, J(x) \Delta_R(x',x)\,,
\end{split}
\ee
where we suppressed the spatial integrations and $\Delta_R(x,x')$
is the retarded two-point function,
\be
\label{eq:deltar}
\Delta_R(x',x) = -i \theta(x'-x) \vev{[\op(x'),\op(x)]}\,,
\ee
which vanishes outside the future lightcone. The response is thus causal, as expected.

\subsection{$i\epsilon$-insertions}
In this section, we briefly review the analytic properties of
two-point functions and corresponding $i\epsilon$-insertions.

We start with the Wightman function
\be
\vev{\psi(x)\psi(0)}\,,
\ee
which is analytic in the lower half of the complex $t$ plane
\cite{Streater:1989vi}. The Wightman function can be obtained by the
replacement $-i\tau = t - i\epsilon$ in the Euclidean correlator,
because then the poles along the real $t$ axis are shifted into the
upper half of the complex $t$ plane. Its Fourier transform,
\be
\int dt d^{d-1} x \, e^{i\omega t - i k x} \vev{\psi(x)\psi(0)}\,,
\ee
vanishes for negative frequencies, since we can close the contour for
the $t$-integral via the lower half plane. Positivity of the spectral
density also implies that the Fourier transform is a real and positive
distribution for positive frequencies \cite{Streater:1989vi}. The
Fourier transform thus maps a function (or distribution) that is
analytic in a upper or lower half plane to a function that vanishes on
the right or the left real axis.

Next, the time-ordered two-point function is defined as
\be
\vev{\cT\psi(x)\psi(0)} = \theta(t)\vev{\psi(x)\psi(0)} + \theta(-t)\vev{\psi(0)\psi(x)}\,,
\ee
which can be obtained from the Euclidean correlator by the replacement $-i\tau = t - i\epsilon t$. Its poles are shifted into the upper half of the complex $t$ plane for $\text{Re }t > 0$ and in the lower half plane for $\text{Re }t < 0$. To obtain the Fourier transform, we close the contour in the appropriate half plane in the complex time plane. Picking up the poles, we find a sum over positive frequencies for $t > 0$ and one over negative frequencies for $t < 0$. This implies that we need the usual Feynman contour around the poles to define the inverse Fourier transform. One may replace $\omega \to \omega + i\epsilon \omega$ in the Fourier-space expression to explicitly indicate such a contour. Obviously
\be
-(\omega + i\epsilon\omega)^2 = -\omega^2 -i\epsilon
\ee
and for example the propagator \eqref{eq:feynmanprop} indeed has the required analyticity properties.

The retarded two-point function was defined in \eqref{eq:deltar}. Causality of the field theory determines that it vanishes completely outside the future lightcone. We may write it as an inverse Fourier transform:
\be
\label{eq:deltarfourier}
\Delta_R(x,0) = \frac{1}{(2\pi)^d}\int d\omega d^{d-1}k \, e^{-i\omega t+ i k x}\Delta_R(\omega,k)\,.
\ee
Notice that $\Delta_R(x,0)$ vanishes for $t < 0$. Since we can then close the $\omega$ integral in \eqref{eq:deltarfourier} in the upper half plane, we find that $\Delta_R(\omega,k)$ must be analytic in the upper half of the complex frequency plane. Finally, the advanced two-point function is the reversed retarded two-point function:
\be
\Delta_A(x,x') = \Delta_R(x',x)\,.
\ee
It therefore vanishes outside of the past lightcone and is analytic in the lower half of the complex frequency plane.

\providecommand{\href}[2]{#2}\begingroup\raggedright\endgroup


\begin{thebibliography}{10}

\bibitem{Maldacena:1997re}
J.~M. Maldacena, ``The large {N} limit of superconformal field theories and
  supergravity,'' {\em Adv. Theor. Math. Phys.} {\bf 2} (1998)  231--252,
\href{http://arxiv.org/abs/hep-th/9711200}{{\tt hep-th/9711200}}.

\bibitem{Gubser:1998bc}
S.~S. Gubser, I.~R. Klebanov, and A.~M. Polyakov, ``{Gauge theory correlators
  from non-critical string theory},'' {\em Phys. Lett.} {\bf B428} (1998)
  105--114,
\href{http://arxiv.org/abs/hep-th/9802109}{{\tt hep-th/9802109}}.

\bibitem{Witten:1998qj}
E.~Witten, ``{Anti-de Sitter space and holography},'' {\em Adv. Theor. Math.
  Phys.} {\bf 2} (1998)  253--291,
\href{http://arxiv.org/abs/hep-th/9802150}{{\tt hep-th/9802150}}.

\bibitem{Aharony:1999ti}
O.~Aharony, S.~S. Gubser, J.~M. Maldacena, H.~Ooguri, and Y.~Oz, ``{Large N
  field theories, string theory and gravity},''
  \href{http://dx.doi.org/10.1016/S0370-1573(99)00083-6}{{\em Phys. Rept.} {\bf
  323} (2000)  183--386},
\href{http://arxiv.org/abs/hep-th/9905111}{{\tt arXiv:hep-th/9905111}}.

\bibitem{D'Hoker:2002aw}
E.~D'Hoker and D.~Z. Freedman, ``{Supersymmetric gauge theories and the AdS/CFT
  correspondence},''
\href{http://arxiv.org/abs/hep-th/0201253}{{\tt arXiv:hep-th/0201253}}.

\bibitem{Skenderis:2002wp}
K.~Skenderis, ``Lecture notes on holographic renormalization,'' {\em Class.
  Quant. Grav.} {\bf 19} (2002)  5849--5876,
\href{http://arxiv.org/abs/hep-th/0209067}{{\tt hep-th/0209067}}.

\bibitem{us}
K.~Skenderis and B.~C. van Rees, ``{Real-time gauge/gravity duality},''
  \href{http://dx.doi.org/10.1103/PhysRevLett.101.081601}{{\em Phys. Rev.
  Lett.} {\bf 101} (2008)  081601},
\href{http://arxiv.org/abs/0805.0150}{{\tt arXiv:0805.0150 [hep-th]}}.

\bibitem{Son:2007vk}
D.~T. Son and A.~O. Starinets, ``{Viscosity, Black Holes, and Quantum Field
  Theory},''
  \href{http://dx.doi.org/10.1146/annurev.nucl.57.090506.123120}{{\em Ann. Rev.
  Nucl. Part. Sci.} {\bf 57} (2007)  95--118},
\href{http://arxiv.org/abs/0704.0240}{{\tt arXiv:0704.0240 [hep-th]}}.

\bibitem{Balasubramanian:1998sn}
V.~Balasubramanian, P.~Kraus, and A.~E. Lawrence, ``{Bulk vs. boundary dynamics
  in Anti-de Sitter spacetime},'' {\em Phys. Rev.} {\bf D59} (1999)  046003,
\href{http://arxiv.org/abs/hep-th/9805171}{{\tt hep-th/9805171}}.

\bibitem{Balasubramanian:1998de}
V.~Balasubramanian, P.~Kraus, A.~E. Lawrence, and S.~P. Trivedi, ``{Holographic
  probes of Anti-de Sitter space-times},'' {\em Phys. Rev.} {\bf D59} (1999)
  104021,
\href{http://arxiv.org/abs/hep-th/9808017}{{\tt hep-th/9808017}}.

\bibitem{Maldacena:2001kr}
J.~M. Maldacena, ``{Eternal black holes in Anti-de-Sitter},'' {\em JHEP} {\bf
  04} (2003)  021,
\href{http://arxiv.org/abs/hep-th/0106112}{{\tt hep-th/0106112}}.

\bibitem{Son:2002sd}
D.~T. Son and A.~O. Starinets, ``{Minkowski-space correlators in AdS/CFT
  correspondence: Recipe and applications},'' {\em JHEP} {\bf 09} (2002)  042,
\href{http://arxiv.org/abs/hep-th/0205051}{{\tt hep-th/0205051}}.

\bibitem{Herzog:2002pc}
C.~P. Herzog and D.~T. Son, ``{Schwinger-Keldysh propagators from AdS/CFT
  correspondence},'' {\em JHEP} {\bf 03} (2003)  046,
\href{http://arxiv.org/abs/hep-th/0212072}{{\tt hep-th/0212072}}.

\bibitem{Satoh:2002bc}
Y.~Satoh and J.~Troost, ``{On time-dependent AdS/CFT},'' {\em JHEP} {\bf 01}
  (2003)  027,
\href{http://arxiv.org/abs/hep-th/0212089}{{\tt hep-th/0212089}}.

\bibitem{Kraus:2002iv}
P.~Kraus, H.~Ooguri, and S.~Shenker, ``{Inside the horizon with AdS/CFT},''
  {\em Phys. Rev.} {\bf D67} (2003)  124022,
\href{http://arxiv.org/abs/hep-th/0212277}{{\tt hep-th/0212277}}.

\bibitem{Marolf:2004fy}
D.~Marolf, ``{States and boundary terms: Subtleties of Lorentzian AdS/CFT},''
  {\em JHEP} {\bf 05} (2005)  042,
\href{http://arxiv.org/abs/hep-th/0412032}{{\tt hep-th/0412032}}.

\bibitem{Festuccia:2005pi}
G.~Festuccia and H.~Liu, ``{Excursions beyond the horizon: Black hole
  singularities in Yang-Mills theories. I},'' {\em JHEP} {\bf 04} (2006)  044,
\href{http://arxiv.org/abs/hep-th/0506202}{{\tt arXiv:hep-th/0506202}}.

\bibitem{Lawrence:2006ze}
A.~Lawrence and A.~Sever, ``{Holography and renormalization in Lorentzian
  signature},'' {\em JHEP} {\bf 10} (2006)  013,
\href{http://arxiv.org/abs/hep-th/0606022}{{\tt arXiv:hep-th/0606022}}.

\bibitem{Iqbal:2008by}
N.~Iqbal and H.~Liu, ``{Universality of the hydrodynamic limit in AdS/CFT and
  the membrane paradigm},''
\href{http://arxiv.org/abs/0809.3808}{{\tt arXiv:0809.3808 [hep-th]}}.

\bibitem{Hartle:1976tp}
J.~B. Hartle and S.~W. Hawking, ``{Path Integral Derivation of Black Hole
  Radiance},''
\href{http://dx.doi.org/10.1103/PhysRevD.13.2188}{{\em Phys. Rev.} {\bf D13}
  (1976)  2188--2203}.

\bibitem{Unruh:1976db}
W.~G. Unruh, ``Notes on black hole evaporation,''
{\em Phys. Rev.} {\bf D14} (1976)  870.

\bibitem{deHaro:2000xn}
S.~de~Haro, S.~N. Solodukhin, and K.~Skenderis, ``{Holographic reconstruction
  of spacetime and renormalization in the AdS/CFT correspondence},''
  \href{http://dx.doi.org/10.1007/s002200100381}{{\em Commun. Math. Phys.} {\bf
  217} (2001)  595--622},
\href{http://arxiv.org/abs/hep-th/0002230}{{\tt arXiv:hep-th/0002230}}.

\bibitem{Bianchi:2001de}
M.~Bianchi, D.~Z. Freedman, and K.~Skenderis, ``{How to go with an RG flow},''
  {\em JHEP} {\bf 08} (2001)  041,
\href{http://arxiv.org/abs/hep-th/0105276}{{\tt arXiv:hep-th/0105276}}.

\bibitem{Bianchi:2001kw}
M.~Bianchi, D.~Z. Freedman, and K.~Skenderis, ``{Holographic
  renormalization},'' {\em Nucl. Phys.} {\bf B631} (2002)  159--194,
\href{http://arxiv.org/abs/hep-th/0112119}{{\tt arXiv:hep-th/0112119}}.

\bibitem{Karch:2005ms}
A.~Karch, A.~O'Bannon, and K.~Skenderis, ``{Holographic renormalization of
  probe D-branes in AdS/CFT},'' {\em JHEP} {\bf 04} (2006)  015,
\href{http://arxiv.org/abs/hep-th/0512125}{{\tt arXiv:hep-th/0512125}}.

\bibitem{Skenderis:2006uy}
K.~Skenderis and M.~Taylor, ``{Kaluza-Klein holography},'' {\em JHEP} {\bf 05}
  (2006)  057,
\href{http://arxiv.org/abs/hep-th/0603016}{{\tt arXiv:hep-th/0603016}}.

\bibitem{Schwinger:1960qe}
J.~S. Schwinger, ``{Brownian motion of a quantum oscillator},''
{\em J. Math. Phys.} {\bf 2} (1961)  407--432.

\bibitem{Bakshi:1962dv}
P.~M. Bakshi and K.~T. Mahanthappa, ``{Expectation value formalism in quantum
  field theory. 1},''
{\em J. Math. Phys.} {\bf 4} (1963)  1--11.

\bibitem{Bakshi:1963bn}
P.~M. Bakshi and K.~T. Mahanthappa, ``{Expectation value formalism in quantum
  field theory. 2},''
{\em J. Math. Phys.} {\bf 4} (1963)  12--16.

\bibitem{Keldysh:1964ud}
L.~V. Keldysh, ``{Diagram technique for nonequilibrium processes},''
{\em Zh. Eksp. Teor. Fiz.} {\bf 47} (1964)  1515--1527 [{\it Sov.\ Phys.\ JETP}
  {\bf 20} (1965) 1018].

\bibitem{Hartle:1983ai}
J.~B. Hartle and S.~W. Hawking, ``{Wave Function of the Universe},''
{\em Phys. Rev.} {\bf D28} (1983)  2960--2975.

\bibitem{Halliwell:1989dy}
J.~J. Halliwell and J.~B. Hartle, ``{Integration Contours for the No Boundary
  Wave Function of the Universe},''
{\em Phys. Rev.} {\bf D41} (1990)  1815.

\bibitem{Henningson:1998gx}
M.~Henningson and K.~Skenderis, ``{The holographic Weyl anomaly},'' {\em JHEP}
  {\bf 07} (1998)  023,
\href{http://arxiv.org/abs/hep-th/9806087}{{\tt arXiv:hep-th/9806087}}.

\bibitem{Henningson:1998ey}
M.~Henningson and K.~Skenderis, ``{Holography and the Weyl anomaly},'' {\em
  Fortsch. Phys.} {\bf 48} (2000)  125--128,
\href{http://arxiv.org/abs/hep-th/9812032}{{\tt arXiv:hep-th/9812032}}.

\bibitem{Balasubramanian:1999re}
V.~Balasubramanian and P.~Kraus, ``{A stress tensor for Anti-de Sitter
  gravity},'' \href{http://dx.doi.org/10.1007/s002200050764}{{\em Commun. Math.
  Phys.} {\bf 208} (1999)  413--428},
\href{http://arxiv.org/abs/hep-th/9902121}{{\tt arXiv:hep-th/9902121}}.

\bibitem{Papadimitriou:2004ap}
I.~Papadimitriou and K.~Skenderis, ``{AdS / CFT correspondence and geometry},''
  {\em Proceedings Strasbourg 2003, AdS/CFT correspondence, ed. O. Biquard}
  73,
\href{http://arxiv.org/abs/hep-th/0404176}{{\tt hep-th/0404176}}.

\bibitem{Papadimitriou:2005ii}
I.~Papadimitriou and K.~Skenderis, ``{Thermodynamics of asymptotically locally
  AdS spacetimes},'' {\em JHEP} {\bf 08} (2005)  004,
\href{http://arxiv.org/abs/hep-th/0505190}{{\tt hep-th/0505190}}.

\bibitem{Hayward:1993my}
G.~Hayward, ``{Gravitational action for space-times with nonsmooth
  boundaries},''
\href{http://dx.doi.org/10.1103/PhysRevD.47.3275}{{\em Phys. Rev.} {\bf D47}
  (1993)  3275--3280}.

\bibitem{Hawking:1996ww}
S.~W. Hawking and C.~J. Hunter, ``{The Gravitational Hamiltonian in the
  Presence of Non- Orthogonal Boundaries},''
  \href{http://dx.doi.org/10.1088/0264-9381/13/10/012}{{\em Class. Quant.
  Grav.} {\bf 13} (1996)  2735--2752},
\href{http://arxiv.org/abs/gr-qc/9603050}{{\tt arXiv:gr-qc/9603050}}.

\bibitem{Brown:2000dz}
J.~D. Brown, S.~R. Lau, and J.~York, ``{Action and Energy of the Gravitational
  Field},''
\href{http://arxiv.org/abs/gr-qc/0010024}{{\tt arXiv:gr-qc/0010024}}.

\bibitem{Papadimitriou:2004rz}
I.~Papadimitriou and K.~Skenderis, ``{Correlation functions in holographic RG
  flows},'' \href{http://dx.doi.org/10.1088/1126-6708/2004/10/075}{{\em JHEP}
  {\bf 10} (2004)  075},
\href{http://arxiv.org/abs/hep-th/0407071}{{\tt arXiv:hep-th/0407071}}.

\bibitem{Birmingham:2002ph}
D.~Birmingham, I.~Sachs, and S.~N. Solodukhin, ``{Relaxation in conformal field
  theory, Hawking-Page transition, and quasinormal/normal modes},''
  \href{http://dx.doi.org/10.1103/PhysRevD.67.104026}{{\em Phys. Rev.} {\bf
  D67} (2003)  104026},
\href{http://arxiv.org/abs/hep-th/0212308}{{\tt arXiv:hep-th/0212308}}.

\bibitem{Imbimbo:1999bj}
C.~Imbimbo, A.~Schwimmer, S.~Theisen, and S.~Yankielowicz, ``{Diffeomorphisms
  and holographic anomalies},''
  \href{http://dx.doi.org/10.1088/0264-9381/17/5/322}{{\em Class. Quant. Grav.}
  {\bf 17} (2000)  1129--1138},
\href{http://arxiv.org/abs/hep-th/9910267}{{\tt arXiv:hep-th/9910267}}.

\bibitem{Skenderis:2000in}
K.~Skenderis, ``{Asymptotically Anti-de Sitter spacetimes and their stress
  energy tensor},'' {\em Int. J. Mod. Phys.} {\bf A16} (2001)  740--749,
\href{http://arxiv.org/abs/hep-th/0010138}{{\tt hep-th/0010138}}.

\bibitem{Mack:1973mq}
G.~Mack and I.~T. Todorov, ``{Conformal-invariant green functions without
  ultraviolet divergences},''
\href{http://dx.doi.org/10.1103/PhysRevD.6.1764}{{\em Phys. Rev.} {\bf D6}
  (1973)  1764--1787}.

\bibitem{birrell}
N.~Birrel and P.~Davies, {\em Quantum Fields in Curved Space}.
\newblock Cambridge University Press, 1982.

\bibitem{Landsman:1986uw}
N.~P. Landsman and C.~G. van Weert, ``{Real and Imaginary Time Field Theory at
  Finite Temperature and Density},''
\href{http://dx.doi.org/10.1016/0370-1573(87)90121-9}{{\em Phys. Rept.} {\bf
  145} (1987)  141}.

\bibitem{KeskiVakkuri:1998nw}
E.~Keski-Vakkuri, ``{Bulk and boundary dynamics in BTZ black holes},'' {\em
  Phys. Rev.} {\bf D59} (1999)  104001,
\href{http://arxiv.org/abs/hep-th/9808037}{{\tt hep-th/9808037}}.

\bibitem{vanRees:2009rw}
B.~C. van Rees, ``{Real-time gauge/gravity duality and ingoing boundary
  conditions},''
\href{http://arxiv.org/abs/0902.4010}{{\tt arXiv:0902.4010 [hep-th]}}.

\bibitem{btz1992}
M.~Ba{\~n}ados, C.~Teitelboim, and J.~Zanelli, ``The black hole in
  three-dimensional space-time,'' {\em Phys. Rev. Lett.} {\bf 69} (1992)
  1849--1851,
\href{http://arxiv.org/abs/hep-th/9204099}{{\tt hep-th/9204099}}.

\bibitem{bhtz1993}
M.~Ba{\~n}ados, M.~Henneaux, C.~Teitelboim, and J.~Zanelli, ``Geometry of the
  2+1 black hole,'' {\em Phys. Rev. D} {\bf 48} (1993)  1506--1525,
\href{http://arxiv.org/abs/gr-qc/9302012}{{\tt gr-qc/9302012}}.

\bibitem{Carter:1968ks}
B.~Carter, ``{Hamilton-Jacobi and Schrodinger separable solutions of Einstein's
  equations},''
{\em Commun. Math. Phys.} {\bf 10} (1968)  280.

\bibitem{Hawking:1998kw}
S.~W. Hawking, C.~J. Hunter, and M.~Taylor, ``{Rotation and the AdS/CFT
  correspondence},'' \href{http://dx.doi.org/10.1103/PhysRevD.59.064005}{{\em
  Phys. Rev.} {\bf D59} (1999)  064005},
\href{http://arxiv.org/abs/hep-th/9811056}{{\tt arXiv:hep-th/9811056}}.

\bibitem{Gibbons:2004js}
G.~W. Gibbons, H.~Lu, D.~N. Page, and C.~N. Pope, ``{Rotating black holes in
  higher dimensions with a cosmological constant},''
  \href{http://dx.doi.org/10.1103/PhysRevLett.93.171102}{{\em Phys. Rev. Lett.}
  {\bf 93} (2004)  171102},
\href{http://arxiv.org/abs/hep-th/0409155}{{\tt arXiv:hep-th/0409155}}.

\bibitem{Weinberg:1995mt}
S.~Weinberg, ``{The Quantum theory of fields. Vol. 1: Foundations},''.
  Cambridge, UK: Univ. Pr. (1995) 609 p.

\bibitem{Hartle:1984ke}
J.~B. Hartle, ``{Ground state wave function of linearized gravity},''
\href{http://dx.doi.org/10.1103/PhysRevD.29.2730}{{\em Phys. Rev.} {\bf D29}
  (1984)  2730--2737}.

\bibitem{Peskin:1995ev}
M.~E. Peskin and D.~V. Schroeder, ``{An Introduction to quantum field
  theory},''. Reading, USA: Addison-Wesley (1995) 842 p.

\bibitem{Streater:1989vi}
R.~F. Streater and A.~S. Wightman, ``{PCT, spin and statistics, and all
  that},''. Redwood City, USA: Addison-Wesley (1989) 207 p. (Advanced book
  classics).

\end{thebibliography}
\end{document}